\documentclass[reprint,final,nofootinbib]{revtex4-1}
\pdfoutput=1

\usepackage{amssymb} 
\usepackage{amsthm}
\usepackage{amsmath}
\usepackage{graphicx}
\usepackage{subcaption}
\usepackage{dcolumn}
\usepackage{bm}
\usepackage{cancel}
\usepackage{float}
\usepackage{epstopdf}
\usepackage{footnote}
\usepackage{url}
\usepackage[svgnames] {xcolor}
\usepackage{hyperref}

\hypersetup{colorlinks=true, linkcolor=blue, urlcolor=blue, citecolor=blue}

\begin{document}

\title{Global Antineutrino Modeling for a Web Application}

\author{S.T. Dye and A.M. Barna}
\affiliation{Department of Physics and Astronomy, University of Hawaii, Honolulu, HI, 96822 USA}

\date{\today}

\begin{abstract}
\vspace{1mm}
\noindent
Antineutrinos stream freely from rapidly decaying fission products within the cores of nuclear reactors and from long-lived natural radioactivity within the rocky layers of the Earth. These global antineutrinos produce detectable signals in large ultra-clear volumes of water- or hydrocarbon-based target liquids, which are viewed by inward-facing photomultiplier tubes. Detected antineutrinos provide information about their shrouded sources and about the fundamental properties of neutrinos themselves. This paper presents the input data, formulae, and plots resulting from the calculations, which, in addition to the time-dependent reaction rates and energy spectra, model the directions of the antineutrinos from IAEA-registered nuclear power reactors and of the neutrinos from $^8$B decay in the Sun. The model includes estimates of the steady state reaction rates and energy spectra of the antineutrinos from the crust and mantle of the Earth. Results are available for any location near the surface of the Earth and comprise both quasi-elastic scattering on free protons and elastic scattering on atomic electrons. This paper compares model results for two underground locations, the Boulby Mine in the United Kingdom and the Morton Salt Mine in the United States. Operational nuclear power reactors are within about $20$ kilometers of these mines, making them candidate sites for antineutrino detectors capable of identifying, monitoring, and locating remote nuclear activity. The model, which is implemented in a web application at https://geoneutrinos.org/reactors/, provides references for the input data and the formulae, as well as an interactive calculator of the significance of the rate of any of the neutrino sources relative to other sources taken as background.

\end{abstract}

\maketitle

\section{Introduction} 
Global antineutrinos arise predominantly from naturally-occurring radioactivity within the Earth and from operating nuclear power reactors \cite{agm15}. Much smaller, and herein neglected, contributions come from nuclear research reactors and spent nuclear fuel. Information about Earth and nuclear power reactors, as well as the properties of neutrinos themselves, presently derives from measuring the rate and energy spectrum of the reactions of antineutrinos. Measuring the directions of antineutrinos represents an important but largely unrealized complement. 

Detecting antineutrinos from nuclear reactors at short \cite{nucifer15,songs07} and long \cite{nudar13,snif10} distances monitors the operation and identifies the location and power of the reactor with important applications for nuclear non-proliferation \cite{adam10}. These detections also contribute to the fundamental understanding of neutrinos \cite{reines53,reines76,jgl08}. Detecting antineutrinos from the nuclear cascades of thorium-232 and uranium-238 within the Earth \cite{kl05} estimates terrestrial radiogenic heating \cite{gando13,agostini15}, leading to a more complete understanding of the composition, structure, and thermal evolution of the Earth \cite{dye_etal15}.

Global antineutrinos emerge from nuclear beta-minus decays, which produce a characteristic energy spectrum for each parent isotope. Decay energies are typically less than $10$ MeV. While the mixture of isotopes decaying within a source uniquely determines the energy spectrum of the emitted antineutrinos, neutrino oscillations distort the spectrum of the detected antineutrinos in a pattern determined by the distance from the source. The rate and energy spectrum of global antineutrino reactions varies dramatically with the near surface location of existing and potential detector sites. Recent modeling projects provide static maps of the surface flux of antineutrinos \cite{baldoncini,agm15} with revisions possible as the inventory of nuclear power reactors changes, the exposure of Earth's antineutrino observations increases, and the precision of geological modeling improves. Global antineutrino modeling for a web application, which is available at https://geoneutrinos.org/reactors/, supplements the static mapping projects: It interactively estimates and displays the reaction rates, energy spectra, and directions of the antineutrinos from nuclear power reactors and from the crust and mantle of the Earth, as well as of the neutrinos from $^8$B decay in the Sun.

\section{Neutrinos}
A neutrino ($\nu$), or little neutral one, is a nearly massless, electrically neutral subatomic particle. Neutrinos travel almost unimpeded through matter, effectively interacting only via the short-range weak nuclear force. During weak interactions, neutrinos associate with one of three electrically charged and far more massive particles: an electron, muon, or tau (e, $\mu$, or $\tau$). These associations represent the neutrino flavor states: $\nu_\mathrm{e}$, $\nu_\mu$, and $\nu_\tau$. Antineutrinos, with symbols $\overline{\nu}_\mathrm{e}$, $\overline{\nu}_\mu$, and $\overline{\nu}_\tau$, are the antiparticles corresponding to the neutrino flavor states. While traveling, neutrinos and antineutrinos have associations with all flavors. The amount of each association varies along the neutrino path with a predictable amplitude and an energy-dependent wavelength. These flavor oscillations follow from a changing mixture of the three neutrino mass states: $\nu_1$, $\nu_2$, and $\nu_3$. The equations in this section express mass and momentum in eV units with implied factors of the speed of light in the vacuum $c$.

Electron antineutrinos ($\overline{\nu}_\mathrm{e}$) appear during nuclear transmutation by beta-minus ($\beta^-$) decay,
\begin{equation}
[A,Z]\rightarrow [A,Z+1] + \beta^- + \overline{\nu}_\mathrm{e},
\label{beta_decay}
\end{equation}
where $A$ is the atomic mass number and $Z$ is the atomic charge number. The three-body decay produces a continuous spectrum of $\overline{\nu}_\mathrm{e}$ energies. A greater difference in rest mass between the parent $[A,Z]$ and the daughter $[A,Z+1]$ radioisotopes gives a greater maximum energy of the $\overline{\nu}_\mathrm{e}$ spectrum and a shorter lifetime of the parent. 

Detecting electron antineutrinos traditionally exploits a scattering reaction on targets in a water- or hydrocarbon-based liquid. The dominant reaction is a quasi-elastic scattering process, which is the inverse of $\beta^-$ decay \eqref{beta_decay}, involving a free proton (p) \cite{reines53}, or hydrogen nucleus. In this case, the reaction 
\begin{equation}
\overline{\nu}_\mathrm{e} + [A,Z+1] \rightarrow [A,Z] + \beta^+ 
\end{equation}
is typically written as
\begin{equation}
\overline{\nu}_\mathrm{e} + \mathrm{p} \rightarrow \mathrm{n} + \mathrm{e}^+
\label{pscat}
\end{equation}
or more compactly as $\overline{\nu}_\mathrm{e}\mathrm{p}$. The model currently ignores antineutrino-proton elastic scattering \cite{fisch77}. Accordingly, the compact notation refers only to \eqref{pscat} herein.

The products of inverse beta decay (IBD) \eqref{pscat} form a coincidence of detectable signals, initially from the positron (e$^+$ or $\beta^+$) and quickly thereafter ($\lesssim 1$ ms) from the nearby ($\lesssim 1$ m) capture of the neutron (n). While the positron acquires most of the kinetic energy of the antineutrino, the neutron gets most of the momentum. Therefore, the positron scattering angles are distributed nearly isotropically with a slight backward asymmetry and the neutron initially follows the direction of the incoming antineutrino \cite{vogel99}. Due to collisions with atomic nuclei the neutron stealthily wanders away from the original path until it is moving slowly enough to be captured. Directional information is available from the vector connecting the positions of the start of the prompt positron track and the delayed capture of the neutron. Resolution of the neutrino source direction with \eqref{pscat} depends on reconstruction efficiencies and statistics \cite{paloverde00,chooz00}. 

\begin{figure}
\includegraphics[trim = 0mm 0mm 0mm 0mm, clip, scale=0.45]{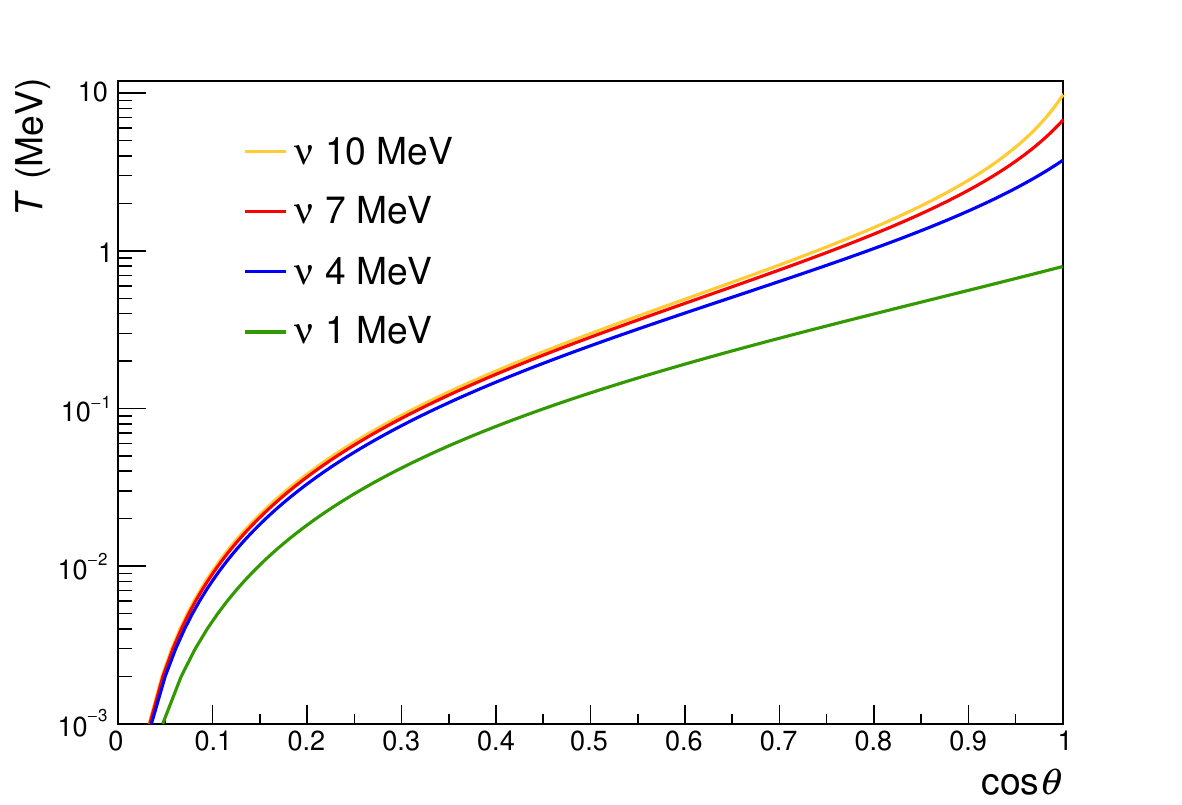}
\caption{The relationship between kinetic energy $T$ and scattering angle $\cos\theta$ for electrons in \eqref{escat} is shown for selected values of the incident neutrino energy.}
\label{fig:Te_cos}
\end{figure}

The subdominant reaction is an elastic scattering (ES) process, which is written as
\begin{equation}
\nu + \mathrm{e} \rightarrow \nu + \mathrm{e}
\label{escat} 
\end{equation}
or more compactly as $\nu \mskip 1mu \mathrm{e}$. In \eqref{escat} $\nu$ represents both neutrinos and antineutrinos.
Standard electroweak theory describes the elastic scattering of neutrinos on electrons \eqref{escat}. The detectable particle is the electron, which always scatters somewhat forward relative to the direction of the neutrino. Elastic scattering provides important information on the direction to the neutrino source \cite{gnudir}. Conservation of energy and momentum define the cosine of the electron scattering angle $\cos\theta$  in terms of the neutrino energy $E_\nu$ and the electron mass $m_\mathrm{e}$ and kinetic energy $T_\mathrm{e} \equiv T$. The relationship is
\begin{equation}
\cos\theta=\frac{1+m_\mathrm{e}/E_\nu}{(1+2m_\mathrm{e}/T)^{{}^1{\mskip -5mu/\mskip -3mu}_2}}.
\label{eleccos}
\end{equation}
As expected the scattering angle decreases with increasing kinetic energy. Setting a minimum electron kinetic energy $T_\mathrm{min}$ removes large angle scatterings, helping to resolve the direction to the neutrino source. This feature of the reaction kinematics is shown in Figure~\ref{fig:Te_cos}. 
Solving \eqref{eleccos} for the electron kinetic energy gives
\begin{equation}
T = \frac{2m_\mathrm{e} \cos^2 \theta} {(1 + m_\mathrm{e}/E_\nu)^2 - \cos^2 \theta}.
\label{electe}
\end{equation} 
The electron kinetic energy is maximum for scattering in the same direction as the incident neutrino (cos$\, \theta=1$),
\begin{equation}
\label{temax}
T_\mathrm{max}=E_\nu / (1+\frac{m_\mathrm{e}}{2E_\nu}).
\end{equation}

\subsection{Cross Sections}
The quasi-elastic scattering cross section for global antineutrinos with $E_{\overline{\nu}_\mathrm{e}} < 10$ MeV is well described by
\begin{equation}
\label{vb99}
\sigma^\mathrm{IBD}(E_\mathrm{e}) = (9.52 \times10^{-44} \mathrm{cm}^2 \mathrm{MeV}^{-2} )p_\mathrm{e} E_\mathrm{e},
\end{equation}
where $E_\mathrm{e}$ and $p_\mathrm{e}= \sqrt{E^2_\mathrm{e}-m^2_\mathrm{e}}$ are the positron energy and momentum, respectively \cite{vogel99}. Antineutrinos with energy greater than the threshold energy
\begin{equation}
\label{ethr}
E_\mathrm{thr} = \frac{(m_\mathrm{n}+m_\mathrm{e})^2 - m_\mathrm{p}^2} {2m_\mathrm{p}}
\end{equation}
($1.80607$ MeV), where $m_\mathrm{n}$ and $m_\mathrm{p}$ are the masses of the neutron and proton, respectively, initiate this reaction. Assuming the antineutrino is massless, then the energy of the incident electron antineutrino $E_{\overline{\nu}_\mathrm{e}}$ relates to the energy of the positron $E_\mathrm{e}$ by
\begin{equation}
E_{\overline{\nu}_\mathrm{e}} = E_\mathrm{e}+E_\mathrm{thr}-m_\mathrm{e}.
\label{thrsh}
\end{equation}
This direct relationship facilitates an estimate of the energy spectrum of antineutrinos by measuring the positron energy.  An alternative formulation of the cross section for quasi-elastic scattering \eqref{pscat}, which avoids an overestimation at high energy, is given by
\begin{equation}
\label{sv03}
\sigma^\mathrm{IBD}(E_\mathrm{e}) = (10.0 \times10^{-44} \mathrm{cm}^2 \mathrm{MeV}^{-2} )p_\mathrm{e} E_\mathrm{e} E_{\overline{\nu}_\mathrm{e}}^{\alpha\mathstrut},
\end{equation}
where $\alpha = -0.07056+0.02018\,\mathrm{ln}E_{\overline{\nu}_\mathrm{e}}-0.001953\,\mathrm{ln}^3E_{\overline{\nu}_\mathrm{e}}$ \cite{strumia03}. The particle masses used for calculating the inverse beta decay cross sections are given in Table~\ref{tab:ibdcons}.

\begin{table}[h]
\caption{Physical Constants in Quasi-elastic Scattering \cite{pdg2020}.} 
\begin{tabular}{c c c}
\hline\noalign{\smallskip}
 $m_\mathrm{n}$ (MeV) & $m_\mathrm{p}$ (MeV) & $m_\mathrm{e}$ (MeV)  \\
\hline\noalign{\smallskip}
 939.56542052 & 938.27208816 & $0.5109989500$  \\
\hline\noalign{\smallskip}
\end{tabular}
\label{tab:ibdcons}
\end{table}

The elastic scattering differential cross section as a function of kinetic energy of the electron $T$ at a given neutrino energy $E_\nu$ is
\begin{equation}
\frac{d\sigma^\mathrm{ES}}{dT}=\frac{2G_F^2m_\mathrm{e}}{\pi}[c_\mathrm{L}^2+c_\mathrm{R}^2(1-\frac{T}{E_\nu})^2-c_\mathrm{L}c_\mathrm{R}\frac{m_\mathrm{e}T}{E_\nu^2}],
\label{dsdt}
\end{equation}
where $G_F$ is the Fermi coupling constant, and the coefficients $c_\mathrm{L}$, $c_\mathrm{R}$ are functions of the weak-mixing angle sin$^2\theta_W$ \cite{fukuyana}. The values for the physical constants and the conversion factor used to calculate elastic scattering cross sections are given in Table \ref{tab:escons}. 
\begin{table}[h]
\caption{Physical Constants in Elastic Scattering. The constant $(\hbar c)^2$ converts inverse energy squared to length squared as needed for expressing the elastic scattering cross section in units of cm$^2$. The electron rest mass $m_\mathrm{e}$ is given in Table~\ref{tab:ibdcons}.}
\begin{tabular}{c c c}
\hline\noalign{\smallskip}
 $G_F$ (MeV$^{-2}$)   \cite{pdg2020} & $(\hbar c)^2$ (MeV-cm)$^2$  \cite{pdg2020} & $\sin^2\theta_W$ \cite{erler05} \\
\hline\noalign{\smallskip}
 $1.1663787\!\!\times\!\!10^{-\!11}$ & $3.893793721\!\!\times\!\!10^{-\!22}$ & 0.23867 \\
\hline\noalign{\smallskip}
\end{tabular}
\label{tab:escons}
\end{table}
Different neutrino flavors have different values of the coefficients $c_\mathrm{L}$, $c_\mathrm{R}$ as given in Table~\ref{tab:clcr}. Muon and tau antineutrinos, which interact identically, are designated as $\overline{\nu}_x$. Solar neutrinos are a source of background to the elastic scattering signal of global antineutrinos, motivating the listing of $c_\mathrm{L}$ and $c_\mathrm{R}$ for $\nu_\mathrm{e}$e.
\begin{table}
\caption{Elastic Scattering Cross Section Coefficients. The coefficients $c_\mathrm{L}$ and $c_\mathrm{R}$ are expressed in terms of the electroweak-mixing angle $\sin^2\theta_W$ \cite{fukuyana}. The value of $\sin^2\theta_W$ is given in Table~\ref{tab:escons}.}
\begin{tabular}{l c c c c}
\hline\noalign{\smallskip}
                                          &  $\overline{\nu}_\mathrm{e}$e   & $\overline{\nu}_x$e & $\nu_\mathrm{e}$e  & $\nu_\mathrm{x}$e \\
\hline\noalign{\smallskip}
     $c_\mathrm{L}$ & sin$^2\theta_W$ & sin$^2\theta_W$ & ${{}^1{\mskip -5mu/\mskip -3mu}_2}$\:\!+\:\!sin$^2\theta_W$ & -${{}^1{\mskip -5mu/\mskip -3mu}_2}$\:\!+\:\!sin$^2\theta_W$ \\
     $c_\mathrm{R}$ & ${{}^1{\mskip -5mu/\mskip -3mu}_2}$\:\!+\:\!sin$^2\theta_W$ & -${{}^1{\mskip -5mu/\mskip -3mu}_2}$\:\!+\:\!sin$^2\theta_W$  & sin$^2\theta_W$ &  sin$^2\theta_W$ \\ [1ex]
\hline\noalign{\smallskip}
\end{tabular}
\label{tab:clcr}
\end{table}

The total elastic scattering cross section obtains from \eqref{dsdt} upon separating variables and integrating from $0$ to $T_\mathrm{max}$. Changing the variable of integration from $T$ to $y=T/E_\nu$ simplifies the final expression. Using $dT = E_\nu dy$, the cross section is given as
\begin{equation}
\label{esxsec}
\begin{split}
\sigma_{T \ge T_\mathrm{min}}^\mathrm{ES}(E_{\nu})=\int_{y_\mathrm{min}}^{y_\mathrm{max}}\frac{d\sigma^\mathrm{ES}(E_\nu)} {dy}dy=\frac{2G_F^2m_\mathrm{e}E_\nu}{\pi} \\
\bigg[\Big(c_\mathrm{L}^2 y_{\mathrm{max}} + c_\mathrm{R}^2\frac{1}{3}\big(1-(1- y_{\mathrm{max}})^3\big) - c_\mathrm{L}c_\mathrm{R}\frac{m_\mathrm{e}}{2E_\nu}y_{\mathrm{max}}^2\Big) - \\
\Big(c_\mathrm{L}^2 y_{\mathrm{min}} + c_\mathrm{R}^2\frac{1}{3}\big(1-(1- y_{\mathrm{min}})^3\big) - c_\mathrm{L}c_\mathrm{R}\frac{m_\mathrm{e}}{2E_\nu}y_{\mathrm{min}}^2\Big)\bigg],
\end{split}
\end{equation}
where $y_{\mathrm{max}}=T_\mathrm{max}/E_\nu$ and $y_\mathrm{min} = T_\mathrm{min}/E_\nu$. A nonzero minimum electron kinetic energy $T_\mathrm{min}$, possibly representing the sensitivity threshold of a detector, reduces the total elastic scattering cross section. Ensuring that $T_\mathrm{max}$ is greater than  $T_\mathrm{min}$ establishes a minimum neutrino energy such that
\begin{equation}
E_\nu \ge \frac{T_\mathrm{min} + (T_\mathrm{min}^2 + 2m_\mathrm{e}T_\mathrm{min})^{{}^1{\mskip -5mu/\mskip -3mu}_2}} {2}.
\end{equation} 

\begin{figure}
\centering
\includegraphics[trim = 0mm 0mm 0mm 0mm, clip, scale=0.45]{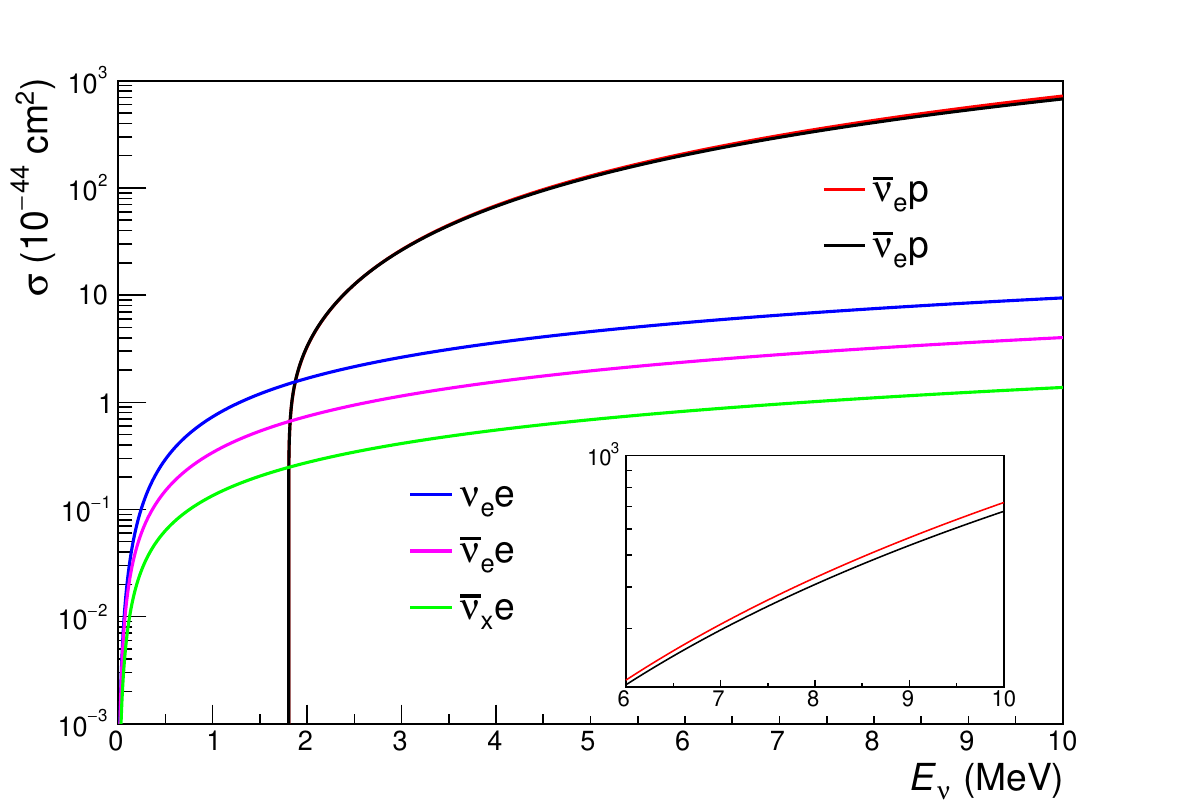}
\caption{Total cross sections as a function of neutrino energy for quasi-elastic ($\overline{\nu}_\mathrm{e}\mathrm{p}$) and elastic ($\nu_\mathrm{e}\mathrm{e}$ in blue, $\overline{\nu}_\mathrm{e}\mathrm{e}$ in purple, $\overline{\nu}_x\mathrm{e}$ in green) scattering. Two versions of the quasi-elastic cross section are shown; \eqref{vb99} in red and \eqref{sv03} in black. The inset compares these versions at the upper end of the energy scale.}
\label{fig:sigma}
\end{figure}
\begin{figure}
\centering
\includegraphics[trim = 0mm 0mm 0mm 0mm, clip, scale=0.45]{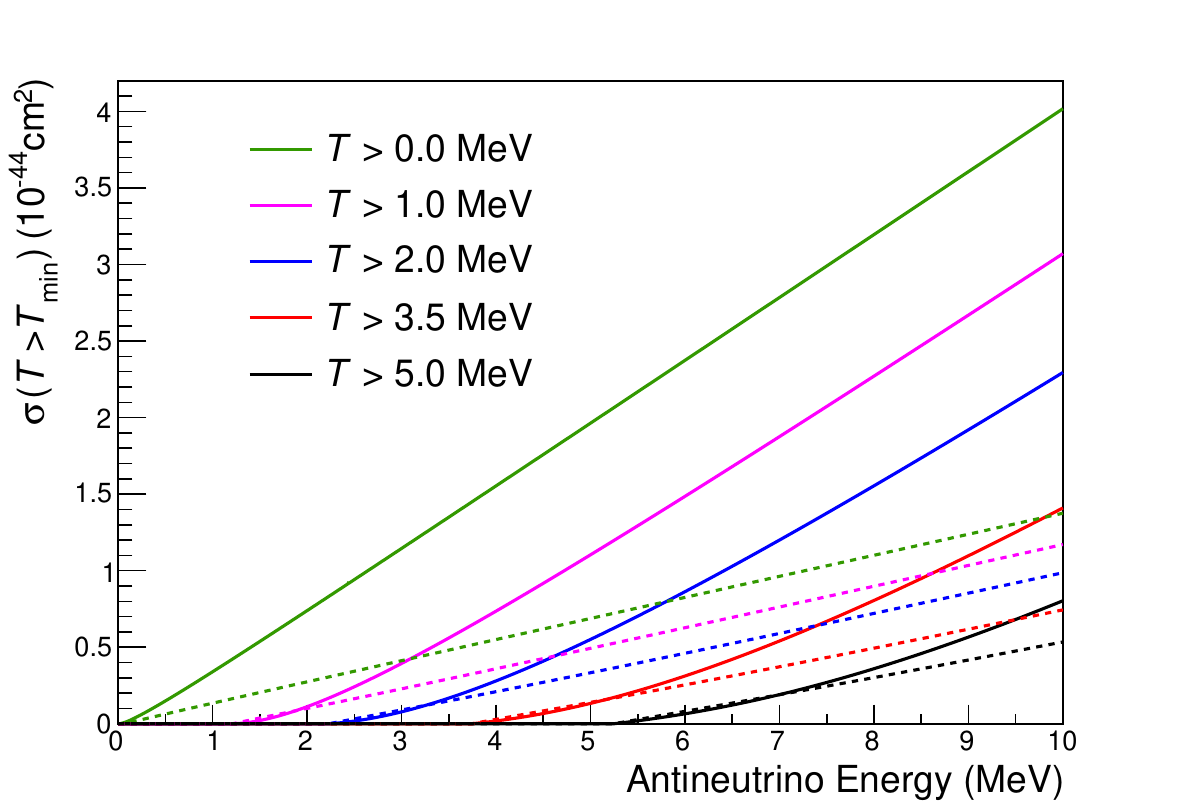}
\caption{Total elastic scattering cross sections as a function of antineutrino energy \eqref{esxsec} for $\overline{\nu}_\mathrm{e}\mathrm{e}$ (solid) and for $\overline{\nu}_x\mathrm{e}$ (dotted) at selected values of the minimum kinetic energy of the electron $T_\mathrm{min}$.}
\label{fig:es_xsec_tmin}
\end{figure}

The total cross sections for quasi-elastic scattering ($\overline{\nu}_\mathrm{e}$p) \eqref{vb99} \eqref{sv03}, and for elastic scattering ($\overline{\nu}_\mathrm{e}$e, $\overline{\nu}_x$e, $\nu_\mathrm{e}$e) \eqref{esxsec}, over the energy range relevant to the geo-neutrino and reactor antineutrino spectra are shown in Fig.~\ref{fig:sigma}. Note that elastic scattering is sensitive to neutrinos with energy less than the threshold \eqref{ethr} for quasi-elastic scattering. The reactor antineutrino flux is poorly known below this threshold energy and elastic scattering offers an interesting measurement opportunity. The decrease in the elastic scattering cross section with increasing values of the minimum kinetic energy of the electron $T_\mathrm{min}$ is shown in Figure~\ref{fig:es_xsec_tmin}.  

\begin{figure}
\centering
\begin{subfigure}[b]{0.23\textwidth}
\includegraphics[trim = 0mm 0mm 0mm 0mm,width=\textwidth]{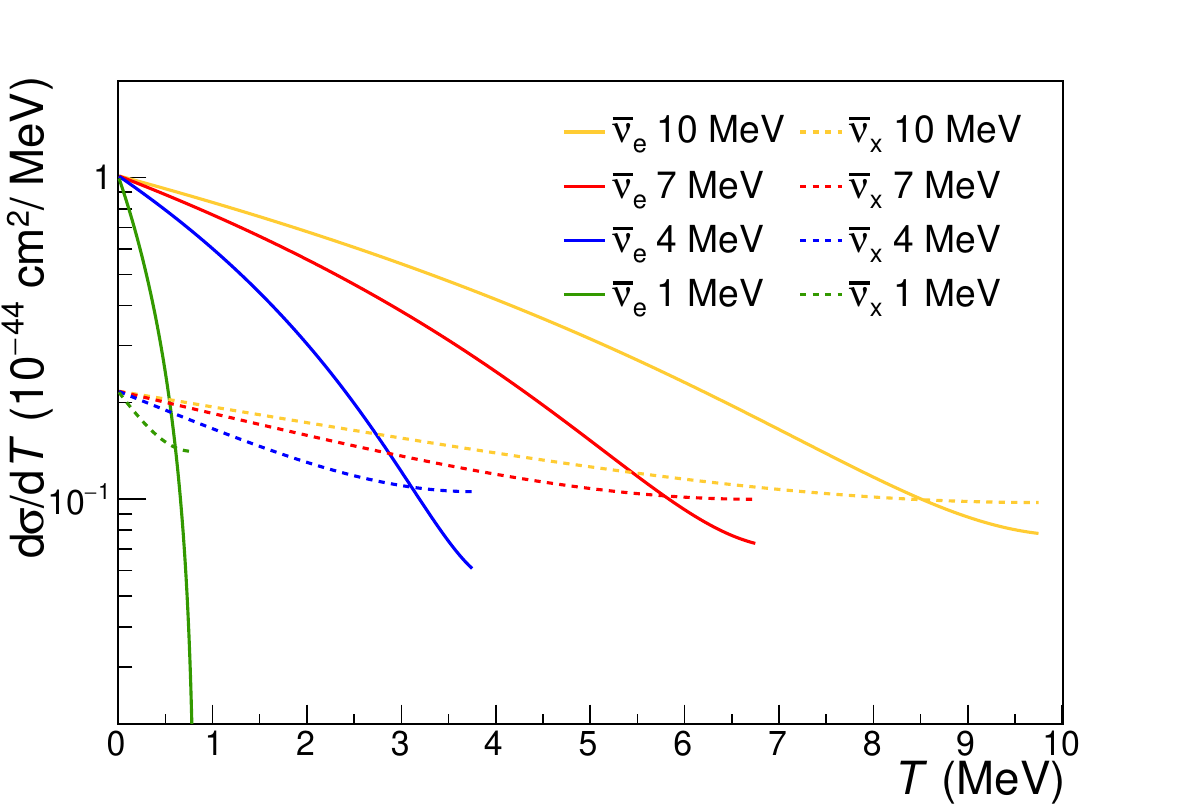}
\end{subfigure}
\begin{subfigure}[b]{0.23\textwidth}
\includegraphics[trim = 0mm 0mm 0mm 0mm,width=\textwidth]{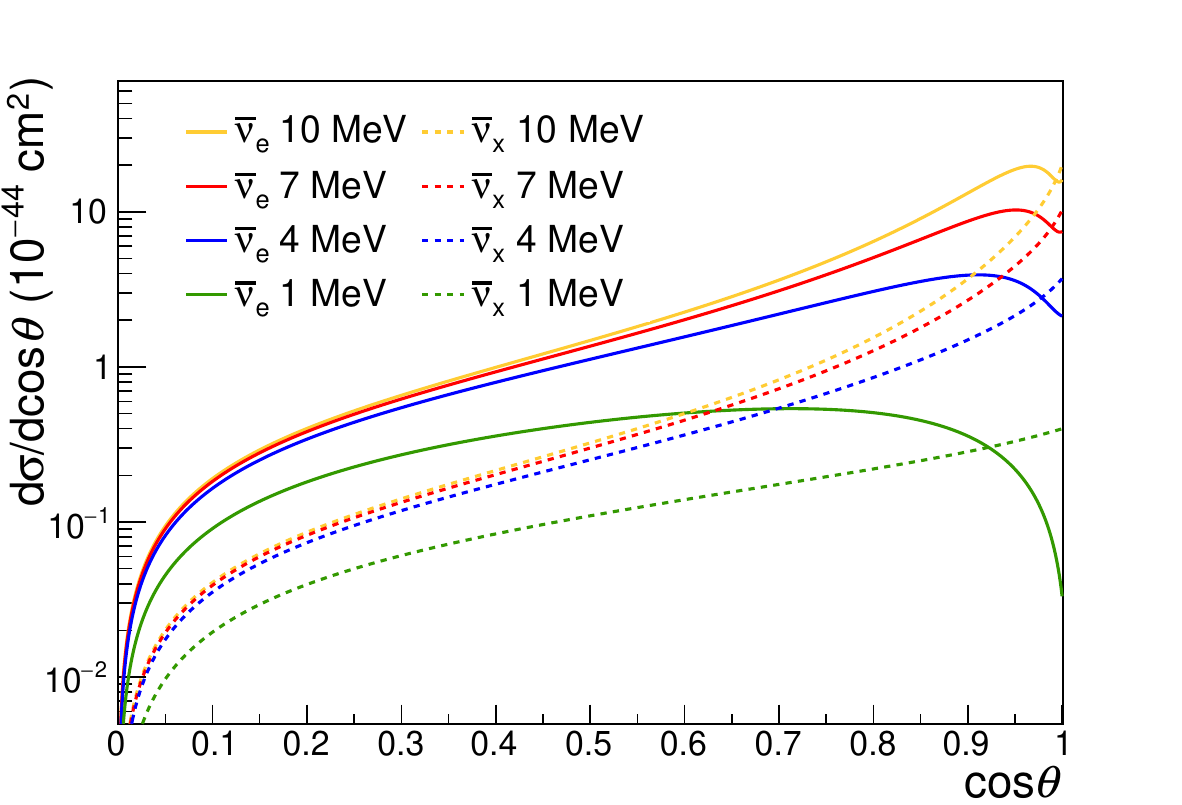}
\end{subfigure}
\caption{Elastic scattering differential cross section, at selected energies, for $\overline{\nu}_\mathrm{e}$ and $\overline{\nu}_x$ as a function \eqref{dsdt} of the electron kinetic energy $T$ (left) and as a function \eqref{dsdcos} of the cosine of the electron scattering angle $\cos \theta$ (right).}
\label{fig:sigte}
\end{figure}

Differentiating \eqref{electe} with respect to $\cos \theta$ gives
\begin{equation}
\frac{dT}{d\cos \theta} = \frac{4m_\mathrm{e} E_\nu^2 (E_\nu + m_\mathrm{e})^2\cos \theta} {[(E_\nu + m_\mathrm{e})^2 - E_\nu^2 \cos^2 \theta]^2}.
\label{dtdcos}
\end{equation}
Multiplying \eqref{dsdt} by \eqref{dtdcos} and using \eqref{electe} gives the differential cross section as a function of $\cos \theta$ at a given neutrino energy $E_\nu$. The result is
\begin{equation}
\begin{split}
\frac{d\sigma^\mathrm{ES}}{d\cos \theta}= \frac {8G_F^2m_\mathrm{e}^2} {\pi} \frac {E_\nu^2 (E_\nu + m_\mathrm{e})^2\cos \theta} {[(E_\nu+m_\mathrm{e})^2 - E_\nu^2 \cos^2 \theta]^2} \\
\Big(c_\mathrm{L}^2+c_\mathrm{R}^2 \big( 1- \frac {2m_\mathrm{e} E_\nu \cos^2 \theta}  {(E_\nu + m_\mathrm{e})^2 - E_\nu^2 \cos^2 \theta } \big)^2 - \\
c_\mathrm{L}c_\mathrm{R} \frac {2m_\mathrm{e}^2 \cos^2 \theta} {(E_\nu + m_\mathrm{e})^2 -E_\nu^2 \cos^2 \theta} \Big).
\end{split}
\label{dsdcos}
\end{equation}
Integrating \eqref{dsdcos} to arrive at the total cross section looks somewhat tedious. Presumably, the result is equivalent to \eqref{esxsec}. Figure \ref{fig:sigte} shows the differential cross section as a function of $T$ and of $\cos \theta$, respectively, for several values of $E_\nu$.

\begin{figure}
\centering
\begin{subfigure}[b]{0.23\textwidth}
\includegraphics[trim = 0mm 0mm 0mm 0mm,width=\textwidth]{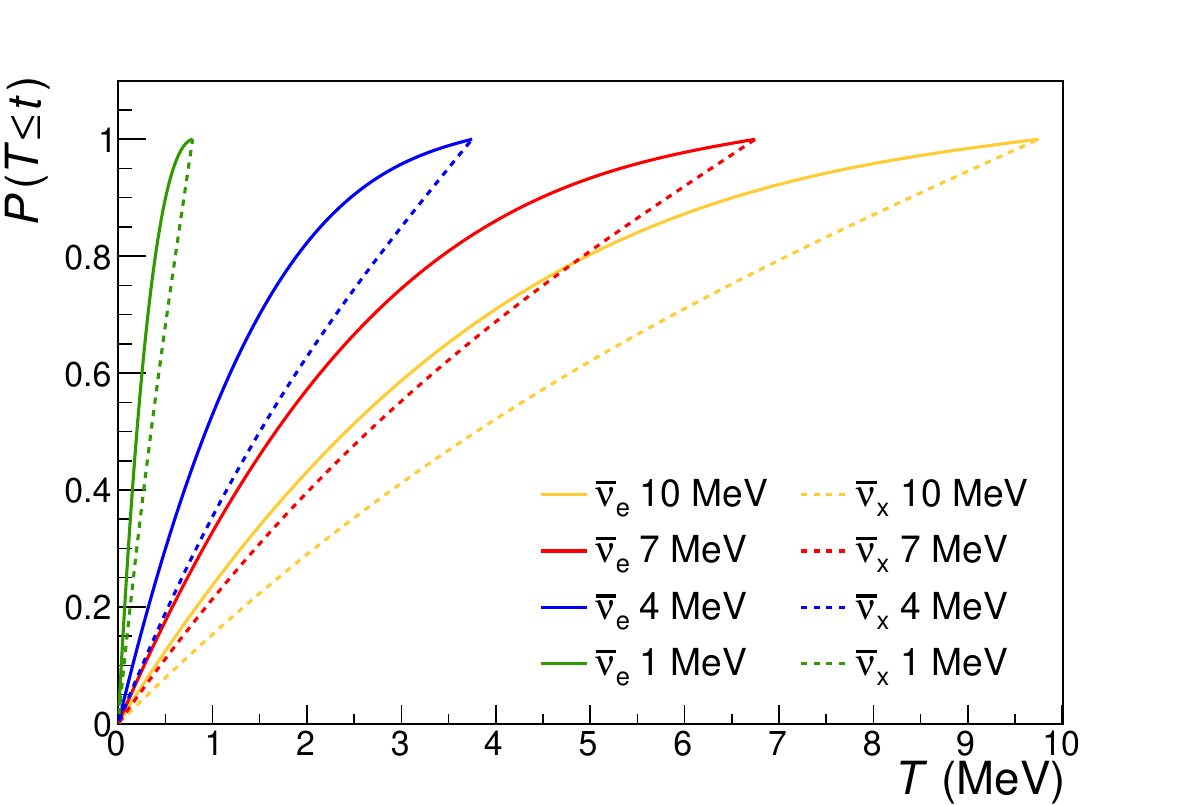}
\end{subfigure}
\begin{subfigure}[b]{0.23\textwidth}
\includegraphics[trim = 0mm 0mm 0mm 0mm,width=\textwidth]{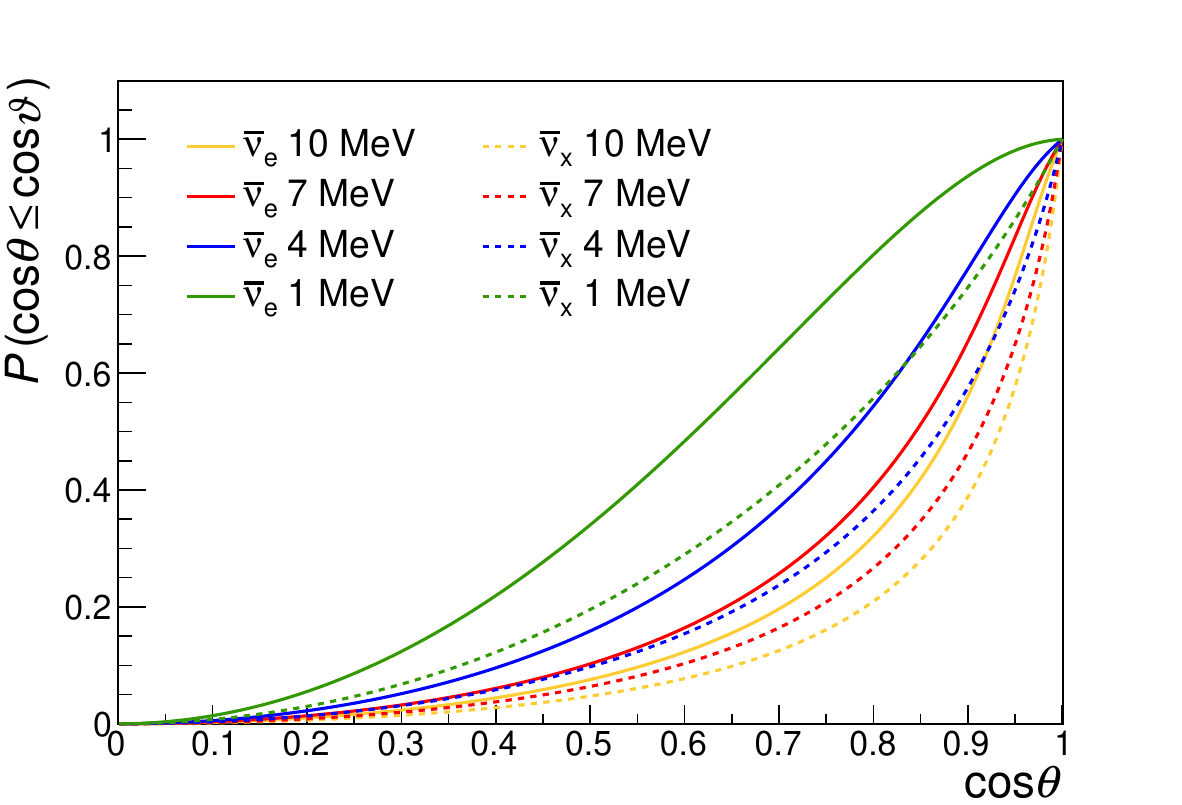}
\end{subfigure}
\caption{Cumulative distribution functions, at selected energies of $\overline{\nu}_\mathrm{e}$ and $\overline{\nu}_x$, for the elastic scattering differential cross section as a function \eqref{cumte} of the electron kinetic energy $T$ (left) and as a function \eqref{cumcos} of the cosine of the electron scattering angle $\cos \theta$ (right).}
\label{fig:pdfte}
\end{figure}

The cumulative distribution function for \eqref{dsdt} at fixed neutrino energy for a value of $T$ is
\begin{equation}\label{cumte}
P(T \leq t) = \frac {\int_{0}^{T} (d\sigma^\mathrm{ES} / dt) \, dt} {\int_{0}^{T_\mathrm{max}} (d\sigma^\mathrm{ES} / dt) \, dt} \text{,}
\end{equation}
where the upper limit of the integral in the denominator is given by \eqref{temax}.
Similarly, the cumulative distribution function for \eqref{dsdcos} at fixed neutrino energy for a value of $\cos\theta$ is
\begin{equation}\label{cumcos}
P(\cos\theta \leq \cos\vartheta) = \frac {\int_{0}^{\cos\theta} (d\sigma^\mathrm{ES} / d\cos\vartheta) \, d\cos\vartheta} {\int_{0}^{1} (d\sigma^\mathrm{ES} / d\cos\vartheta) \, d\cos\vartheta} \text{.}
\end{equation}
The cumulative distribution functions for the two elastic scattering differential cross sections, one as a function of $T$ and the other as a function of $\cos \theta$, at several values of $E_\nu$ are shown in Fig.~\ref{fig:pdfte}. These distributions transform one to the other by converting $T$ to $\cos \theta$ using \eqref{electe}, or $\cos \theta$ to $T$ using \eqref{eleccos}, as appropriate. This demonstrates the equivalence of the differential cross sections \eqref{dsdt} and \eqref{dsdcos}. Cumulative distribution functions are readily compared with random numbers $[0,1]$ to help select the $T$ and $\cos \theta$ pairs for elastic scattering reactions within simulated event samples. 

\subsection{Neutrino oscillations}
Neutrino flavors, which associate with the charged leptons e, $\mu$, and $\tau$, are quantum mechanical mixtures of three neutrino mass states $\nu_1$, $\nu_2$, $\nu_3$ with masses $m_1$, $m_2$, and $m_3$. Mixture varies with distance travelled as a function of energy, according to the well established phenomenon of neutrino oscillations. The probability that an electron neutrino or an electron antineutrino, assuming $CPT$ invariance, of energy $E_{{\nu}_\mathrm{e}}$ in MeV changes flavor after traveling a distance $L$ in meters is
\begin{equation}
\label{nuosc}
\begin{split}
P_{\mathrm{e}x}(L,E_{{\nu}_\mathrm{e}})=\cos^4\theta_{13}\sin^22\theta_{12}\sin^2(1.27\delta m^2_{21}L/E_{{\nu}_\mathrm{e}})\,\\
+\cos^2\theta_{12}\sin^22\theta_{13}\sin^2(1.27\delta m^2_{31}L/E_{{\nu}_\mathrm{e}})\,\\
+\sin^2\theta_{12}\sin^22\theta_{13}\sin^2(1.27\delta m^2_{32}L/E_{{\nu}_\mathrm{e}}),\\
\end{split}
\end{equation}
where $\delta m_{ji}^2=m_j^2-m_i^2$ is the neutrino mass-squared difference in eV$^2$ and $\theta_{12}$, $\theta_{13}$ are the solar, reactor mixing angles, respectively. The complementary probability, $P_\mathrm{ee}(L,E_{{\nu}_\mathrm{e}}) = 1 - P_{\mathrm{e}x}$, gauges survival of electron flavor. Table~\ref{tab:nuosc} lists the neutrino oscillation parameter values \cite{pdg2016} used to estimate the spectral distortion of reactor antineutrino reactions and the overall suppression of geo-neutrino reactions. An average survival probability, given by 
\begin{equation} \label{avgposc}
\begin{array} [b] {l}
<\!\!{P}_\mathrm{ee}\!\!> = 1- \frac{1} {2} \big(\cos^4 \theta_{13} \sin^2(2\theta_{12}) + \sin^2(2\theta_{13}) \big), \\
\end{array}
\end{equation}
accounts for the effect of oscillations on geo-neutrinos. A constant suppression factor of $0.5473$ ($0.5471$) for normal (inverted) mass ordering follows from the adopted mixing angle values in Table~\ref{tab:nuosc} and the trigonometric identities $\sin(2u)=2\sin u \cos u$ and $\cos^2u=1-\sin^2u$.
\begin{table}
\caption{Neutrino Oscillation Parameters. Values shown are for normal and inverted neutrino mass ordering \cite{pdgnuosc2020}: solar mixing angle $\theta_{12}$, reactor mixing angle $\theta_{13}$, and mass-squared differences $\delta m_{ij}^2$.}
\begin{tabular}{l c c c c c}
\hline\noalign{\smallskip}
 & $\dfrac{\sin^2 \theta_{12}}{10^{-1}}$  & $\dfrac{\sin^2 \theta_{13}}{10^{-2}}$ & $\dfrac{\delta m_{21}^2}{10^{-5}\mathrm{eV}^2}$ &$\dfrac{\delta m_{32}^2}{10^{-3}\mathrm{eV}^2}$ & $\dfrac{\delta m_{31}^2}{10^{-3}\mathrm{eV}^2}$ \\ [1.5ex]
\hline\noalign{\smallskip}
Normal & 3.10 & 2.241 & $7.39$  &  $\quad\!2.449$ & $\quad\!2.523$ \\
Inverted & 3.10 & 2.261 &  $7.39$  & $-2.509$ & $-2.435$ \\
\hline\noalign{\smallskip}
\end{tabular}
\label{tab:nuosc}
\end{table}

Neutrino oscillations reduce the reaction rate and distort the energy spectrum of the detected charged leptons. The reduction and distortion are more pronounced for inverse beta decay \eqref{pscat} than for elastic scattering \eqref{escat}. Global antineutrinos which convert to $\overline\nu_{\mu}$ or $\overline\nu_{\tau}$ lack the energy required to initiate quasi-elastic scattering,
\begin{equation}
\overline{\nu}_\mu + \mathrm{p} \rightarrow \mathrm{n} + \mu^+ \: \mathrm{or} \;\: \overline{\nu}_\tau + \mathrm{p} \rightarrow \mathrm{n} + \tau^+.
\label{muscat}
\end{equation}
This is easily confirmed by substituting the rest mass of the $\mu$ ($105.66$ MeV) or the $\tau$ ($1776.9$ MeV) for the rest mass of the electron in the equation for $E_\mathrm{thr}$ \eqref{ethr}. The resulting energy spectrum for quasi-elastic scattering carries the full disappearance imprint. The spectral distortion of inverse beta decay \eqref{pscat} reactions provides important information on the distance to the source of antineutrinos \cite{dye09,nudar13}. On the other hand, global antineutrinos which convert to $\overline\nu_{\mu}$ or $\overline\nu_{\tau}$ are free to initiate elastic scattering \eqref{escat}, which has no energy threshold. Although the cross section for $\overline\nu_{\mu}$ or $\overline\nu_{\tau}$ is smaller than for $\overline\nu_\mathrm{e}$ (Fig.~\ref{fig:sigma}), the recovered reactions mitigate the reduction and distortion of the energy spectrum. A distinct aspect of this subdominant reaction is retention of the incident neutrino direction by the scattered electron (Fig.~\ref{fig:sigte}). This provides important pointing information, which aids in the identification and characterization of neutrino sources \cite{gnudir}. 


\section{Reactor antineutrinos} \label{reactor}
\begin{figure}
\includegraphics[trim = 0mm 0mm 0mm 0mm, clip, scale=0.45]{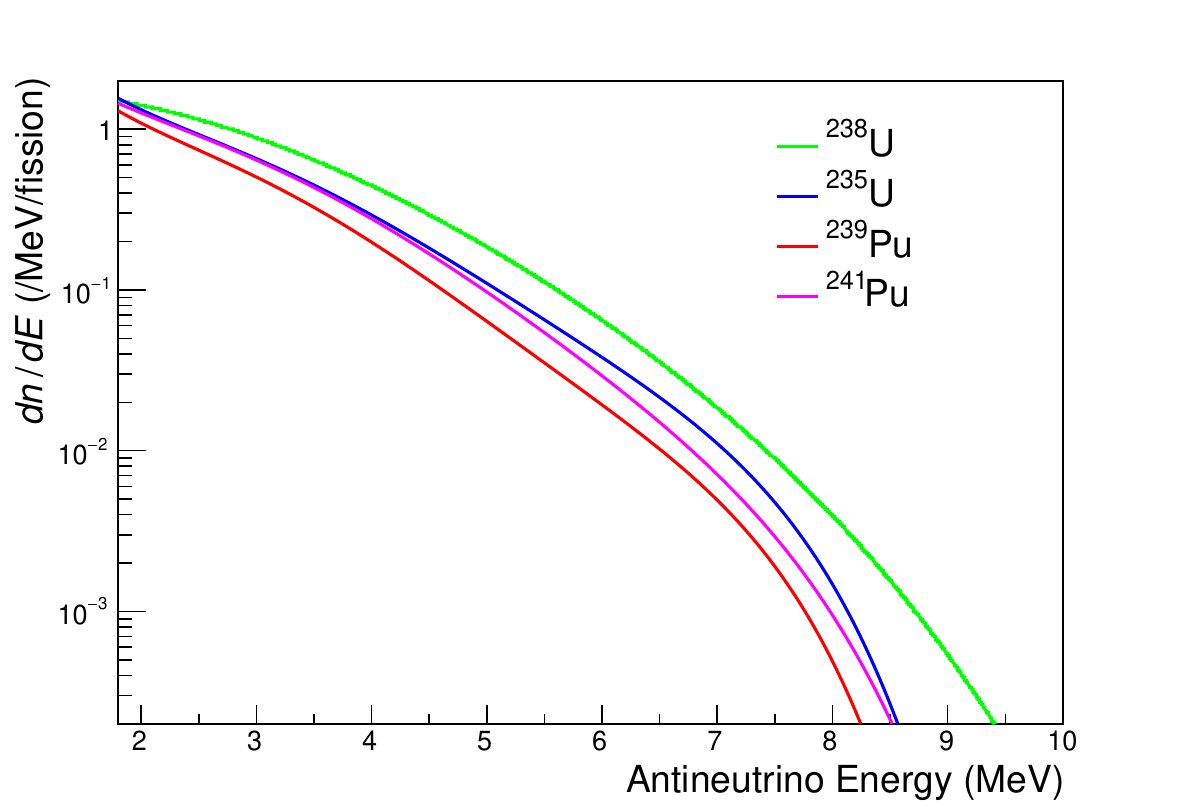}
\caption{Energy spectra of reactor antineutrino emissions \eqref{nuspec} from the fissile isotopes $^{235}$U, $^{239}$Pu, $^{241}$Pu \cite{huber}, and  $^{238}$U \cite{mueller}.  Spectra are estimated using the parameters given in Table~\ref{tab:params} and are valid only above the energy threshold \eqref{ethr} for the quasi-elastic scattering reaction \eqref{pscat}.}
\label{fig:iso_spectra}
\end{figure}

Nuclear reactors generate heat by controlled fission of uranium and plutonium isotopes. Antineutrinos emerge from the $\beta^-$ decays \eqref{beta_decay} of the many fission fragments. The main fissile isotopes are $^{235}$U, $^{238}$U, $^{239}$Pu, and $^{241}$Pu.  An estimate of the differential energy spectrum $\lambda(E_{\overline{\nu}_\mathrm{e}})$ of antineutrinos emitted by the fission of a given isotope is the exponential of a degree-five polynomial of antineutrino energy \cite{mueller,huber}. Specifically, in units of antineutrinos per fission, the spectral estimate is
\begin{equation}
\lambda(E_{\overline{\nu}_\mathrm{e}})=\mathrm{exp}\bigg(\sum_{j=0}^5 {a_j E_{\overline{\nu}_\mathrm{e}}^j}\bigg),
\label{nuspec}
\end{equation}
where $E_{\overline{\nu}_\mathrm{e}}$ is in MeV and the coefficients $a_j$ are fit parameters. Table~\ref{tab:params} lists the values of the coefficients of the main fissile isotopes. 
\begin{table}
\caption{Reactor Antineutrino Emission Spectra. Coefficients $a_j$ are for estimating the differential energy spectra of antineutrinos for the main fissile isotopes: $^{235}$U \cite{huber}, $^{238}$U \cite{mueller}, $^{239}$Pu, and $^{241}$Pu \cite{huber}.}
\begin{tabular}{l c c c c}
\hline\noalign{\smallskip}
                                 & $^{235}$U & $^{238}$U  & $^{239}$Pu & $^{241}$Pu \\
\hline\noalign{\smallskip}
$a_0$                       & 4.367          & 0.4833       & 4.757           & 2.990 \\
$a_1$ (MeV$^{-1}$) & -4.577         & 0.1927       & -5.392         & -2.882 \\
$a_2$ (MeV$^{-2}$) & 2.100          & -0.1283      & 2.563          & 1.278 \\
$a_3$ (MeV$^{-3}$) & -0.5294       & -0.006762  & -0.6596       & -0.3343 \\
$a_4$ (MeV$^{-4}$) & 0.06186      & 0.002233    & 0.07820     & 0.03905 \\
$a_5$ (MeV$^{-5}$) & -0.002777   & -0.0001536 & -0.003536  & -0.001754 \\
\hline\noalign{\smallskip}
\end{tabular}
\label{tab:params}
\end{table}
Figure~\ref{fig:iso_spectra} shows the estimated emission spectra of these isotopes for $E_{\overline{\nu}_\mathrm{e}}>1.8$ MeV. Each isotope ($i=1, 2,3,4$) releases an average thermal energy per fission $Q_i$, which depends on the average energy of the emitted antineutrinos \cite{kopeikin_etal}. 

The estimated reaction rate spectrum per target per unit time, ignoring neutrino oscillations, from the $i^\mathrm{th}$ isotope in a reactor with thermal power $P_{th}$, and at a distance $d$ is
\begin{equation}
R_i(E_{\overline{\nu}_\mathrm{e}})= P_{th} \frac{p_i \lambda_i(E_{\overline{\nu}_\mathrm{e}})}{Q_i} \frac{\sigma(E_{\overline{\nu}_\mathrm{e}})}{4\pi d^2},
\label{isorate}
\end{equation}
where  $p_i$ is the fraction of the thermal power produced by the $i^\mathrm{th}$ isotope, $\lambda_i(E_{\overline{\nu}_\mathrm{e}})$ is given by the spectral estimate \eqref{nuspec}, and $Q_i$ is the fission energy. The isotopic power fractions $p_i$ relate to the fractions of the total fissions and the fission energies of the main isotopes according to
\begin{equation}
p_i=\frac{f_iQ_i}{\sum_i{f_iQ_i}}.
\end{equation}
Summing the contributions from the main fissile isotopes gives the reaction rate spectrum
\begin{equation}
R(E_{\overline{\nu}_\mathrm{e}})= \sum_i {R_i(E_{\overline{\nu}_\mathrm{e}})}
\label{totrate}
\end{equation}
for a specified reactor core. 

\begin{figure}
\begin{subfigure}[b]{0.23\textwidth}
\includegraphics[trim = 0mm 0mm 0mm 0mm,width=\textwidth]{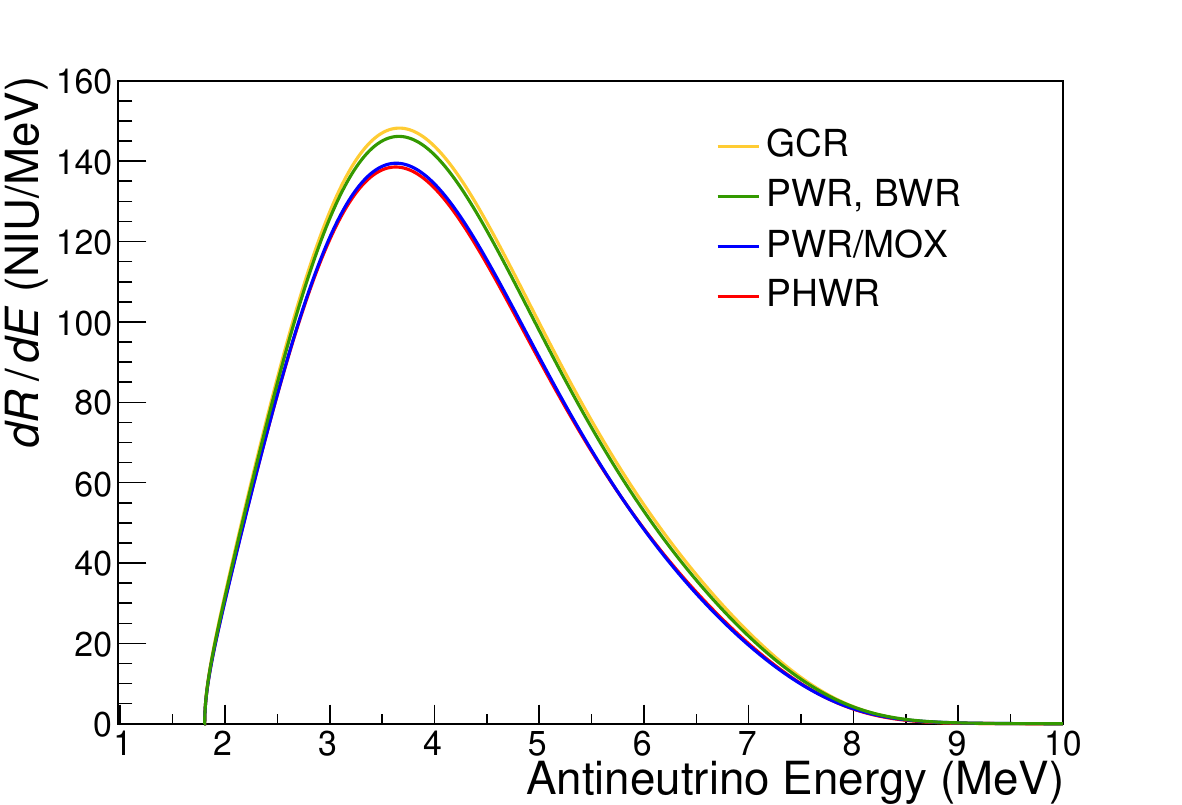}
\end{subfigure}
\begin{subfigure}[b]{0.23\textwidth}
\includegraphics[trim = 0mm 0mm 0mm 0mm,width=\textwidth]{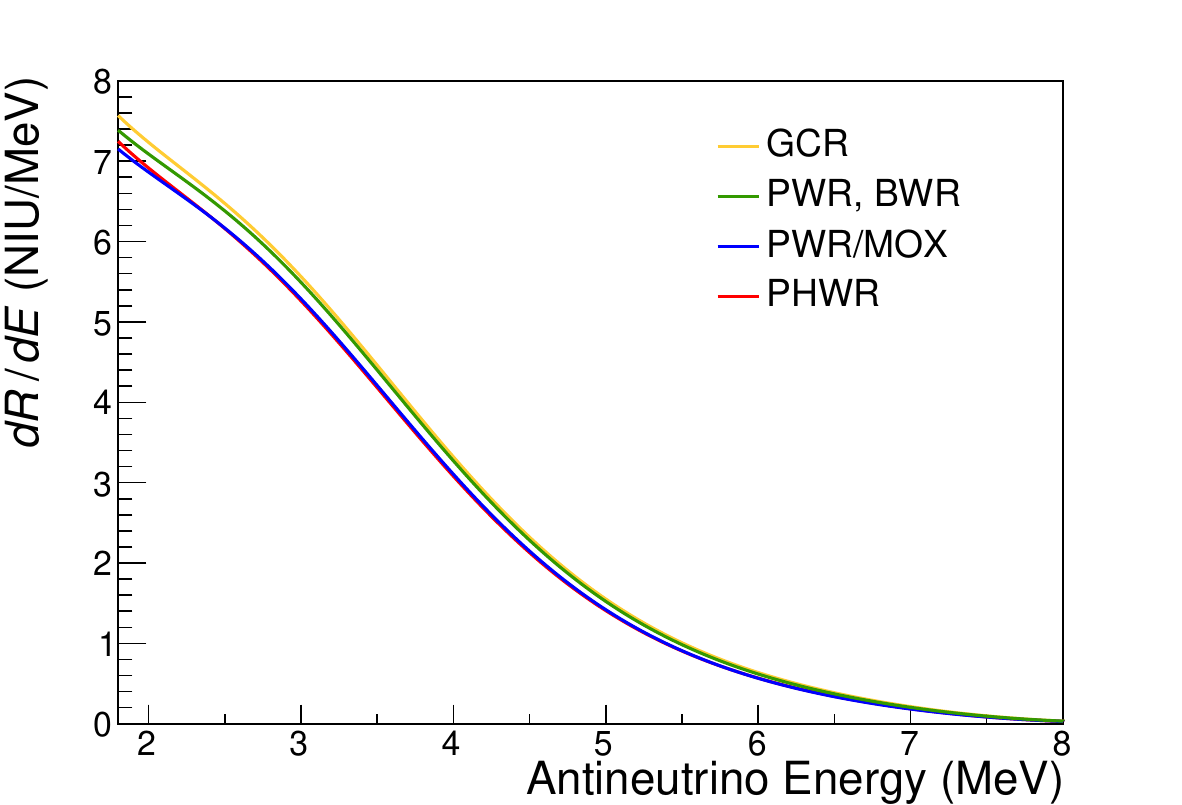}
\end{subfigure}
\caption{The rate of reactor antineutrino quasi-elastic scattering reactions using \eqref{sv03} (left) and elastic scattering reactions using \eqref{esxsec} (right) by reactor type as a function of energy. Spectra are without oscillations and are normalized to $1$ MW of thermal power at a standoff distance of $1$ km. Elastic scattering spectra are cut off at low energy because the parameterized fits to the isotopic spectra \eqref{nuspec} are not reliable below the quasi-elastic scattering reaction threshold energy \eqref{ethr}.}
\label{fig:fuels_ibd}
\end{figure}

The several types of nuclear power reactors \cite{prisgloss} produce thermal energy with differening proportions of fissions from the main isotopes. These different fission or power fractions give rise to distinct antineutrino spectra. Table~\ref{tab:rspec} lists values for the power fractions  $p_i$ for PWR, BWR, MOX \cite{baldoncini}, the fission fractions $f_i$ for GCR \cite{mills}, PHWR \cite{mchen}, and the thermal energies per fission $Q_i$ \cite{kopeikin_etal} for the main fissile isotopes. All values correspond to the midpoint of the reactor operating period. Of the $455$ operating power reactor cores listed by the Power Reactor Information System (PRIS) \cite{pris} in $2019$, $30$ burn fuel containing mixed oxides. These cores are geographically clustered in the European countries of Belgium, France, Germany, and Switzerland. The power fractions labeled PWR/MOX assume a mixture of $70$\% PWR and $30$\% MOX. The resulting energy spectra of the antineutrino quasi-elastic and elastic scattering reaction rates without oscillations at a distance of $1$ km due to $1$ MW thermal power for the four types of nuclear power reactors considered herein are shown in Fig.~\ref{fig:fuels_ibd}. PHWR and PWR/MOX have lower rates than GCR and PWR, BWR types. This suggests possible verification of burning mixed oxides by measuring the antineutrino rate \cite{jaffke_huber17}.

\begin{table}[h]
\caption{Fissile Isotope Data. Fission energies $Q_i$ \cite{kopeikin_etal}, power fractions $p_i$ \cite{baldoncini}, and fission fractions $f_i$ \cite{mills,mchen} at the middle of the fuel cycle for $^{235}$U, $^{238}$U, $^{239}$Pu, and $^{241}$Pu. Power and fission fractions vary by reactor type and fuel mixture: PWR, BWR, MOX \cite{baldoncini}; GCR \cite{mills};  PHWR \cite{mchen}.The entry PWR/MOX assumes a mixture of $70$\% PWR and $30$\% MOX.}
\begin{tabular}{l c c c c}
\hline\noalign{\smallskip}
                     & $^{235}$U & $^{238}$U & $^{239}$Pu & $^{241}$Pu \\
\hline\noalign{\smallskip}
$Q_i$ (MeV) & 201.92 & 205.52 & 209.99 & 213.60 \\
\hline\noalign{\smallskip}
$p_i$ (PWR, BWR) & .560 & .080 & .300 & .060 \\
$p_i$ (MOX) & .000 & .081 & .708 & .212 \\
$p_i$ (PWR/MOX) & .392 & .080 & .422 & .106 \\
\hline\noalign{\smallskip}
$f_i$ (GCR) & .7248 & .0423 & .2127 & .0202 \\
$f_i$ (PHWR) & .520 & .050 & .420 & .010 \\
\hline\noalign{\smallskip}
\end{tabular}
\label{tab:rspec}
\end{table}

Reconciling the disparate units of the quantities encountered in the reaction rate spectrum estimate \eqref{isorate} requires several conversion factors. Because the $Q_i$ values are given in MeV and $P_{th}$ is typically given in MW,  a factor of $1/e$ (eV/J), where $e$ is the elementary charge, reconciles the energy units. With $\sigma(E_{\overline{\nu}_\mathrm{e}})$ in cm$^2$/target and $d$ in km, then a factor of $10^{-10}$ (km$^2$/cm$^2$) reconciles the distance units. The smallness of the cross sections suggests expressing the reaction rates per $10^{32}$ targets per year or in Neutrino Interaction Units (NIU). For free proton inverse beta decay \eqref{pscat} the NIU is equivalent to the terrestrial neutrino unit (TNU) \cite{mantovani}.

Table~\ref{tab:const} lists the value for the elementary charge and the numbers of free protons (p$^+$) for \eqref{pscat} and atomic electrons (e$^-$) for \eqref{escat} in a kilo-tonne (kT) of water (H$_2$O). These numbers convert NIU to interactions per kT of water per year and follow from the molar mass of water given as $18.01528$ g/mole. For water loaded with $0.2\%$ by mass of gadolinium sulfate (Gd$_2$(SO$_4$)$_3$), which has no free protons and a molar mass of $603.183$ g/mol, there are $0.6672221\times10^{32}$ ($0.9980000\times0.6685592$) free protons and $3.3482274\times10^{32}$ ($1.001625\times3.342796$) atomic electrons per kT. 

\begin{table}[h]
\caption{Constants for Reactor Antineutrino Rates. The value of the elementary charge $e$ is exact by definition of the coulomb.}
\begin{tabular}{c c c}
 $e$ (C) & p$^+$/kT(H$_2$O) & e$^-$/kT(H$_2$O) \\
\hline\noalign{\smallskip}
$1.602176634\!\times\!10^{-19}$ & $0.6685592\!\times\!10^{32}$ & $3.342796\!\times\!10^{32}$  \\
\hline\noalign{\smallskip}
\end{tabular}
\label{tab:const}
\end{table}

The thermal power $P_{th}$ in \eqref{isorate} corresponds to the thermal capacity of any given reactor core. Reactor cores require maintenance and refueling, which decrease the thermal power for periods of time. Moreover, reactor operators often run cores in excess of the thermal capacity by up to $4-5$\%. The model accounts for variations in thermal power with monthly-averaged load factors published by IAEA and made publicly available \cite{infn}. 

Fission energies $Q_i$, fission fractions $f_i$, and power fractions $p_i$ evolve with time as the nuclear fuel burns within a reactor. This burn up effect, which accounts for decreases and increases of the fissile isotopes, depends on enrichment and how often and how much fuel is replaced in the reactor core. These time varying quantities are currently considered to be constant and tied to the middle of the fuel cycle. There are several available estimates of fission fractions at the beginning, middle, and end of the reactor core operation cycle. These estimates allow comparison of the antineutrino rate and spectrum at these stages in the cycle. The available fission fractions are listed in Table~\ref{tab:fisfrac_cycle}. Figure

\begin{table}[h]
\caption{Fissile Fractions by Stage in Operation Cycle. Fission fractions $f_i$ and corresponding power fractions $p_i$ at the beginning, middle, and end of the reactor core operation cycle for $^{235}$U, $^{238}$U, $^{239}$Pu, and $^{241}$Pu are shown for an advanced gas-cooled reactor (AGR) \cite{mills} and a pressurized water reactor (VVER-1000) \cite{kopeikin12} reactors.}
\begin{tabular}{l l c c c c}
\hline\noalign{\smallskip}
          &           & $^{235}$U & $^{238}$U & $^{239}$Pu & $^{241}$Pu \\
\hline\noalign{\smallskip}
                     & begin & .7667 & .0407 & .1777 & .0149 \\
 $f_i$ (AGR) & middle & .7248 & .0420 & .2127 & .0205 \\
                     & end & .6829 & .0432 & .2476 & .0263 \\
\cline{1-2}
                      & begin & .65 & .07 & .24 & .04 \\
 $f_i$ (VVER) & middle & .56 & .07 & .31 & .06 \\
                     & end & .48 & .07 & .37 & .08 \\
\hline\noalign{\smallskip}
                      & begin & .7601 & .0411 & .1832 & .0156 \\
 $p_i$ (AGR) & middle & .7173 & .0423 & .2189 & .0215 \\
                      & end & .6747 & .0434 & .2544 & .0275 \\
\cline{1-2}
                      & begin & .641 & .070 & .247 & .042 \\
 $p_i$ (VVER) & middle & .550 & .070 & .318 & .062 \\
                      & end & .470 & .070 & .378 & .082 \\
\end{tabular}
\label{tab:fisfrac_cycle}
\end{table}

\begin{figure}
\begin{subfigure}[b]{0.23\textwidth}
\includegraphics[trim = 0mm 0mm 0mm 0mm,width=\textwidth]{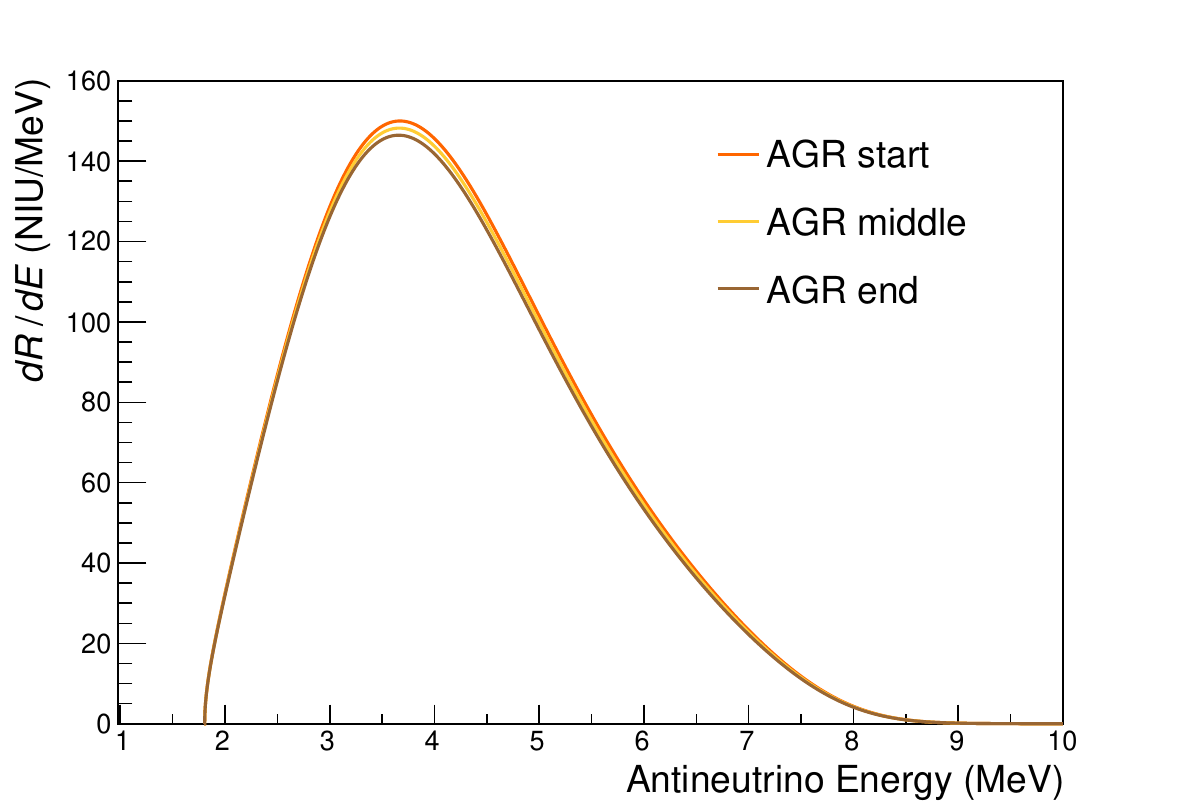}
\end{subfigure}
\begin{subfigure}[b]{0.23\textwidth}
\includegraphics[trim = 0mm 0mm 0mm 0mm,width=\textwidth]{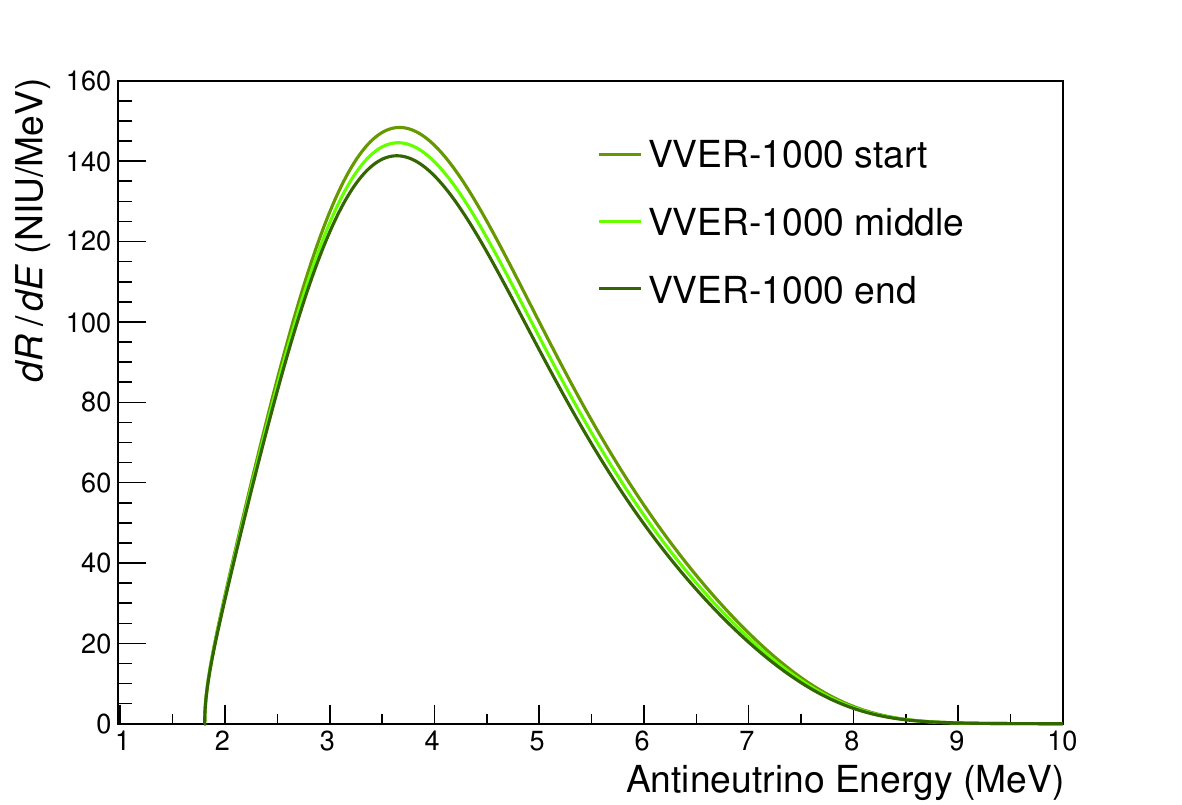}
\end{subfigure}
\caption{Quasi-elastic scattering reaction rates as a function of antineutrino energy are shown at different phases of the operating cycle for an advanced gas-cooled reactor \cite{mills}- AGR (left) and a pressurized water reactor \cite{kopeikin12}- VVER-1000 (right). Spectra use \eqref{sv03} without oscillations and are normalized to $1$ MW of thermal power at a standoff distance of $1$ km.}
\label{fig:rate_type}
\end{figure}

\begin{figure}
\begin{subfigure}[b]{0.23\textwidth}
\includegraphics[trim = 0mm 0mm 0mm 0mm,width=\textwidth]{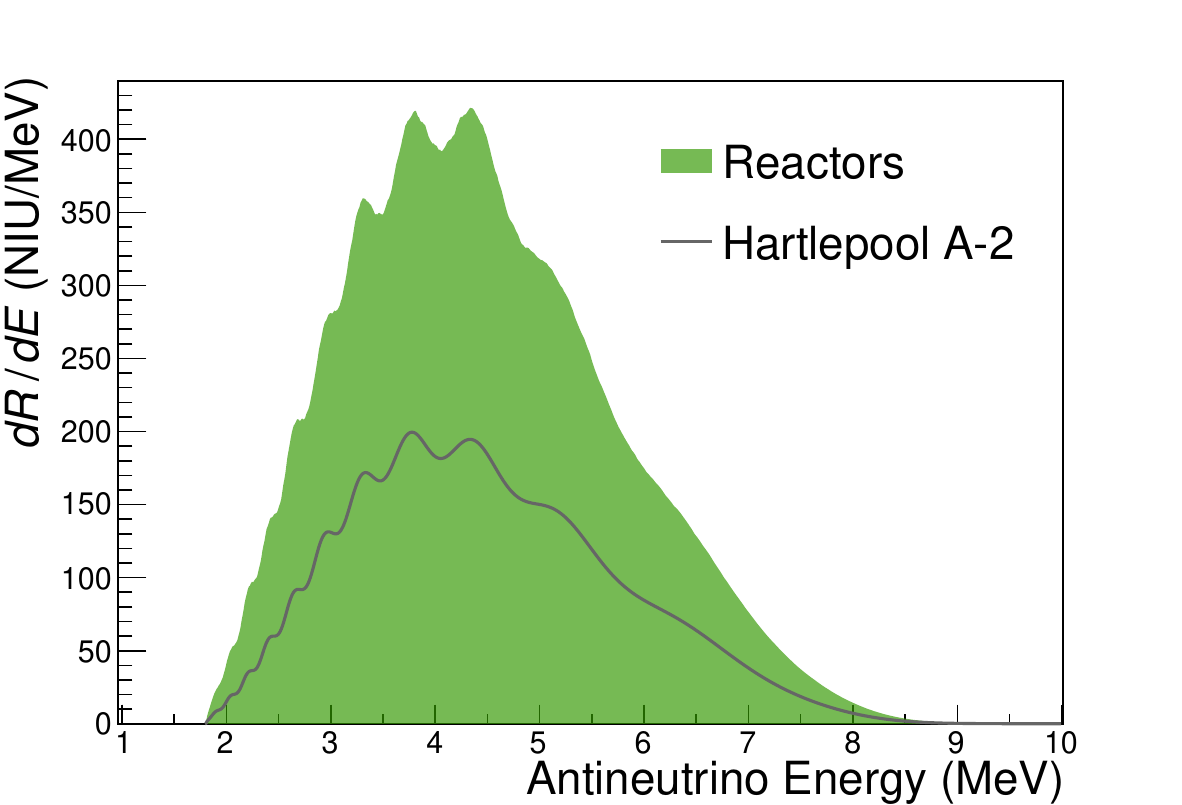}
\end{subfigure}
\begin{subfigure}[b]{0.23\textwidth}
\includegraphics[trim = 0mm 0mm 0mm 0mm,width=\textwidth]{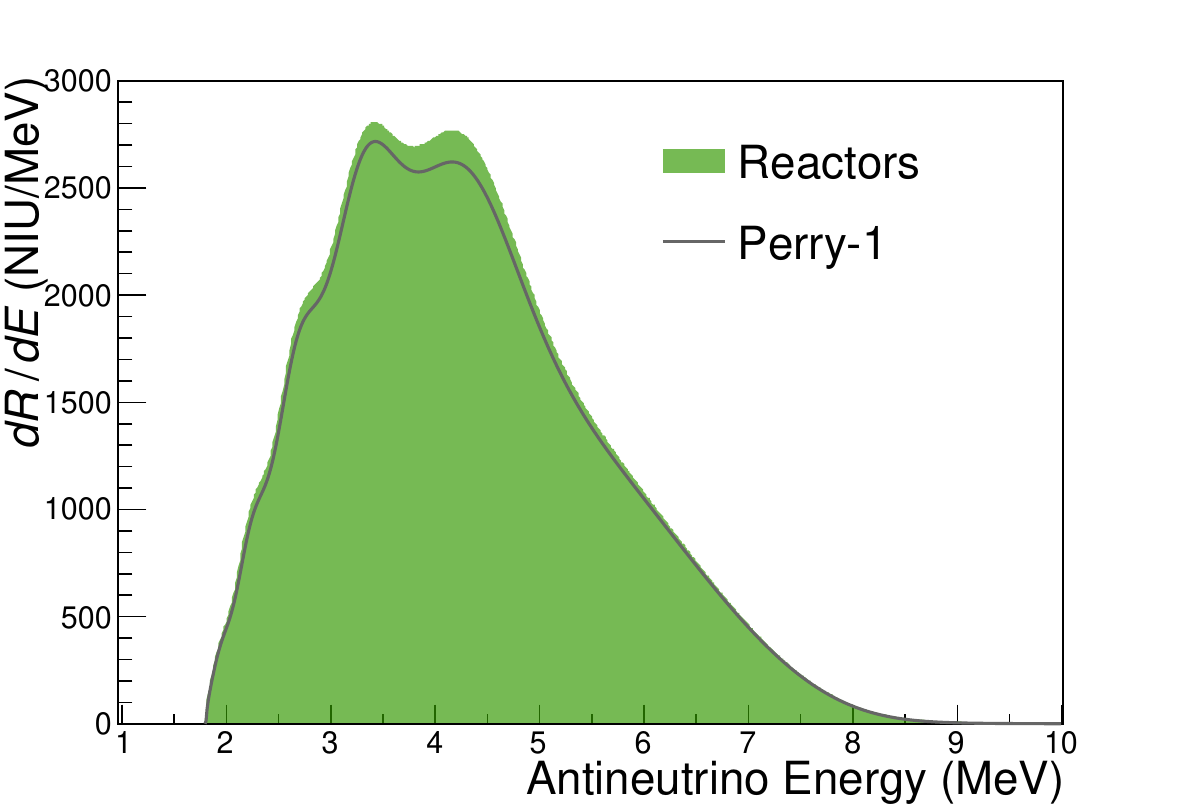}
\end{subfigure}
\caption{Quasi-elastic scattering reaction rates as a function of antineutrino energy from all nuclear power reactor cores are calculated with the 2018 annual average load factors at Boulby (left) and Morton (right). The grey line shows the contribution from the closest reactor core, which is Hartlepool A-2 ($0.877$ LF) $26.0$ km from Boulby and Perry-1 ($0.987$ LF) at $13.2$ km from Morton.}  
\label{fig:rate_spec}
\end{figure}

Calculation of the energy spectrum of quasi-elastic scattering reactions of reactor antineutrinos proceeds using the previously presented equations. Figure~\ref{fig:rate_type} shows the predicted spectrum by reactor type and fuel mixture at the Boulby Mine in the United Kingdom (Boulby) and at the Morton Salt Mine (Morton) in the United States. The rate at Boulby is dominated by GCR cores, while the rate at Morton is dominated by PWR and BWR cores.
\begin{figure}
\begin{subfigure}[b]{0.23\textwidth}
\includegraphics[trim = 0mm 0mm 0mm 0mm,width=\textwidth]{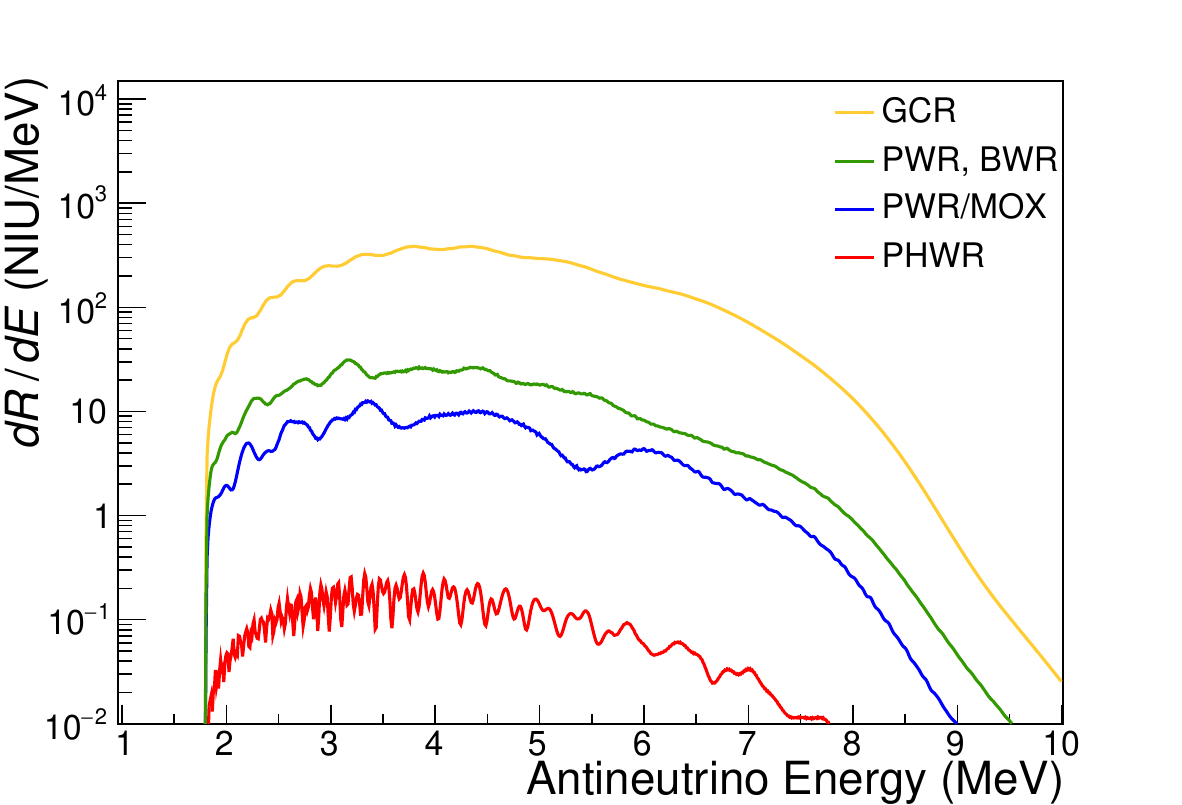}
\end{subfigure}
\begin{subfigure}[b]{0.23\textwidth}
\includegraphics[trim = 0mm 0mm 0mm 0mm,width=\textwidth]{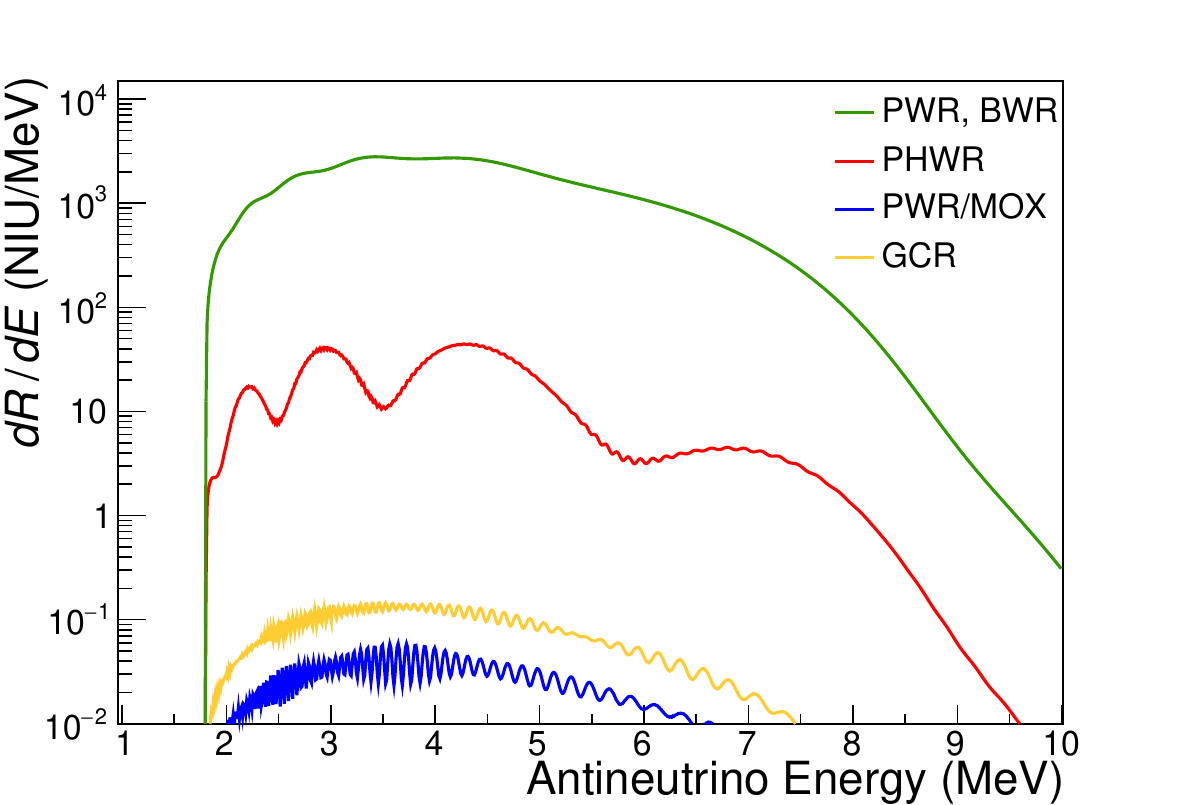}
\end{subfigure}
\caption{Quasi-elastic scattering reaction rates as a function of antineutrino energy from the different nuclear power reactor types (color-coded) are calculated with the 2018 annual average load factors at Boulby (left) and at Morton (right).}
\label{fig:rate_type}
\end{figure}
Figure~\ref{fig:rate_spec} shows the total quasi-elastic scattering reaction rate spectra of reactor antineutrinos at Boulby and at Morton. The rate at Boulby is dominated by the nearby ($\simeq26$ km) Hartlepool (GCR) reactor complex, while the rate at Morton is dominated by the nearby ($\simeq13$ km) Perry-1 (BWR) reactor. 

\begin{figure}
\begin{subfigure}[b]{0.23\textwidth}
\includegraphics[trim = 0mm 0mm 0mm 0mm,width=\textwidth]{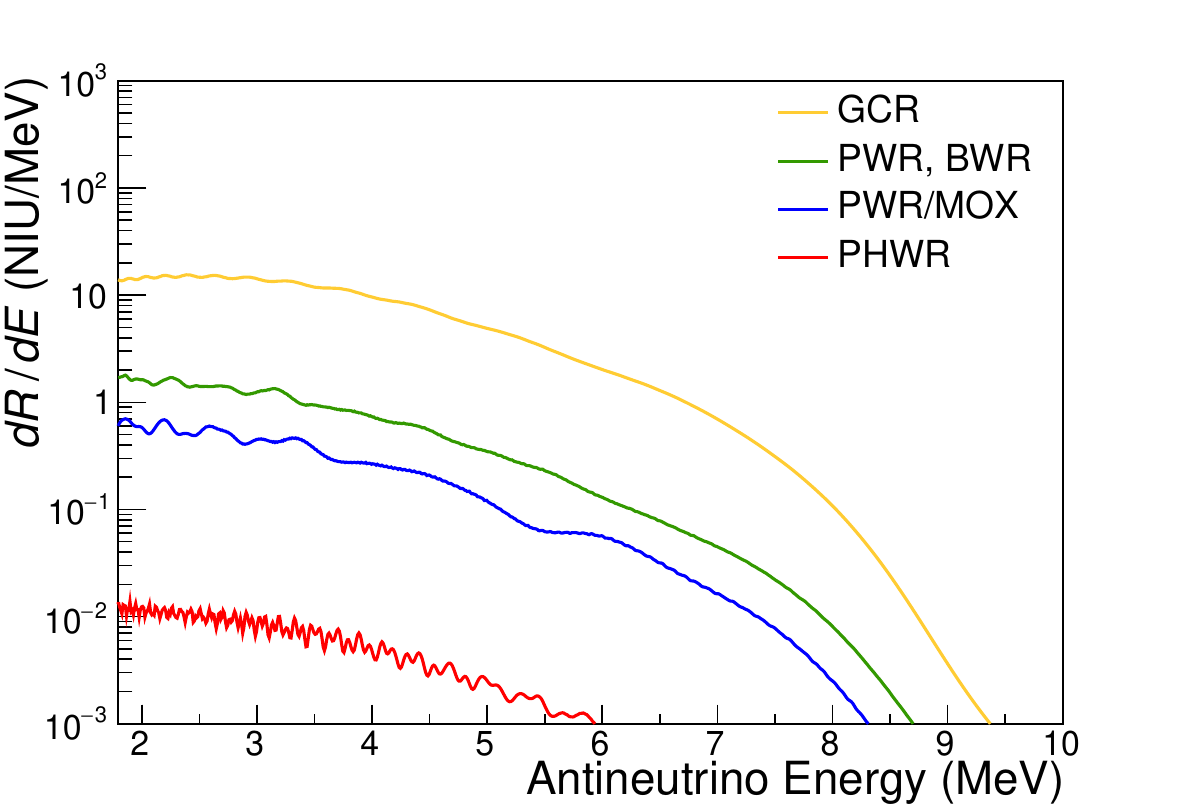}
\end{subfigure}
\begin{subfigure}[b]{0.23\textwidth}
\includegraphics[trim = 0mm 0mm 0mm 0mm,width=\textwidth]{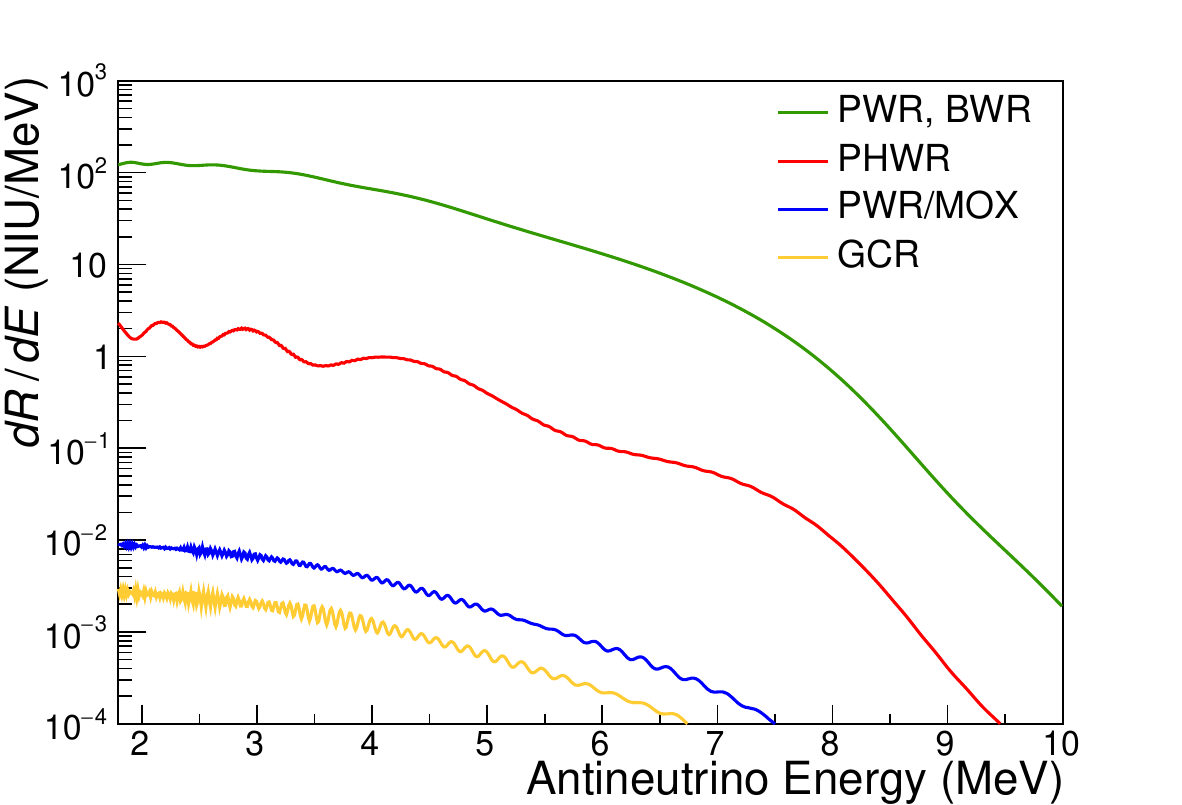}
\end{subfigure}
\caption{Elastic scattering reaction rates as a function of antineutrino energy from the different nuclear power reactor types (color-coded) are calculated with the 2018 annual average load factors at Boulby (left) and at Morton (right).}
\label{fig:rate_type_es}
\end{figure}

\begin{figure}
\begin{subfigure}[b]{0.23\textwidth}
\includegraphics[trim = 0mm 0mm 0mm 0mm,width=\textwidth]{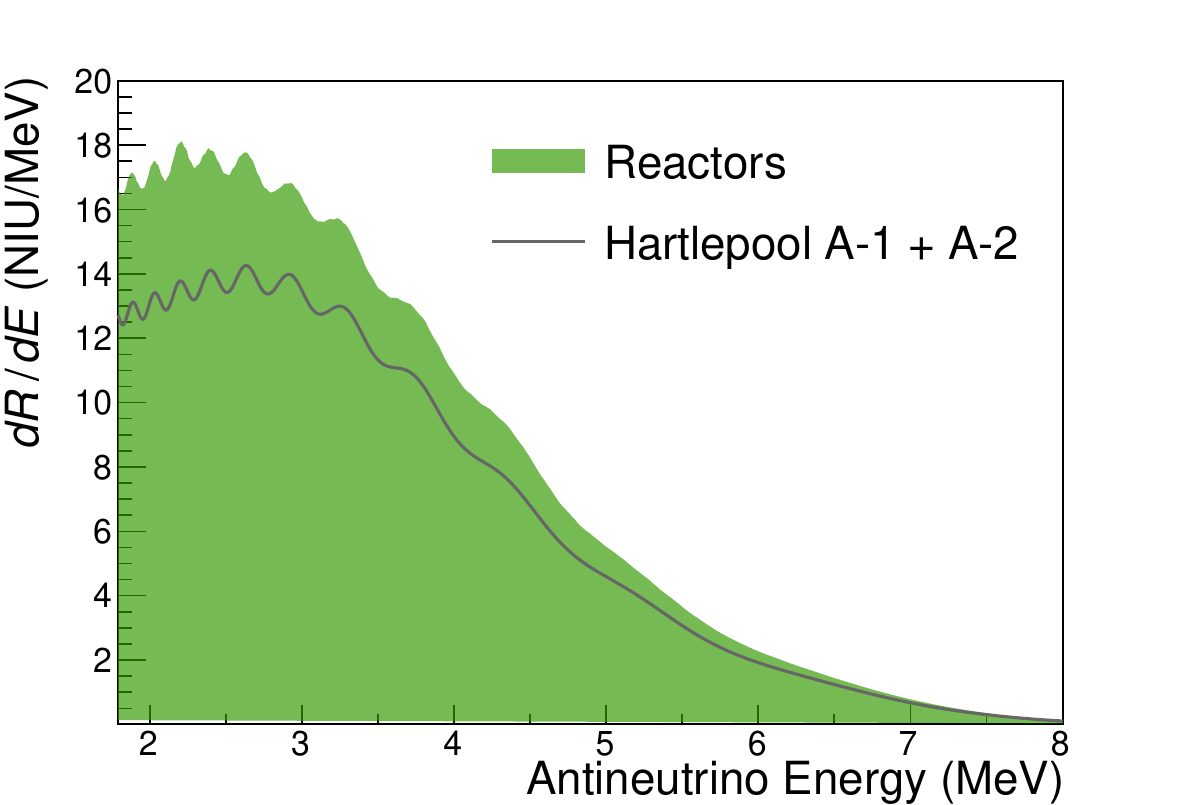}
\end{subfigure}
\begin{subfigure}[b]{0.23\textwidth}
\includegraphics[trim = 0mm 0mm 0mm 0mm,width=\textwidth]{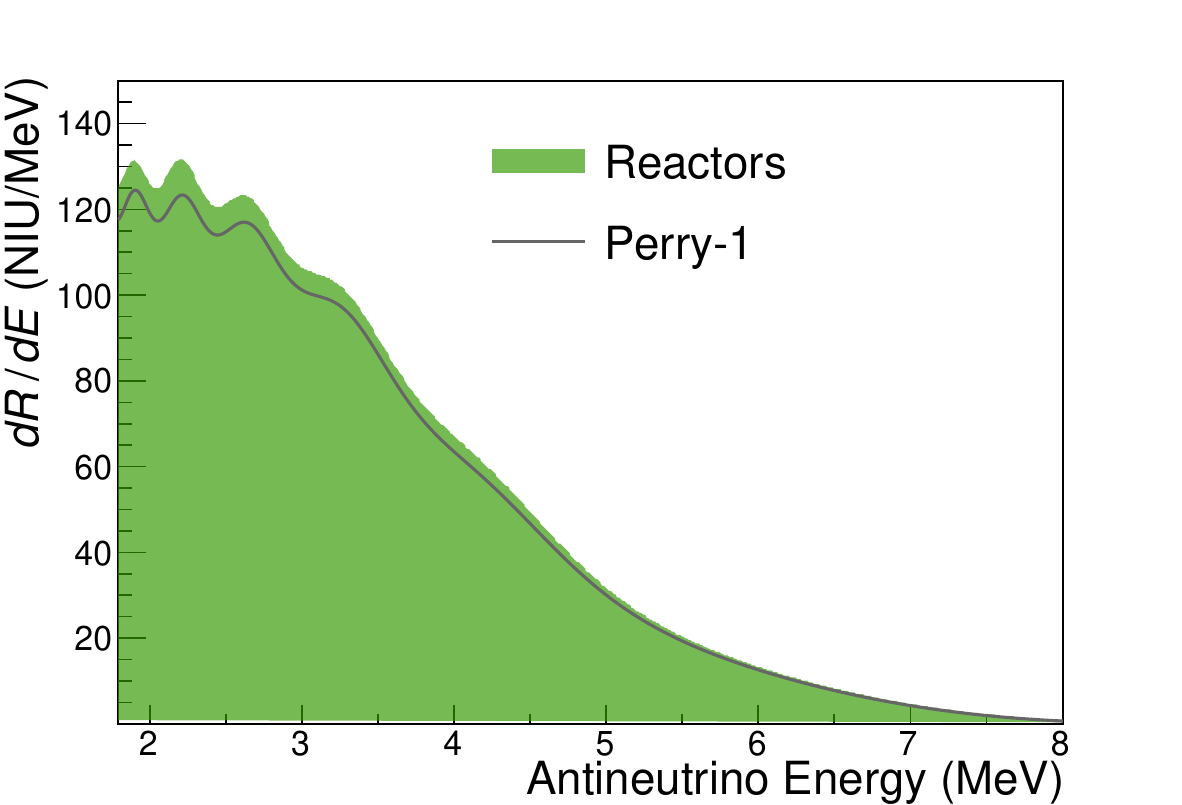}
\end{subfigure}
\caption{Elastic scattering reaction rates as a function of antineutrino energy from all nuclear power reactor cores are calculated with the 2018 annual average load factors at Boulby (left) and Morton (right). The grey line shows the contribution from the closest reactor core, which is Hartlepool A-2 ($0.877$ LF) $26.0$ km from Boulby and Perry-1 ($0.987$ LF) at $13.2$ km from Morton.}\label{fig:rate_spec_es}
\end{figure}

Calculation of the energy spectrum of elastic scattering reactions of reactor antineutrinos proceeds using the previously presented equations. Figure~\ref{fig:rate_type_es} shows the predicted spectrum by reactor type and fuel mixture at Boulby and at Morton. The rate at Boulby is dominated by GCR cores, while the rate at Morton is dominated by PWR and BWR cores. Figure~\ref{fig:rate_spec_es} shows the total elastic scattering reaction rate spectra of reactor antineutrinos at Boulby and at Morton. The rate at Boulby is dominated by the nearby ($\simeq26$ km) Hartlepool reactor complex, while the rate at Morton is dominated by the nearby ($\simeq13$ km) Perry-1 reactor. 

\begin{figure}
\begin{subfigure}[b]{0.23\textwidth}
\includegraphics[trim = 0mm 0mm 0mm 0mm,width=\textwidth]{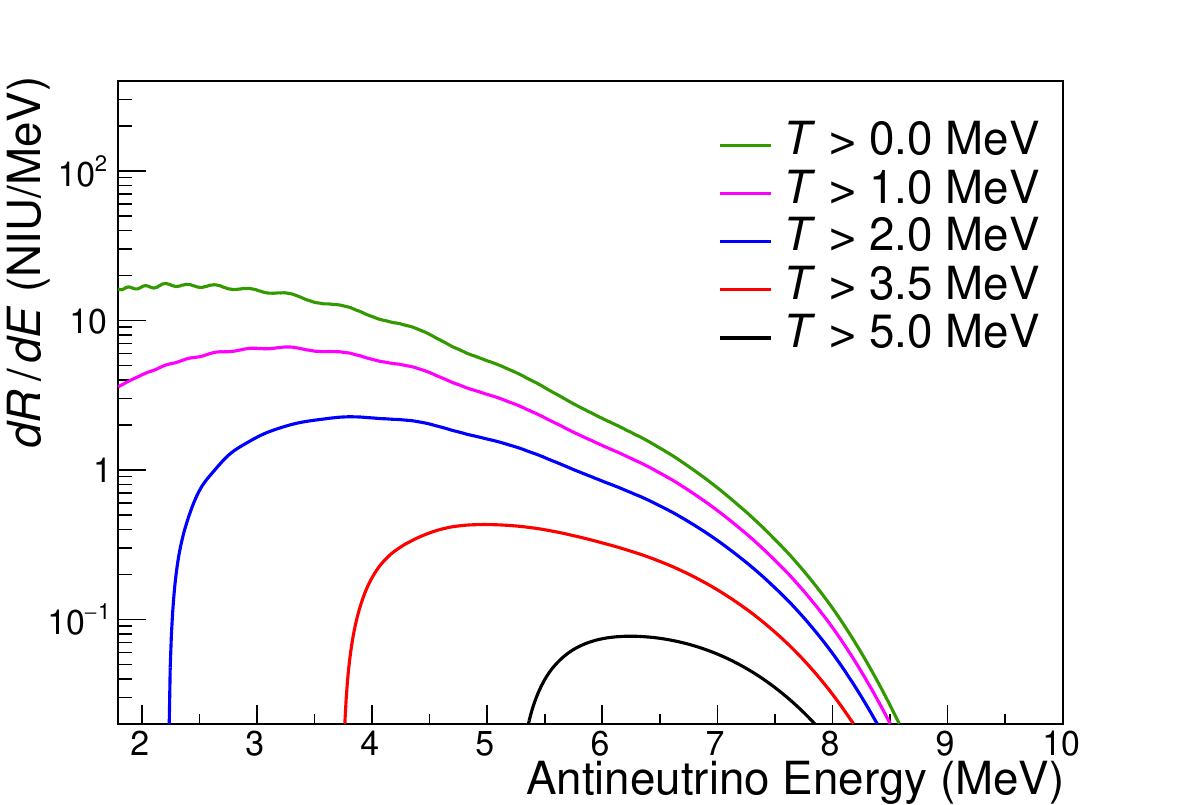}
\end{subfigure}
\begin{subfigure}[b]{0.23\textwidth}
\includegraphics[trim = 0mm 0mm 0mm 0mm,width=\textwidth]{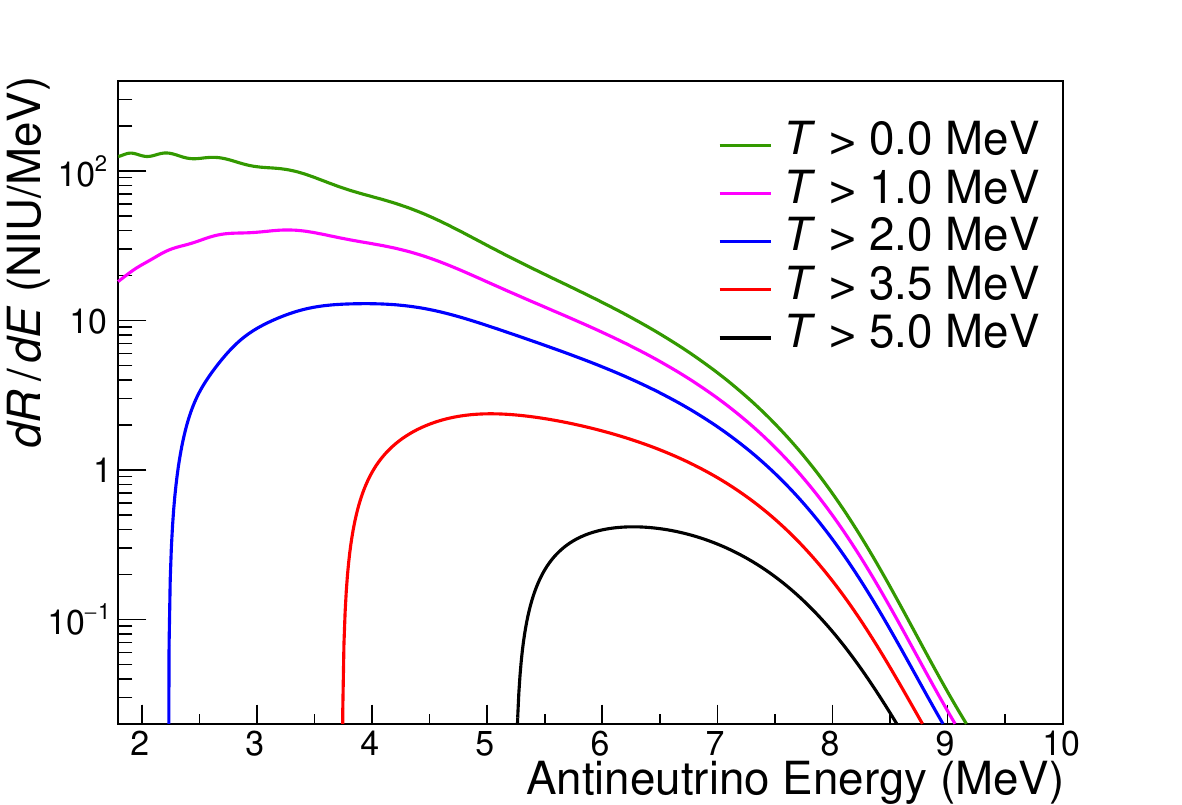}
\end{subfigure}
\caption{Energy spectra for reactor antineutrino elastic scattering reactions for several values of the minimum electron kinetic energy for Boulby (left) and Morton (right), using the annual average load factors from 2018.}
\label{fig:boulby_reac_spec_es_tmin}
\end{figure}

\begin{figure}
\begin{subfigure}[b]{0.23\textwidth}
\includegraphics[trim = 0mm 0mm 0mm 0mm,width=\textwidth]{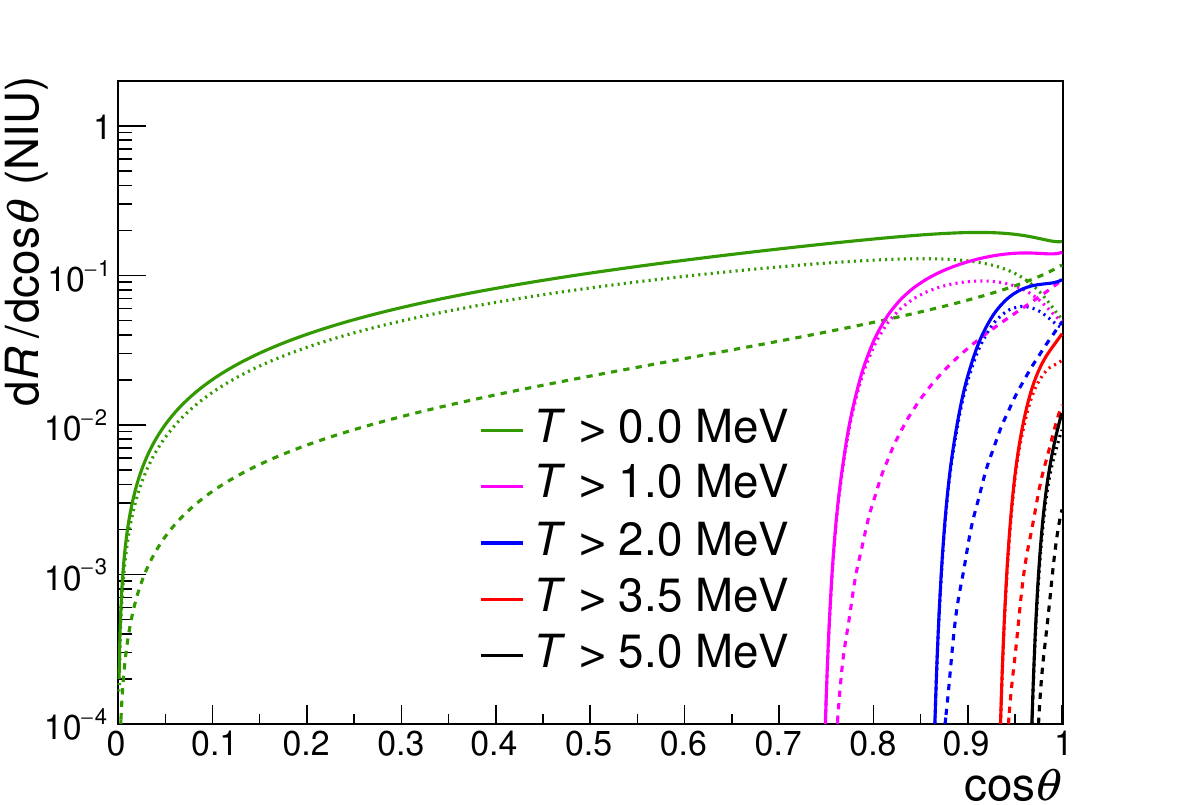}
\end{subfigure}
\begin{subfigure}[b]{0.23\textwidth}
\includegraphics[trim = 0mm 0mm 0mm 0mm,width=\textwidth]{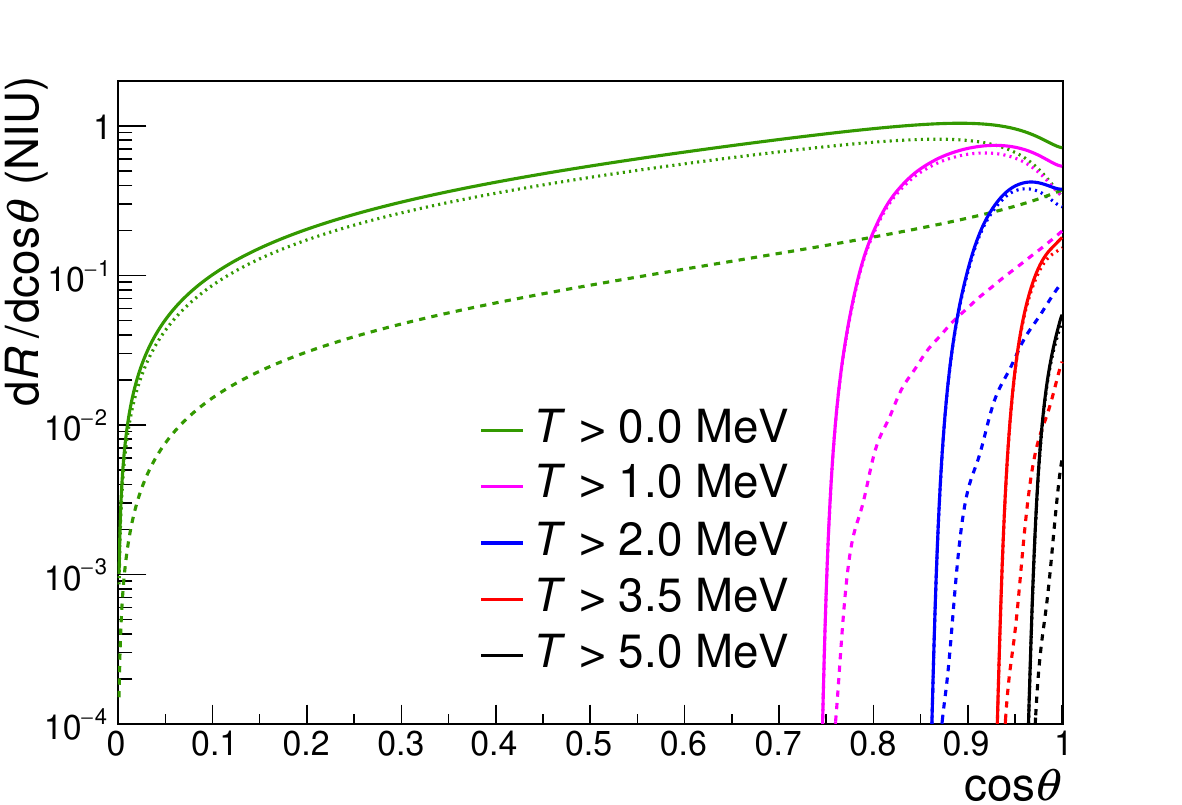}
\end{subfigure}
\caption{Elastic scattering reaction rates as a function of the cosine of the electron scattering angle $\cos \theta$ ($1000$ bins) for selected values of the minimum electron kinetic energy at Boulby (left) from the Hartlepool cores and at Morton (right) from the Perry-1 core are calculated at thermal capacity for $\overline{\nu}_\mathrm{e}$ (dotted), $\overline{\nu}_\mathrm{x}$ (dashed), and $\overline{\nu}_\mathrm{e} + \overline{\nu}_\mathrm{x}$  (solid).}
\label{fig:boulhart_costheta}
\end{figure}

The elastic scattering reaction rate decreases when restricting the electron kinetic energy $T$ to values greater than a minimum value $T_\mathrm{min}$. Figure~\ref{fig:boulby_reac_spec_es_tmin} shows reactor antineutrino reaction rate spectra estimated at Boulby and at Morton for several values of $T_\mathrm{min}$. The angular distributions of scattered electrons from the nearby Hartlepool cores at Boulby and from the nearby Perry-1 reactor at Morton are shown in Fig.~\ref{fig:boulhart_costheta}. The rate of reactions at each minimum kinetic energy are given in Table~\ref{tab:boulhart}. 

\begin{table}
\caption{Elastic Scattering Reaction Rates for Reactor Antineutrinos from Nearby Cores. Reaction rates for selected values of the minimum electron kinetic energy assume the 2018 annual average load factors for the Hartlepool cores nearby Boulby and for the Perry-1 core nearby Morton.}
\begin{tabular} {l c r r r r r}
\hline\noalign{\smallskip}
 && \multicolumn{5}{c}{ $T_\mathrm{min}$ (MeV)} \\
\cline{3-7}
  &                                                               & $0.0$ & $1.0$ & $2.0$ & $3.5$ & $5.0$ \\
\hline\noalign{\smallskip}
 &$\overline{\nu}_\mathrm{e}$                                              & $60.1$ & $14.76$ & $4.26$  & $0.763$  & $0.106$ \\
Boulby $R$ (NIU) &$\overline{\nu}_\mathrm{x}$                  & $23.3$ &  $7.06$  & $1.54$   & $0.199$  & $0.022$ \\
 & $\overline{\nu}_\mathrm{e} + \overline{\nu}_\mathrm{x}$ & $83.4$ & $21.82$ & $5.81$  & $0.963$  & $0.128$ \\
 \hline\noalign{\smallskip}
 &$\overline{\nu}_\mathrm{e}$                                              & $469.6$ & $109.6$ & $33.72$  & $5.48$  & $0.729$ \\
Morton $R$ (NIU) &$\overline{\nu}_\mathrm{x}$                       & $117.7$ & $16.8$   & $4.42$    & $0.64$  & $0.080$ \\
 & $\overline{\nu}_\mathrm{e} + \overline{\nu}_\mathrm{x}$ & $587.2$ & $126.3$  & $38.14$  & $6.12$  & $0.809$ \\
\hline
\end{tabular}
\label{tab:boulhart}
\end{table}

\section{Geo-neutrinos}
The dominant fluxes of geo-neutrinos originate from the radioactive decay of isotopes of uranium, thorium, and potassium. Uranium and thorium emit geo-neutrinos through a series of decays, leading to stable isotopes of lead, while potassium emits geo-neutrinos either by a single beta decay to calcium or electron capture to argon. These processes follow the general form
\begin{equation}
[A,Z] \rightarrow [A',Z'] + N_{\alpha} + N_\mathrm{e} + N_{\nu} + Q_{\mathrm{dk}}.
\end{equation}
Calculation of the decay energy $Q_\mathrm{dk}$ uses established values for the electron mass and the $\alpha$ particle mass as given in Table~\ref{tab:const}, along with atomic masses $M_A$, $M_{A'}$ and the number of $\alpha$ particles $N_\alpha$ as given in Table~\ref{tab:isodat}, according to
\begin{equation} \label{dkener}
Q_\mathrm{dk} = M_A - M_{A'} - N_{\alpha} (m_{\alpha} + 2 m_\mathrm{e}).
\end{equation}
Uncertainties are calculated assuming uncorrelated, Gaussian errors by standard methods, specifically 
\begin{equation} \label{dkunc}
\delta Q_\mathrm{dk} = [(\delta M_A)^2 + (\delta M_{A'})^2 + (N_{\alpha} \delta m_{\alpha})^2 + (2N_{\alpha} \delta m_\mathrm{e})^2]^{1/2}.
\end{equation}

\begin{table}
\caption{Radioisotope Decay Reaction Data. Isotope masses $M_A$ are in atomic mass units ($1\mathrm{u}=931.494102$ MeV) \cite{nist}. Calculations assume $m_\alpha = 3727.379240$ MeV and $m_\mathrm{e}$ as given in Table~\ref{tab:ibdcons}.}
\begin{tabular} {l r r r r r}
\hline\noalign{\smallskip}
$A$ &  $M_A$ (u) & $N_\alpha$  & $A'$  &  $M_{A'}$ (u)  & $Q_\mathrm{dk}$ (MeV)\\
\hline\noalign{\smallskip}
$^{238}$U             & 238.050788  & 8 &  $^{206}$Pb            & 205.974466    & 51.6835938 \\

$^{235}$U             & 235.043930   & 7 &  $^{207}$Pb            & 206.975897    & 46.4042969 \\

$^{232}$Th           & 232.038056   & 6 &  $^{208}$Pb            & 207.976652     & 42.6464844 \\

$^{40}$K$_\beta$ & 39.96399817  & 0 &  $^{40}$Ca       & 39.96259086        & 1.31119275 \\

$^{40}$K$_\mathrm{EC}$ & 39.96238312  & 0 &  $^{40}$Ar       & 39.96238312        & 1.50307465 \\
\hline
\end{tabular}
\label{tab:isodat}
\end{table}

\subsection{Isotopic geo-neutrino luminosity}
The geo-neutrino luminosities per unit mass of the elements uranium, thorium, and potassium depend on the contributions from each of the parent isotopes $^{238}$U, $^{235}$U, $^{232}$Th, and $^{40}$K. The contributions are functions of well known quantities. Specifically, the isotopic luminosity per unit mass is
\begin{equation} \label{lumiso}
l = \frac{\mathrm{ln}(2)} {t_{{}^1{\mskip -5mu/\mskip -3mu}_2}} \frac {n_{\nu} BR} {M_A},
\end{equation}
where $t_{{}^1{\mskip -5mu/\mskip -3mu}_2}$ is the half-life, and $n_{\nu}$ is the number of emitted neutrinos per decay, and $BR$ is the branching ratio. The measured input values and the resulting luminosity per unit mass, according to \eqref{lumiso}, are given in Table~\ref{tab:isolumi} for each isotope. Half-life 
is the source of largest uncertainty in the isotopic neutrino luminosity.

\begin{table}
\caption{Isotopic Neutrino Luminosity per Unit Mass. Uranium half-life data from Jaffey et al. 1971 Phys Rev C 4, 1889.} 
\begin{tabular} {l l r r r}
\hline\noalign{\smallskip}
Isotope $A$          & $BR$ &{    $n_{\nu}$}   &  {    $t_{{}^1{\mskip -5mu/\mskip -3mu}_2}$ (Gy)}  & {    $l$(/kg/$\mu$s)}  \\[2pt]
\hline\noalign{\smallskip}
$^{238}$U    & 1.00 & 6                         & 4.468(3)                  &  74.601  \\

$^{235}$U    & 1.00 & 4                         & .7038(5)                &  319.841  \\

$^{232}$Th  & 1.00 & 4                         & 14.05(6)                  & 16.229  \\


$^{40}$K$_\beta$  & 0.8928 & 1     & 1.277(8)                 &  231.402  \\ 

$^{40}$K$_\mathrm{EC}$    & 0.1072 & 1   & 1.277(8)      &  27.785  \\ 
\hline
\end{tabular}
\label{tab:isolumi}
\end{table}
\subsection{Isotopic geo-neutrino spectra}\label{subsec:gnuspec}
Geo-neutrinos from the actinide elements uranium and thorium are exclusively antineutrinos. These antineutrinos exhibit continuous and, owing to the varied energy states of the daughter nuclei, rather complicated energy spectra. Geo-neutrinos from the alkali metal potassium are both continuous energy antineutrinos and mono-energetic neutrinos, mixed according to the branching ratios given in Table~\ref{tab:isolumi}. 

The electron capture reaction is
\begin{equation}
[A,Z] + \mathrm{e}^- \rightarrow [A,Z-1] + \nu_\mathrm{e} + Q_\beta \text{.}
\end{equation}
This two-body decay leads to mono-energetic neutrinos. Although the decay energy of about $1.5$ MeV is potentially available to the neutrino, this branch is heavily suppressed. We do not further consider the contribution to the geo-neutrino flux from the electron capture reaction of $^{40}$K.

The beta decay reaction for electron antineutrino emission is
\begin{equation}
[A,Z] \rightarrow [A,Z+1] + \mathrm{e}^- + \overline{\nu}_\mathrm{e} + Q_\beta \text{.}
\end{equation}
This three body decay shares energy and momentum between the emitted electron and antineutrino. Neglecting the mass of the antineutrino, conservation of energy dictates 
$w_\mathrm{e}=Q_\beta + m_\mathrm{e} - E_{\overline{\nu}_\mathrm{e}}$,
where $w_\mathrm{e}$ is the electron energy and $m_\mathrm{e}$ is the electron mass. The momentum of the electron is
$p_\mathrm{e}= \sqrt{w_\mathrm{e}^2 - m_\mathrm{e}^2}$.
An approximation of the differential energy spectrum of neutrinos from beta decay \cite{pres62}, which ignores forbidden transitions, specifies the proportionality
\begin{equation} \label{betaspec}
dn(E_{\overline{\nu}_\mathrm{e}}) / dE \propto w_\mathrm{e}E_{\overline{\nu}_\mathrm{e}}^2 p_\mathrm{e}^{2\gamma - 1} \mathrm{e}^{\pi \eta} |\Gamma(\gamma + i\eta)|^2 \text{.}
\end{equation}
The final two factors estimate the Fermi function that corrects for Coulomb effects on the emitted beta, where $\Gamma$ is the gamma function. In terms of the fine structure constant $\alpha$,
$\gamma = \sqrt{1 - \alpha^2(Z + 1)^2}$ 
and
$ \eta = \alpha(Z + 1) w_\mathrm{e}/ p_\mathrm{e}$.

\begin{figure}
\includegraphics[trim = 0mm 0mm 0mm 0mm, clip,scale=0.45]{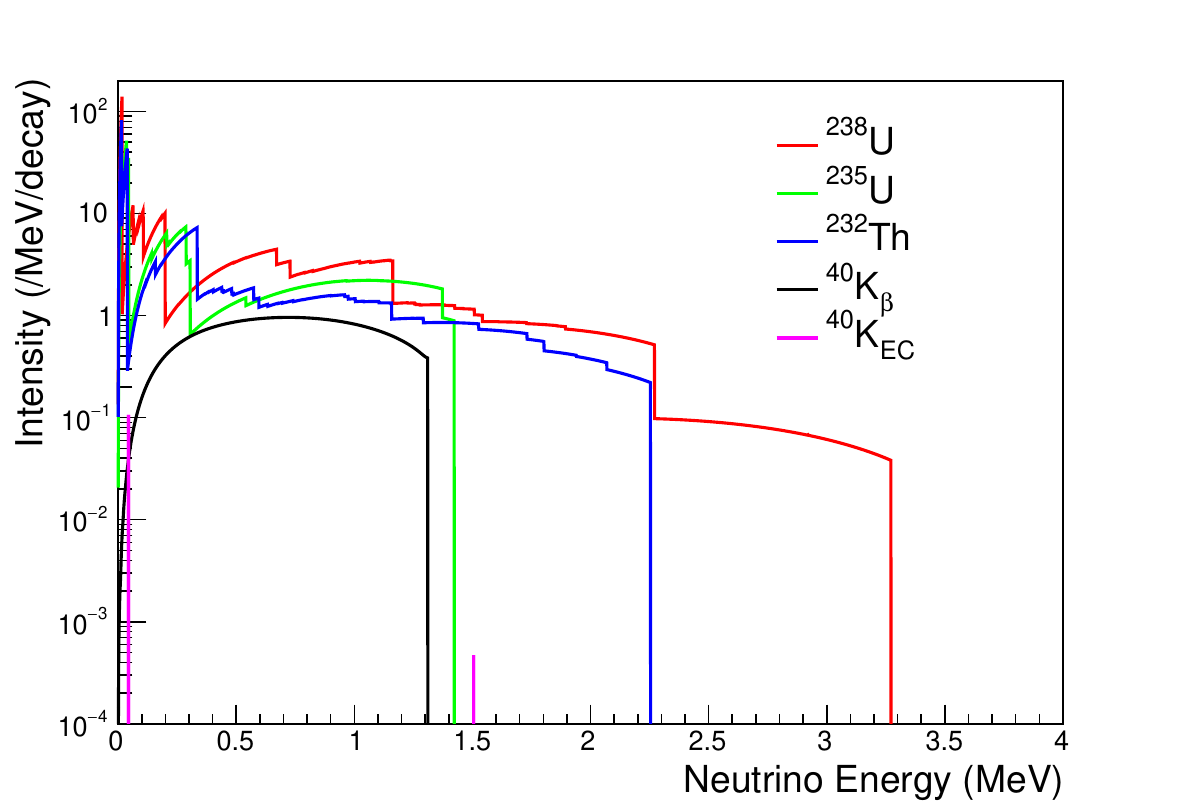}
\caption{Geo-neutrino spectra are normalized to the number of antineutrinos emitted in the cascade of decays of $^{238}$U, $^{235}$U, and $^{232}$Th, and to the branching ratio of $^{40}$K$_\beta$ decay and computed elsewhere \cite{sanshiro_spec}. The emission of neutrinos from $^{40}$K$_\mathrm{EC}$ are split in two lines with total intensity equal to the branching ratio.}
\label{fig:gnuint}
\end{figure}

The preceding equations permit expression of the beta spectrum \eqref{betaspec} in terms of the antineutrino energy. Summing the neutrino spectra of all beta decays in a given decay series, appropriately weighted for branching ratios and intensity fractions for transitions to the various energy states of the daughter nuclei, estimates the antineutrino spectrum for the parent isotope. 
The model uses pre-calculated spectra, which include the forbidden transitions of $^{40}$K and $^{210}$Bi \cite{sanshiro_spec}, for estimating the geo-neutrino reaction rates. Antineutrino energy spectra for $^{238}$U, $^{235}$U, $^{232}$Th, and $^{40}$K are shown in Figure~\ref{fig:gnuint}. These spectra are normalized such that their areas equal the number of neutrinos emitted per decay multiplied by the branching ratio as given in Table~\ref{tab:isolumi}.

\subsection{Isotopic radiogenic heating}
To determine the radiogenic heating per unit mass produced by each of the parent isotopes requires an evaluation of the mean energy carried off by the neutrinos \eqref{avgenu}. All other radiations relinquish their kinetic energy to Earth. Averaging each spectrum in Figure~\ref{fig:gnuint} yields the required values. Subtracting the mean energy carried away by neutrinos from the decay energy computes the average radiogenic heat absorbed by Earth per decay of the parent isotope \eqref{isoheat}. 
\begin{equation}\label{avgenu}
<\!\!{Q}_\nu\!\!> = \frac {\int E_{\overline{\nu}_\mathrm{e}} \big(dn_i(E_{\overline{\nu}_\mathrm{e}}) / dE \big) \, dE} {\int \big(dn_i(E_{\overline{\nu}_\mathrm{e}}) / dE \big)\, dE} \text{,}
\end{equation}

\begin{equation}
\label{isoheat}
Q_h = Q_\mathrm{dk} - <\!\!{Q}_\nu\!\!>.
\end{equation} 
Specifically, the radiogenic heating per unit mass is
\begin{equation} \label{lumele}
h = \frac{\mathrm{ln}(2)} {t_{{}^1{\mskip -5mu/\mskip -3mu}_2}} \frac {Q_h} {M_A} = \frac{l} {n_\nu} Q_h,
\end{equation}
where $t_{{}^1{\mskip -5mu/\mskip -3mu}_2}$ is the half-life as given in Table~\ref{tab:isolumi}, $M_A$ is the isotope mass, and $Q_h$ is the average isotopic heat deposit \eqref{isoheat}. The calculated energies and the resulting heating per unit mass, according to \eqref{lumele}, are given in Table~\ref{tab:isoheat} for each isotope, where $Q_\mathrm{dk}$ values are converted from MeV to pJ using the established value for the electron charge as given in Table~\ref{tab:const}. 
The uncertainty of $h$ is calculated by standard methods, specifically
\begin{equation}
\frac{\delta h} {h} = \bigg [ \bigg ( \frac{\delta M_A} {M_A} \bigg )^2 + \bigg ( \frac{\delta Q_h} {Q_h} \bigg )^2 \bigg ] ^{{}^1{\mskip -5mu/\mskip -3mu}_2},
\end{equation}
with
\begin{equation}
\delta Q_h = [ (\delta Q_\mathrm{dk})^2 + (\delta <\!\!{Q}_\nu\!\!>)^2]^{{}^1{\mskip -5mu/\mskip -3mu}_2}.
\end{equation}

\begin{table}
\caption{Isotopic Radiogenic Heating per Unit Mass}
\begin{tabular} {l r r r}
\hline\noalign{\smallskip}
{Isotope $A$ }          &     {    $<\!\!{Q}_\nu\!\!>$ (MeV)  }  & $Q_h$ (MeV)    & {    $h$ ($\mu$W/kg)} \\ 
\hline\noalign{\smallskip}
$^{238}$U    &   0.674114685 &  51.0094791  &   101.614 \\
$^{235}$U    &   0.502104065 &  45.9019256  &588.051 \\
$^{232}$Th  &   0.561477661 &   42.0850067  & 27.357 \\
$^{40}$K$_\beta$  &    0.722908630 &  0.58828412  &  21.810 \\
$^{40}$K$_{\text{EC}}$   &  0.050438676 & 1.45263597  &   6.467 \\
\hline
\end{tabular}
\label{tab:isoheat}
\end{table}

\subsection{Elemental luminosity and heat production}
Of primary relevance to geology is the amount of uranium, thorium, and potassium in Earth. Calculation of the elemental luminosity and radiogenic heating per unit mass of these elements uses knowledge of the natural abundances of the parent radioactive isotopes as given in Table~\ref{tab:nabund}. These values are presented in Table~\ref{tab:eleluhe}.

\begin{table}
\caption{Percent Natural Abundance of Parent Isotopes}
\begin{tabular} {r r r r r}
\hline\noalign{\smallskip}
               & { $^{238}$U  }  & { $^{235}$U  }  & { $^{232}$Th  }   & { $^{40}$K  } \\
\hline\noalign{\smallskip}
 NA (\%) & 99.2745(60)  & 0.7200(12)     & 100.0(0)          & 0.0117(1) \\
  \hline
  \end{tabular}
  \label{tab:nabund}
  \end{table}

\begin{table}
\caption{Elemental Neutrino Luminosity and Radiogenic Heating per Unit Mass}
\begin{tabular} {r r r r r}
\hline\noalign{\smallskip}
                                  & Uranium    & Thorium        & Potassium \\
\hline\noalign{\smallskip}
 $l$ (/kg/$\mu$s)      &    76.362  &   16.229   & 0.0303   \\
 $h$ ($\mu$W/kg)     &    105.111  &   27.357    & 0.00331 \\
\hline
\end{tabular}
\label{tab:eleluhe}
\end{table}

\subsection{Geo-neutrino reaction rates}
Geo-neutrino reaction rates derive from Earth's crust and mantle. This is consistent with the standard geochemical model \cite{mcd95}, which confines potassium (K), thorium (Th), and uranium (U) almost exclusively in the rocky layers of Earth with negligible amounts in the metallic core. The estimated reaction rate spectrum per target per unit time from the $i^{th}$ isotope is the product of the isotopic geo-neutrino flux $\phi_i$, the cross section, $\sigma^\mathrm{IBD}(E_{\overline{\nu}})$ or $\sigma^\mathrm{ES}(E_{\overline{\nu}})$, and the differential energy spectrum, all scaled by the average oscillation probability $<\!\!{P}_\mathrm{ee}\!\!>$ \eqref{avgposc}. Explicitly,
\begin{equation}
\label{gnuratespec}
\frac {dR_i (E_{\overline{\nu}_\mathrm{e}})} {dE} = <\!\!{P}_\mathrm{ee}\!\!> \frac{\phi_i}{n_{\nu_{i}}} \sigma(E_{\overline{\nu}_\mathrm{e}}) \frac{dn_i(E_{\overline{\nu}_\mathrm{e}})} {dE} .
\end{equation}
Total isotopic reaction rates $R_i$ easily obtain by integrating \eqref{gnuratespec}
\begin{equation}
\label{intrate}
R_i = \int \frac{dR_i(E_{\overline{\nu}_\mathrm{e}})} {dE} dE.
\end{equation}

The model uses a pre-computed prediction of the non-oscillated surface fluxes from $^{40}$K, $^{232}$Th, and $^{238}$U in the crust  \cite{huang13}. It uses a default value for the non-oscillated mantle geo-neutrino flux from $^{238}$U, $\phi(^{238}\mathrm{U}) = 1.00 \times 10^6$ cm$^{-2}$ s$^{-1}$. This flux corresponds to an IBD reaction rate of $7.46$ NIU, including neutrino oscillations. A seismological model \cite{dziew81} defines a mantle density profile with spherical symmetry and leads to the geophysical response \cite{kgs84}. 
At a radius of $6371$ km, the mantle geophysical response given by this model sums to $11.77 \times 10^5$ kg/cm$^2$. Together with the default mantle $^{238}$U flux, this defines a constant isotopic mass abundance of $11.7 \times 10^{-9}$ g($^{238}$U)/g, providing about $4.6$ TW of radiogenic heating. Abundances of the isotopes of a given element follow from the natural abundances $a_i = \mathrm{NA}_i a$. The default values of the element mass abundance ratios are $a_{\mathrm{Th}}/a_{\mathrm{U}}=3.9$ and $a_{\mathrm{K}}/a_{\mathrm{U}}=10^4$. The model accepts alternative values for the $^{238}$U flux and the element mass abundance ratios.

The mantle geo-neutrino fluxes from isotopes scale with the isotopic luminosity and mass abundance ratios, as given by
\begin{equation} \label{fluxes}
\frac {\phi_i } {\phi_j} = \frac {l_i} {l_j} \frac{a_i} {a_j}.
\end{equation}
Specifically,
\begin{equation} \label{fluxus}
\frac {\phi \mathrm{(}^{235}\mathrm{U)}} {\phi \mathrm{(}^{238}\mathrm{U)}} = \frac {l \mathrm{(}^{235}\mathrm{U)}}  {l \mathrm{(}^{238}\mathrm{U)}}
\frac {\mathrm{NA(} ^{235}\mathrm{U)}} {\mathrm{NA(} ^{238}\mathrm{U)}},
\end{equation}
\begin{equation} \label{fluxkappa}
\frac {\phi \mathrm{(}^{232}\mathrm{Th)}} {\phi \mathrm{(}^{238}\mathrm{U)}} = 
\frac {l \mathrm{(}^{232}\mathrm{Th)}} {l \mathrm{(}^{238}\mathrm{U)}} 
\frac {\mathrm{NA(} ^{232}\mathrm{Th)} a_\mathrm{Th}} {\mathrm{NA(} ^{238}\mathrm{U)}a_\mathrm{U}}
\end{equation}
and
\begin{equation} \label{fluxku}
\frac {\phi \mathrm{(}^{40}\mathrm{K)}} {\phi \mathrm{(}^{238}\mathrm{U)}} = 
\frac {l \mathrm{(}^{40}\mathrm{K)}} {l \mathrm{(}^{238}\mathrm{U)}} 
\frac {\mathrm{NA(} ^{40}\mathrm{K)}a_\mathrm{K}} {\mathrm{NA(} ^{238}\mathrm{U)}a_\mathrm{U}} 
\end{equation}


The previous results enable a calculation of factors for converting the signal rates of inverse beta decay and elastic scattering reactions to the detected fluxes of geo-neutrinos ($\phi_i = C_i R_i$). In units of $10^{32}$ target-year cm$^{-2}$ s$^{-1}$ these factors are given by
\begin{equation}\label{rateflux}
C_i = \frac{1}{3.15576\times10^{39}} \frac{\int \big(dn_i(E_{\overline{\nu}_\mathrm{e}}) / dE \big) \,dE} {\int \sigma(E_{\overline{\nu}_\mathrm{e}}) \big(dn_i(E_{\overline{\nu}_\mathrm{e}})/dE \big)\, dE} \text{,}
\end{equation}
where the numerical factor in the denominator is the number of seconds in one year multiplied by $10^{32}$ targets. This calculation, which depends on the reaction cross section \eqref{vb99}\eqref{sv03}\eqref{esxsec}, produces the values listed in Table~\ref{tab:convert}.

\begin{table}
\caption{Rate to Flux Conversion Factors. IBD and ES reaction rates convert to geo-neutrino fluxes, using these factors with the unit $10^{32}$ target-year cm$^{-2}$s$^{-1}$.}
\begin{tabular} {l r r r r}
\hline\noalign{\smallskip}
                                                                          & $^{238}$U  & $^{235}$U & $^{232}$Th & $^{40}$K$_\beta$ \\
\hline\noalign{\smallskip}
IBD- VB99 \cite{vogel99}                                  &    75573.4   &                  &   241727  &  \\
IBD- SV03 \cite{strumia03}                              &    75483.4    &                  &   239787  &  \\
ES- $\overline{\nu}_\mathrm{e}\mathrm{e}$    &   139613.0   & 196267.4  &  170975.9 &  134662.9 \\
ES- $\overline{\nu}_x\mathrm{e}$                    &   356457.6   & 489439.6 &  433063.2  & 333348.7 \\
 \hline
\end{tabular}
\label{tab:convert}
\end{table}
These results enable interpretation of observed geo-neutrino signal rates. Suppose the number of reactions does not permit a direct measurement of separate signals from uranium and thorium. In this case, the reported rate assumes the signal was due to a geological reservoir with a certain thorium to uranium elemental mass ratio ($a_\mathrm{Th}/a_\mathrm{U}$). The thorium to uranium geo-neutrino flux ratio detected from this reservoir is directly found. Using \eqref{fluxkappa}, it is straightforward to calculate the ratio of the signal rates due to different isotopes.
\begin{equation} \label{sigflux}
\frac {R_i} {R_j} = \frac {\phi_i} {\phi_j} \frac  {C_j} {C_i}
\end{equation}
The separate contributions to the IBD signal rate $R\mathrm{(}^{238}\mathrm{U)}$ and $R\mathrm{(}^{232}\mathrm{Th)}$ are found directly from \eqref{sigflux} and the requirement $R_{\mathrm{total}} = R\mathrm{(}^{238}\mathrm{U)} + R\mathrm{(}^{232}\mathrm{Th)}$.

The mass abundance ratios ($a_\mathrm{Th} / a_\mathrm{U}$ and $a_\mathrm{K} / a_\mathrm{U}$) of elements, which produce geo-neutrinos and radioactive heat, vary throughout Earth. Geo-neutrino measurements are sensitive to these variations if the detected reaction rates from the respective elements are separately resolved. Assuming $100\%$ detection efficiency across the energy spectrum, the mass abundance ratios calculated from the reaction rates are
\begin{equation} \label{thurat}
\frac {a_\mathrm{Th}} {a_\mathrm{U}} = \frac {R\mathrm{(}^{232}\mathrm{Th)}} {R\mathrm{(}^{238}\mathrm{U)}}  \frac {C\mathrm{(}^{232}\mathrm{Th)}} {C\mathrm{(}^{238}\mathrm{U)}}   \frac {l\mathrm{(}^{238}\mathrm{U)}} {l\mathrm{(}^{232}\mathrm{Th)}} \frac {\mathrm{NA(}^{238}\mathrm{U)}} {\mathrm{NA(}^{232}\mathrm{Th)}}
\end{equation}
and, exclusively for elastic scattering reactions,
\begin{equation} \label{kurat}
\frac {a_\mathrm{K}} {a_\mathrm{U}} = \frac {R\mathrm{(}^{40}\mathrm{K_\beta)}} {R\mathrm{(}^{238}\mathrm{U)}}  \frac {C\mathrm{(}^{40}\mathrm{K_\beta)}} {C\mathrm{(}^{238}\mathrm{U)}}   \frac {l\mathrm{(}^{238}\mathrm{U)}} {l\mathrm{(}^{40}\mathrm{K_\beta)}} \frac {\mathrm{NA(}^{238}\mathrm{U)}} {\mathrm{NA(}^{40}\mathrm{K)}}.
\end{equation}
To account for an energy dependent detection efficiency, simply include this factor inside the integrals in both the numerator and the denominator of \eqref{rateflux}.



\section{Distance calculation}
The model estimates the reactor antineutrino reaction rate and energy spectrum at any location near the surface of Earth. It calculates the distance between the selected location and each nuclear power reactor core. This distance appears as $d$ in the reaction rate spectra \eqref{isorate} and as $L$ in the oscillation probability equation \eqref{nuosc}. Latitude, longitude, and elevation relative to the reference ellipsoid (WGS84) specify the geocentric rectangular coordinates ($X,Y,Z$) of each position. The elevations of power reactor cores are available \cite{jocher} as are those of existing and potential underground and underwater detector sites. The elevation of all other locations is set by default to the surface of the reference ellipsoid.
With geocentric rectangular coordinates established for each source (core) and field (detector) location the distance calculation is
\begin{equation} \label{distance}
d = \sqrt{(X_s - X_f)^2 + (Y_s - Y_f)^2 + (Z_s - Z_f)^2}.
\end{equation}
Distances less than $100$ km retain the full precision ($64$ bit) offered by the code, while those greater than $100$ km are rounded to the nearest km. Rounding introduces a maximum reaction rate uncertainty of $<\sim1$\% to the contribution from an individual reactor core.

\section{Direction calculation}
\begin{figure}
\begin{subfigure}[b]{0.23\textwidth}
\includegraphics[trim = 0mm 0mm 0mm 0mm,width=\textwidth]{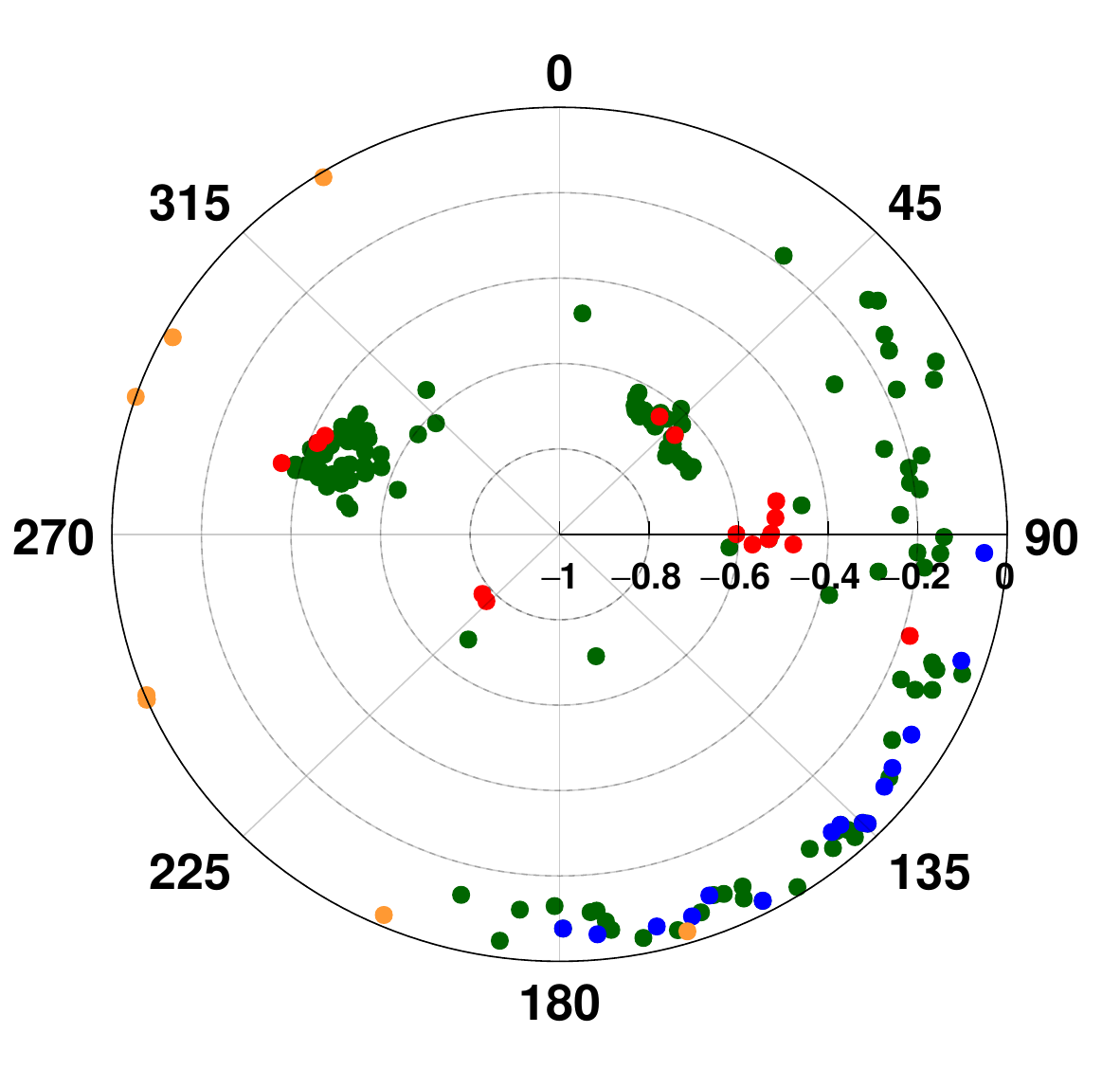}
\end{subfigure}
\begin{subfigure}[b]{0.23\textwidth}
\includegraphics[trim = 0mm 0mm 0mm 0mm,width=\textwidth]{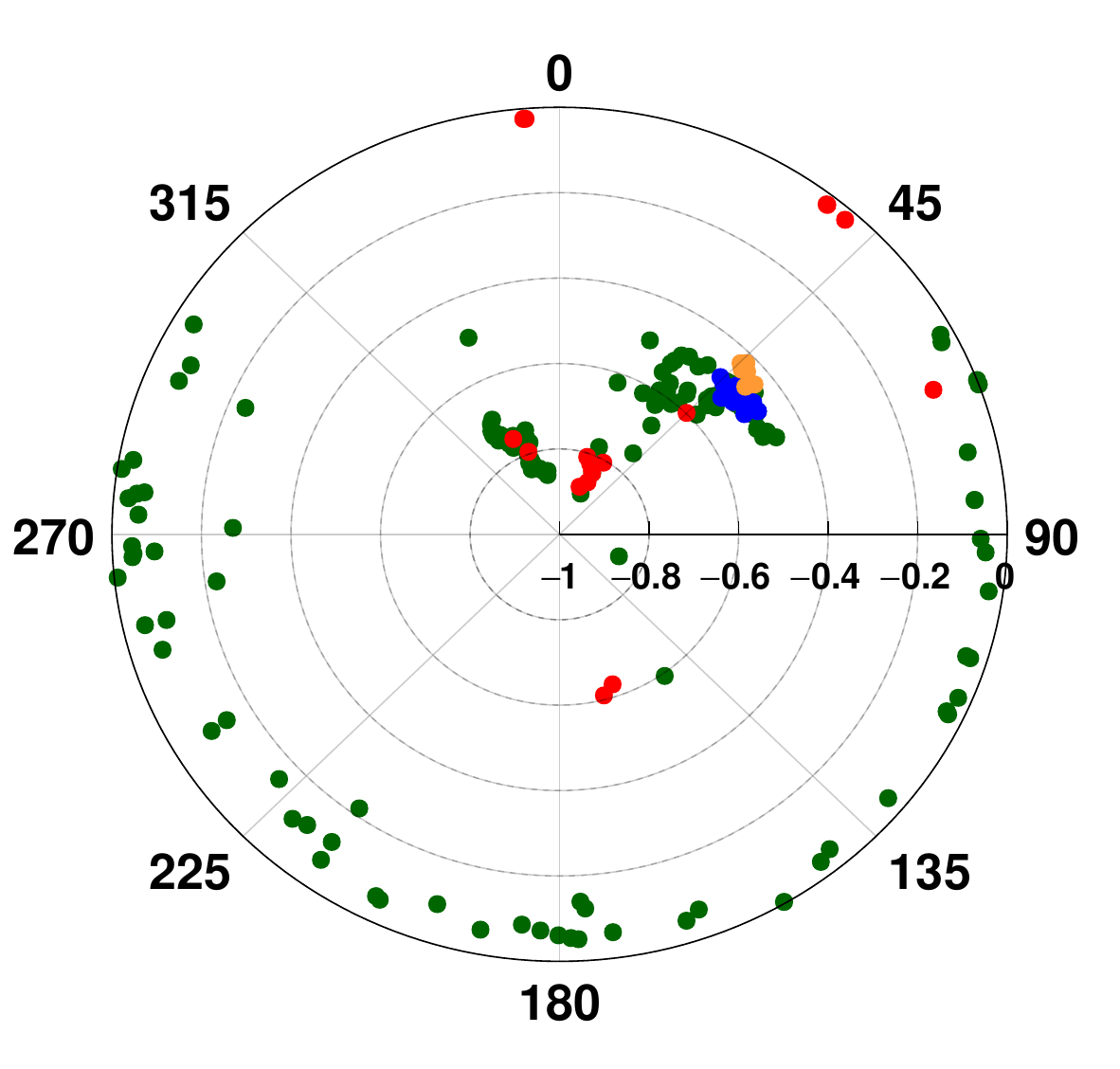}
\end{subfigure}
\caption{Directions to the different nuclear power reactor types (green- PWR, BWR), red- PHWR, orange- GCR, blue- PWR/MOX) as seen at Boulby (left) and Morton (right). Azimuth ($\gamma$ in degrees) is positive CW from north around the rim and elevation ($\alpha$ shown as $\sin\alpha$) is along the radius.}
\label{fig:polar}
\end{figure}

The unit vector pointing from the field point (detector) to the source point (core) has rectangular components given by
\begin{equation} \label{direction}
\frac{X_s - X_f} {d} ; \; \frac{Y_s - Y_f} {d} ; \;  \frac{Z_s - Z_f} {d}.
\end{equation}
This vector points in the direction opposite to the path followed by antineutrinos from the source point traveling to the field point. It readily transforms to local coordinates with east on the local $x$-axis, north on the local $y$-axis, and up on the local $z$-axis, by standard methods. The directions to nuclear reactor cores as observed at Boulby and at Morton are shown in Fig.~\ref{fig:polar}. 

  
\section{$^8$B Solar Neutrinos}
Nuclear fusion in the Sun makes electron neutrinos ($\nu_\mathrm{e}$) not antineutrinos ($\overline{\nu}_\mathrm{e}$). The quasi-elastic scattering process for neutrinos, $\nu_\mathrm{e} + \mathrm{n} \rightarrow \mathrm{p} + \mathrm{e}^-$ corresponding to \eqref{pscat}, is not present due to the absence of free neutron targets. Furthermore, it is assumed that deuteron targets are not present negating the reactions $\nu_\mathrm{e} + \mathrm{d} \rightarrow \mathrm{p} + \mathrm{p} + \mathrm{e}^-$ and $\nu_x + \mathrm{d} \rightarrow \nu_x + \mathrm{n} + \mathrm{p}$. This restricts consideration to neutrino-electron elastic scattering ($\nu_\mathrm{e} \mathrm{e}$ and $\nu_x \mathrm{e}$). Solar neutrinos are well studied with measured fluxes from many of the fusion processes. The flux from the $\beta^+$ decay of $^8\mathrm{B}$ is the dominant background to global antineutrinos. It is at least several orders of magnitude greater than the flux from Hep solar neutrinos and the fluxes from all other solar fusion processes produce neutrinos with energy less than $1.8$ MeV, which is the lower limit to the validity of the reactor antineutrino flux model. 

\begin{figure}
\begin{subfigure}[b]{0.23\textwidth}
\includegraphics[trim = 0mm 0mm 0mm 0mm,width=\textwidth]{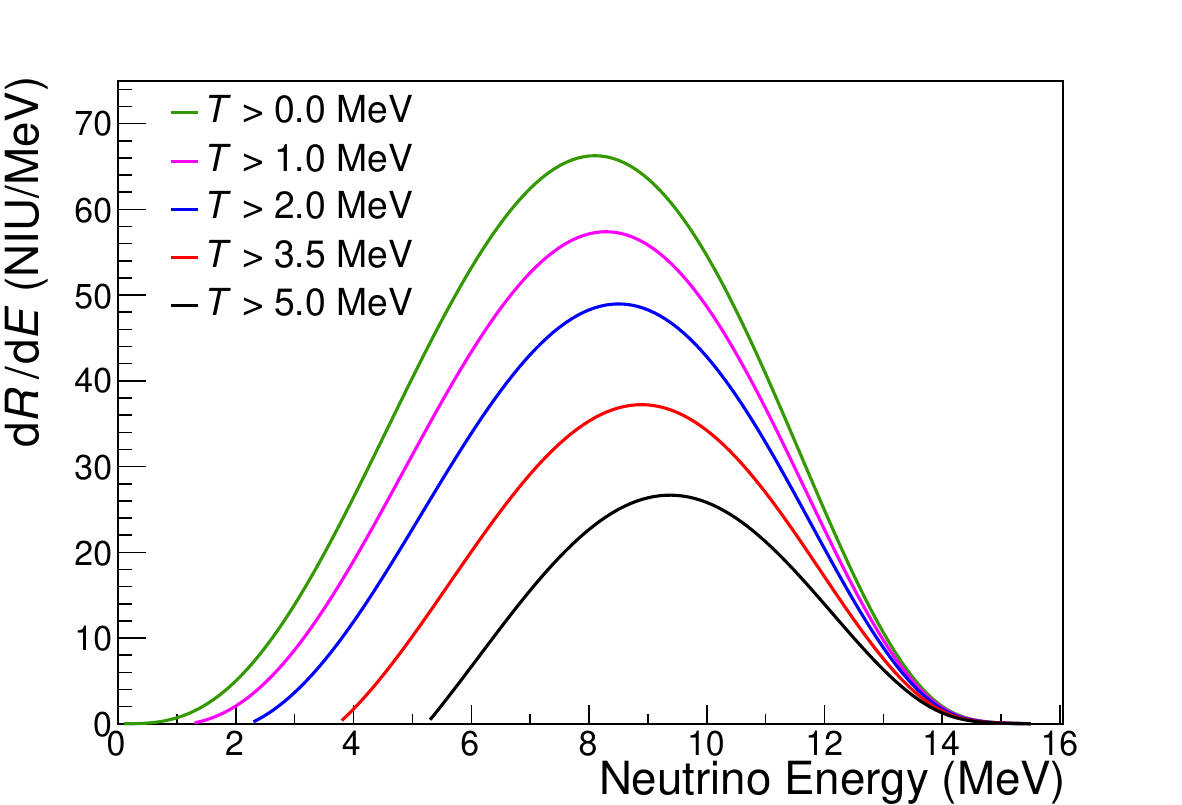}
\end{subfigure}
\begin{subfigure}[b]{0.23\textwidth}
\includegraphics[trim = 0mm 0mm 0mm 0mm,width=\textwidth]{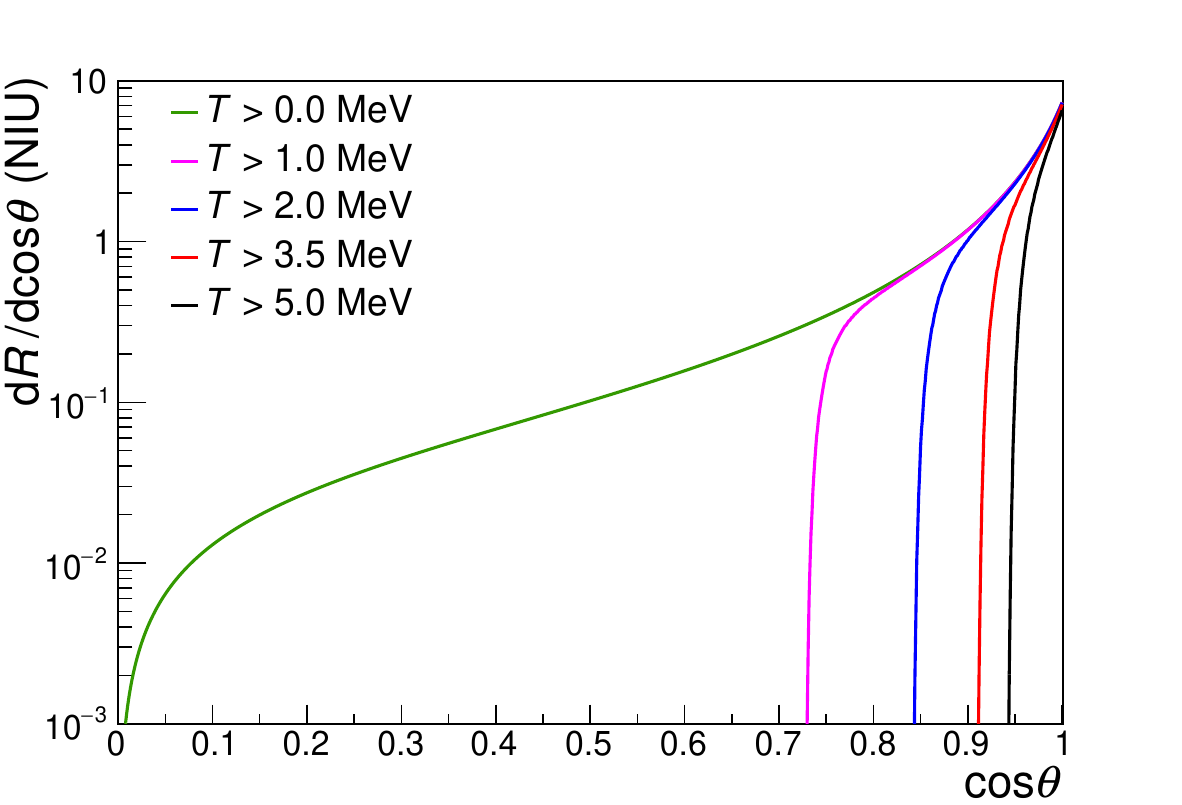}
\end{subfigure}
\caption{Elastic scattering reaction rates as a function of neutrino energy (left) and as a function of the cosine of the scattering angle (right) in $1000$ bins for $^8$B solar neutrinos at selected values of the minimum electron kinetic energy $T_\mathrm{min}$.}
\label{fig:8Brate}
\end{figure}

A time-averaged flux of $<\!\phi_\odot\!> = 2.345 \times 10^6$ cm$^{-2}$s$^{-1}$, which is the central value of the measurement by Super-Kamiokande \cite{abe2016}, is assumed for calculation of the $^8$B solar neutrino background. The reaction rate (/target/s) of solar neutrinos from $^8$B decay is found using \eqref{intrate} with $\mathrm{d}n/\mathrm{d}E$ given by the decay spectrum \cite{winter06}, $<P_\mathrm{ee}>=1$, $n_{\nu_{i}} = 1$, and $\phi_i$ replaced by $\phi(r)$. Figure~\ref{fig:8Brate} shows energy spectra of the elastic scattering reaction rates of $^8$B solar neutrinos for several values of the minimum electron kinetic energy $T_\mathrm{min}$. Even a modest cut on the minimum electron kinetic energy $T_\mathrm{min}$ significantly sharpens the distribution of the scattering angles of the electrons $\cos\theta$ along the direction away from the source. This feature of the reaction kinematics is shown in Fig.~\ref{fig:Te_cos}. The reductions to the $^8$B solar neutrino time-averaged reaction rate associated with the selected minimum electron kinetic energies are listed in Table~\ref{tab:8Brate_tmin}. 

\begin{table}
\caption{Elastic Scattering Reaction Rates of $^8$B Solar Neutrinos for Selected Values of the Minimum Kinetic Energy of the Electron}
\begin{tabular} {l r r r r r}
\hline\noalign{\smallskip}
 $T_\mathrm{min}$ (MeV) & $0.0$ & $1.0$ & $2.0$ & $3.5$ & $5.0$ \\
\hline\noalign{\smallskip}
$R$ (NIU) & $464.9$ & $389.8$ & $318.4$ & $221.0$ & 139.8 \\
 \hline
\end{tabular}
\label{tab:8Brate_tmin}
\end{table}

The solar neutrino flux has an expected annual variation of almost $7$\% in magnitude due to the changing distance between the Earth and the Sun. The maximum flux is at perihelion in early January and the minimum flux is at aphelion in early July. According to Kepler's law of elliptical orbits, the distance between the Earth and the Sun as a function of the angle $\psi$ since perihelion is
\begin{equation}
\label{rorb}
r(\psi) = \frac {a(1 - \epsilon^2)} {1 + \epsilon \cos \psi},
\end{equation}
where $a$ is the semi-major axis ($1$ a.u.) and $\epsilon$ is the eccentricity of the Earth's orbit. The model uses a publicly available calculation of the Sun's position (elevation and azimuth) and distance as a function of time and location on Earth \cite{noaa}. The calculated distance \eqref{rorb} modulates the flux according to
\begin{equation}
\phi(r) = \, <\!\phi_\odot\!> \frac{a^2} {r(\psi)^2}.
\end{equation}

\begin{figure}
\begin{subfigure}[b]{0.23\textwidth}
\includegraphics[trim = 0mm 0mm 0mm 0mm,width=\textwidth]{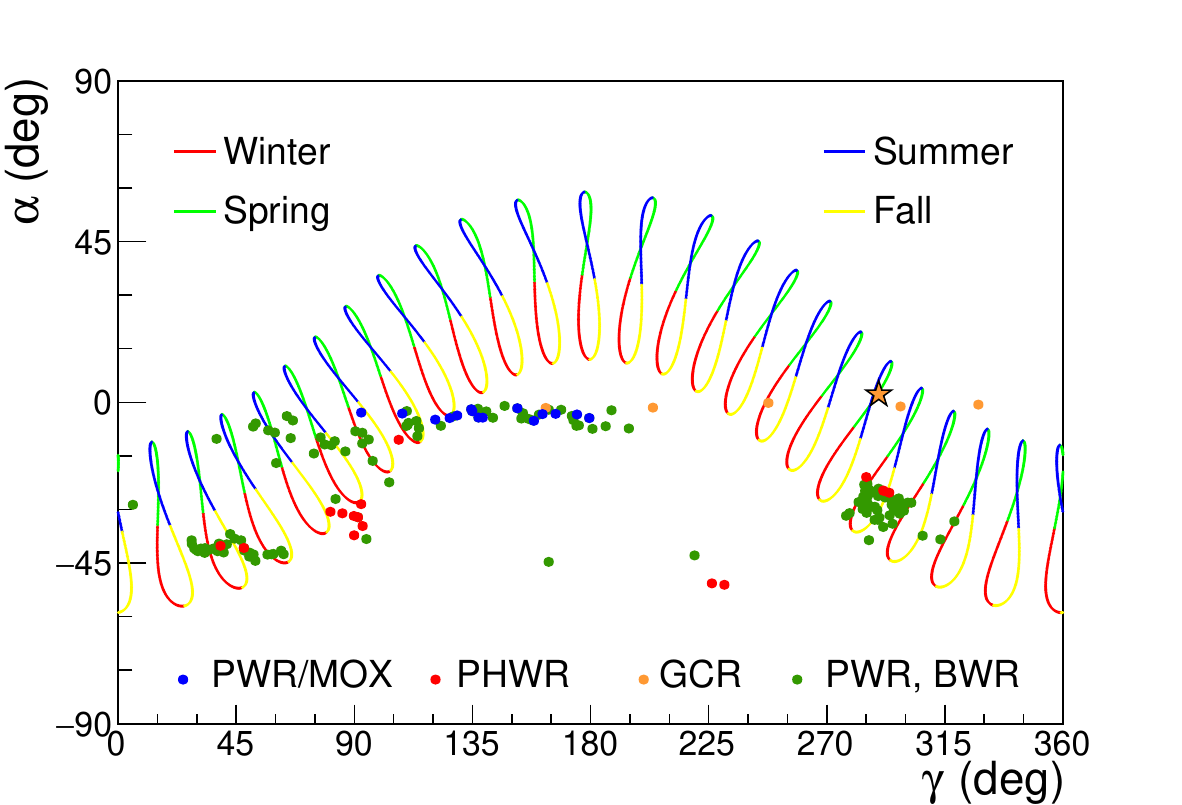}
\end{subfigure}
\begin{subfigure}[b]{0.23\textwidth}
\includegraphics[trim = 0mm 0mm 0mm 0mm,width=\textwidth]{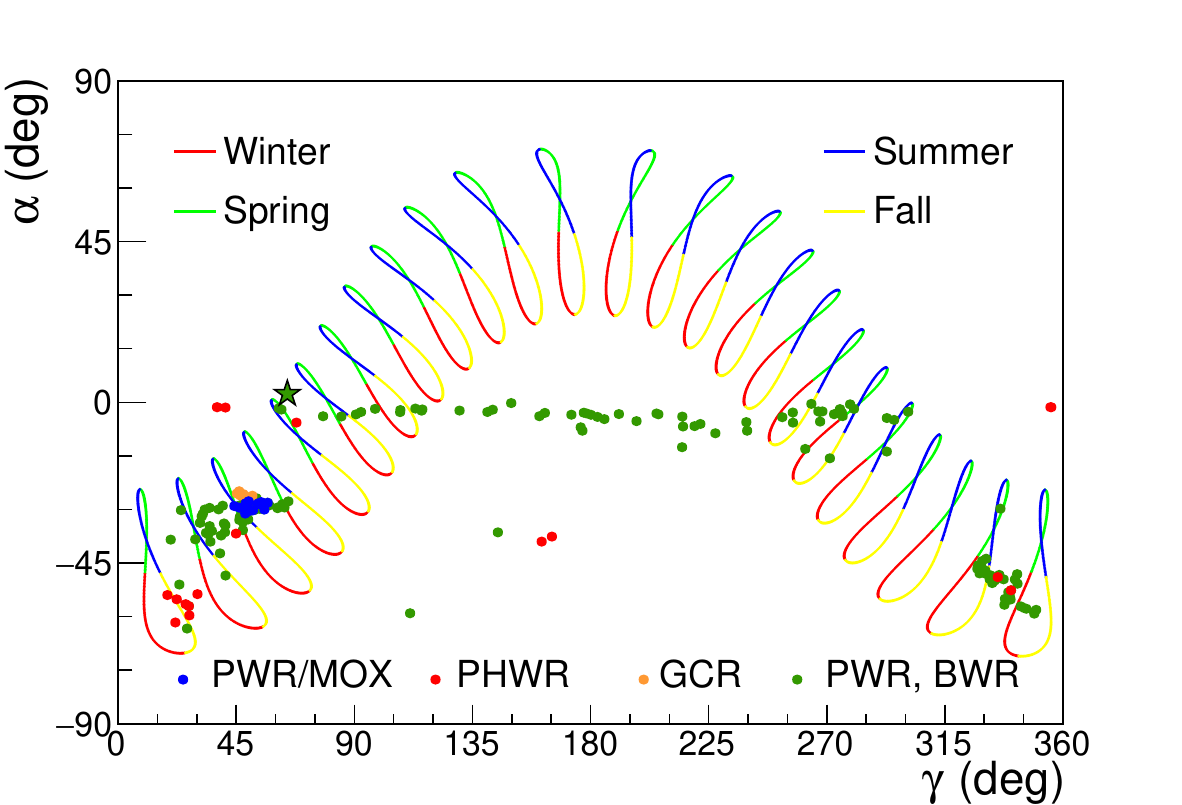}
\end{subfigure}
\caption{Hourly analemma for the year $2020$ as seen at Boulby (left) and at Morton (right) are plotted by elevation $\alpha$ and azimuth $\gamma$ with seasons in different colors. Azimuth is positive CW from north. Midnight is at the left and right edges ($\gamma=0^{\circ}$ and $360^{\circ}$) of the plot with noon at the center. Morning (afternoon) hours are to the left (right) of noon ($\gamma=180^{\circ}$). The positions of $456$ nuclear power reactor cores operational in 2018 are indicated by colored dots except for the closest reactor, which is shown as a star with black outline next to spring near the 19:00 ($\gamma=285^{\circ}$) analemma (left) and near the 05:00 ($\gamma=60^{\circ}$) analemma (right)}
\label{fig:analemma}
\end{figure}

\begin{figure}
\begin{subfigure}[b]{0.23\textwidth}
\includegraphics[trim = 0mm 0mm 0mm 0mm,width=\textwidth]{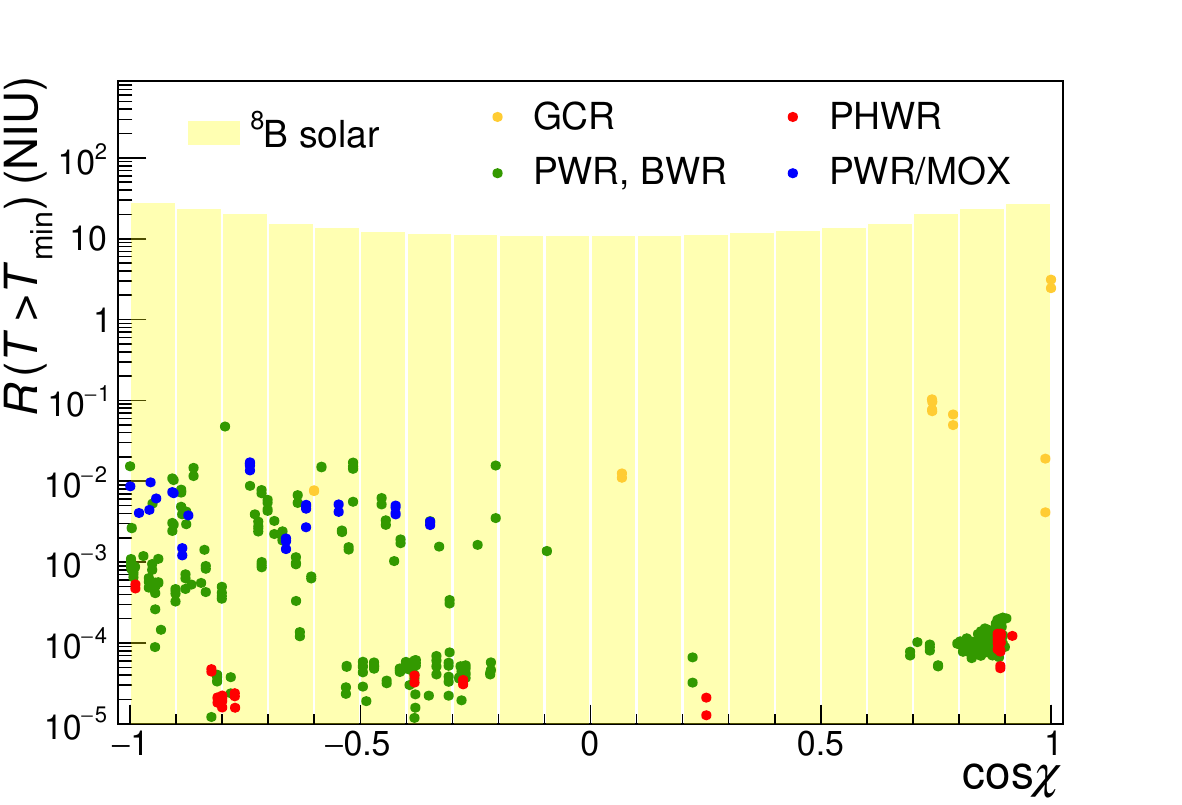}
\end{subfigure}
\begin{subfigure}[b]{0.23\textwidth}
\includegraphics[trim = 0mm 0mm 0mm 0mm,width=\textwidth]{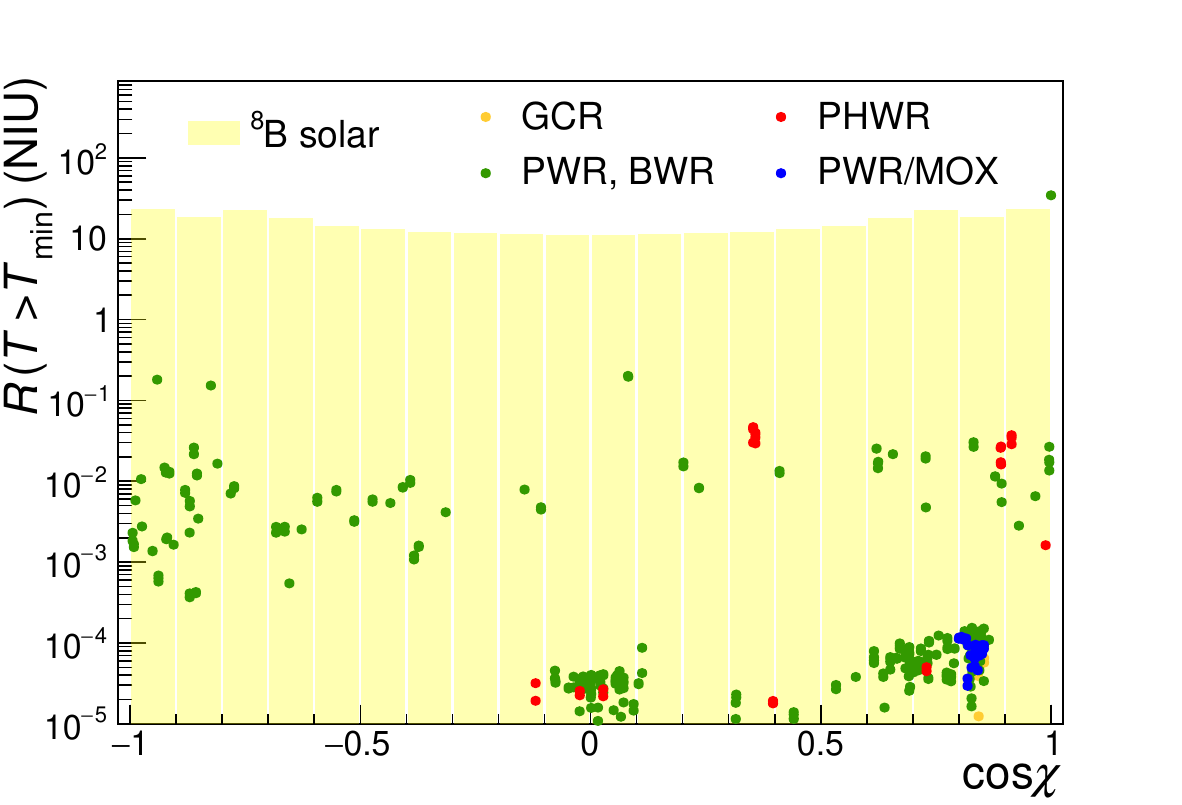}
\end{subfigure}
\caption{Elastic scattering reaction rates with $T_\mathrm{min}= 2$ MeV for $^8$B solar neutrinos (yellow histogram) and reactor antineutrinos (colored dots) as a function of the cosine of the angle $\chi$ of the source relative to the direction to the closest reactor as observed from Boulby (left) and from Morton (right).}
\label{fig:solbkg}
\end{figure}

The directions of solar neutrinos move across the sky with the Earth's rotation and orbit. It is possible to remove a portion of the solar neutrino background by restricting reconstructed electron tracks to lie in a cone of acceptance centered on the signal direction. The size of the cone depends on the global antineutrino source. It is probably much smaller for a nuclear reactor than for Earth's mantle. Hourly analemma at Boulby and at Morton for the year $2021$ overlaid with the positions of nuclear power reactors are shown in Fig.~\ref{fig:analemma}. Using the direction to the closest reactor core as the reference angle, the elastic scattering reaction rates from the stationary nuclear power reactors and those from $^8$B decays in the roving Sun as a function of the cosine of the separation angle $\chi$ at Boulby and at Morton are compared in Fig.~\ref{fig:solbkg}.

\begin{figure}
\centering
\begin{subfigure}[b]{0.23\textwidth}
\includegraphics[trim = 0mm 0mm 0mm 0mm,width=\textwidth]{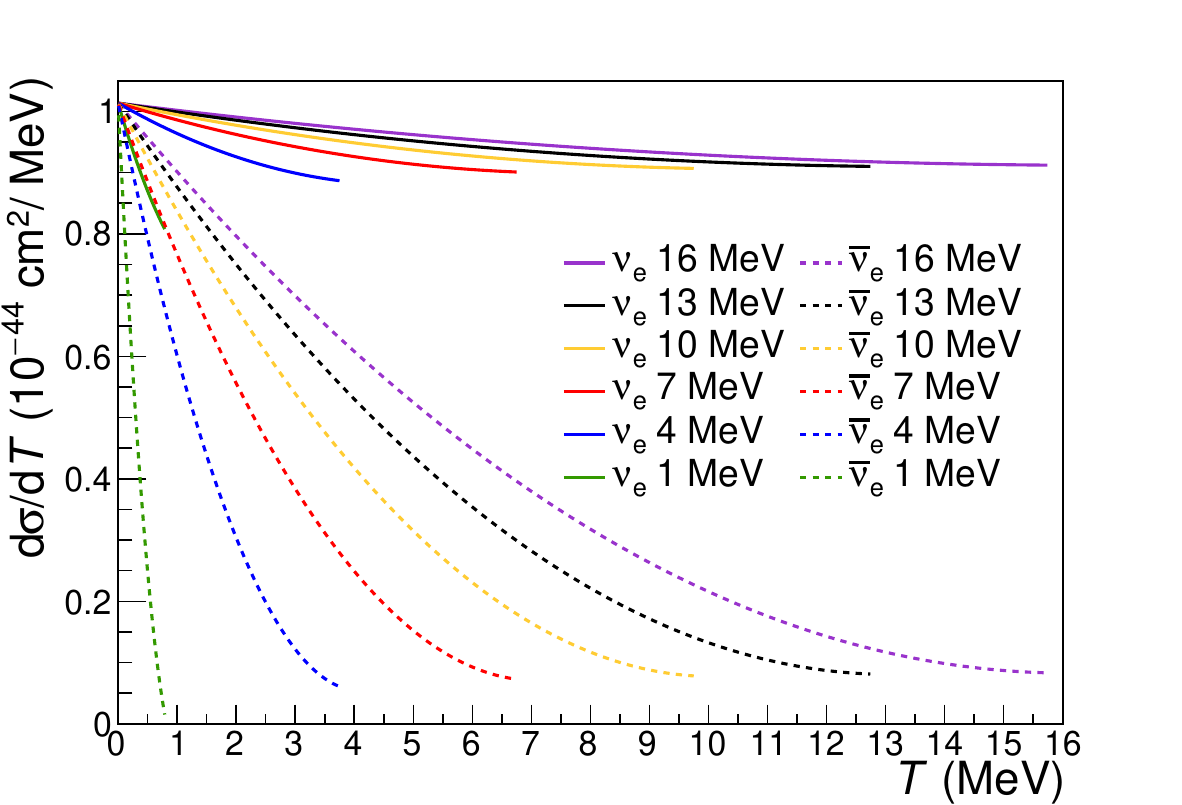}
\end{subfigure}
\begin{subfigure}[b]{0.23\textwidth}
\includegraphics[trim = 0mm 0mm 0mm 0mm,width=\textwidth]{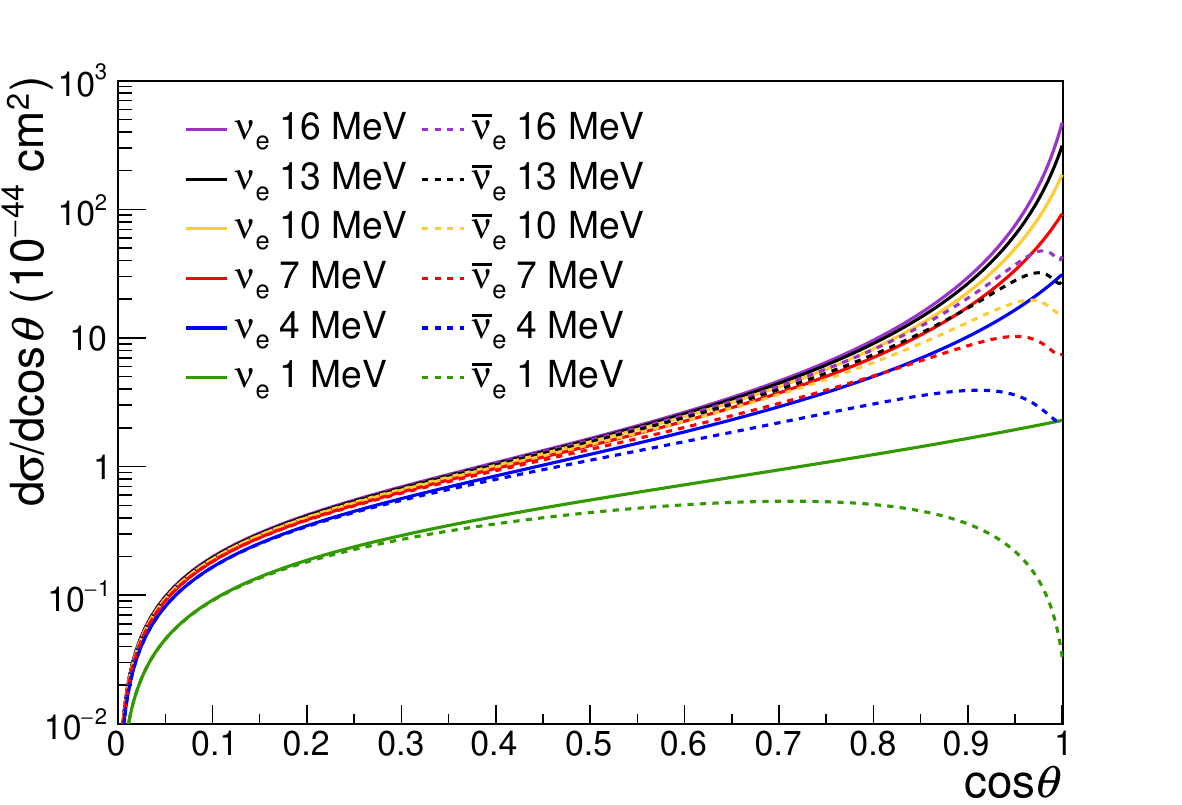}
\end{subfigure}
\caption{Elastic scattering differential cross section, at selected energies, for $\nu_\mathrm{e}$ as a function \eqref{dsdt} of the electron kinetic energy $T$ (left) and as a function \eqref{dsdcos} of the cosine of the electron scattering angle $\cos \theta$ (right). The corresponding distributions for $\overline{\nu}_\mathrm{e}$ are shown for comparison.}
\label{fig:dsdt_nue}
\end{figure}

The two versions of the differential elastic scattering cross section for $\nu_\mathrm{e}$, one as a function of $T$  \eqref{dsdt} and the other as a function of $\cos \theta$ \eqref{dsdcos}, are shown in Fig.~\ref{fig:dsdt_nue}. With the rates of $^8$B solar neutrino-induced elastic scattering reactions for selected values of the minimum electron kinetic energy as given in Fig.~\ref{fig:8Brate} (left), the angular distribution of electron directions relative to the Sun is calculated using \eqref{dsdcos} and shown in Fig.~\ref{fig:8Brate} (right). 

\begin{figure}
\centering
\begin{subfigure}[b]{0.23\textwidth}
\includegraphics[trim = 0mm 0mm 0mm 0mm,width=\textwidth]{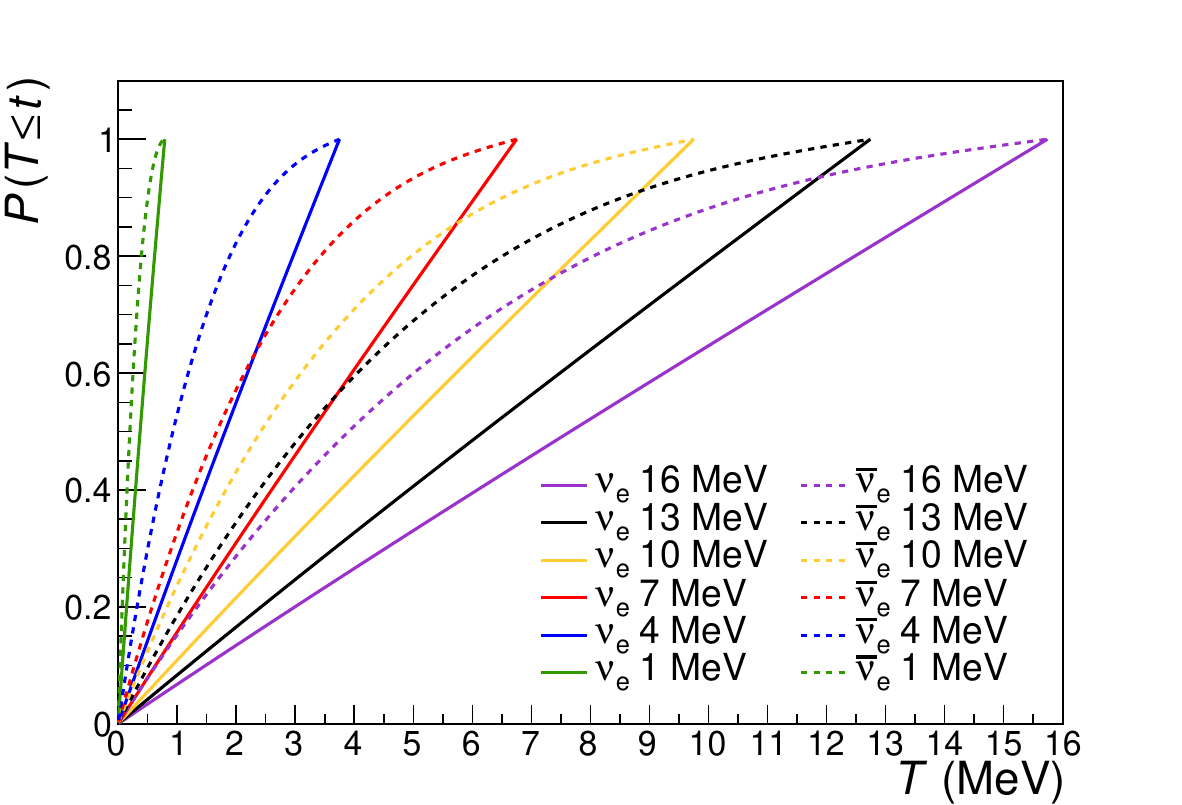}
\end{subfigure}
\begin{subfigure}[b]{0.23\textwidth}
\includegraphics[trim = 0mm 0mm 0mm 0mm,width=\textwidth]{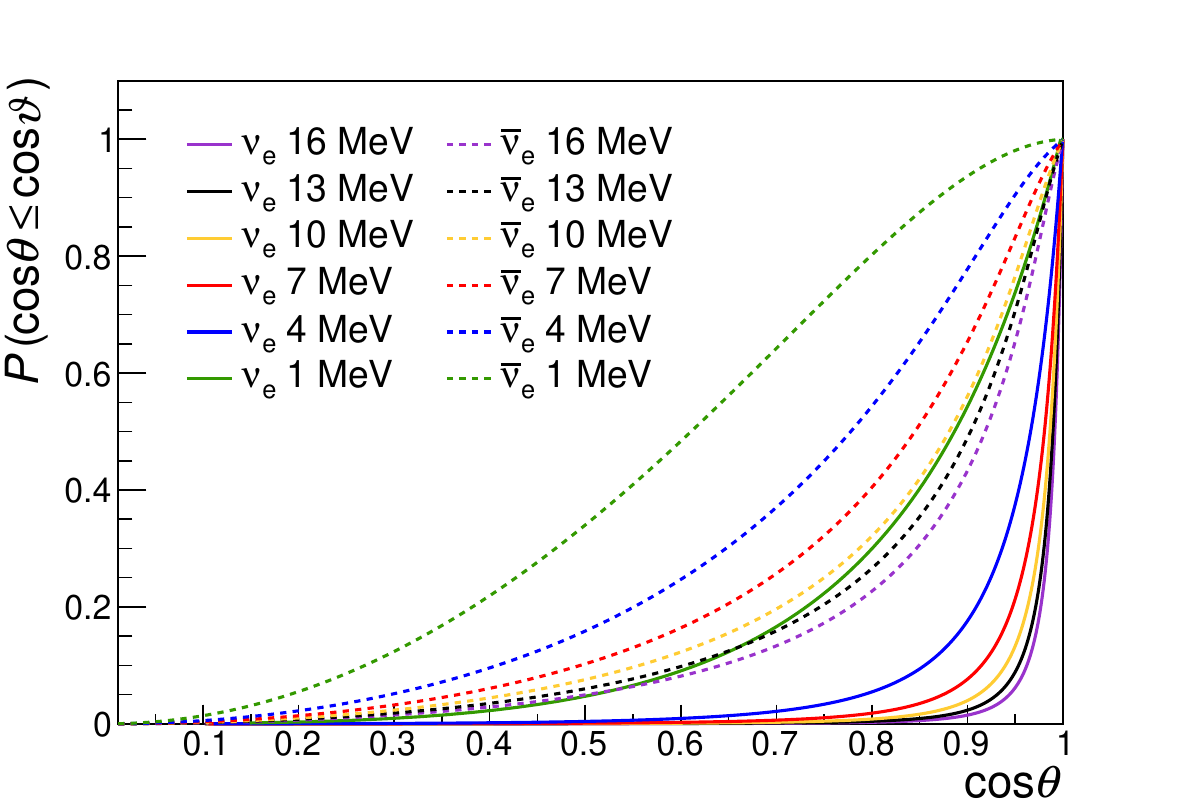}
\end{subfigure}
\caption{Cumulative distribution functions, at selected energies of $\nu_\mathrm{e}$ and $\overline{\nu}_e$, for the elastic scattering differential cross section as a function \eqref{cumte} of the electron kinetic energy $T$ (left) and as a function \eqref{cumcos} of the cosine of the electron scattering angle $\cos \theta$ (right).}
\label{fig:pdfte_nue}
\end{figure}

The cumulative distribution functions for the elastic scattering differential cross sections, \eqref{dsdt} and \eqref{dsdcos}, are calculated for $\nu_\mathrm{e}$ with \eqref{cumte} and \eqref{cumcos} at several values of $E_\nu$, are shown in Fig.~\ref{fig:pdfte_nue}. Cumulative distribution functions are readily compared with random numbers $[0,1]$ to help select the $T$ and $\cos \theta$ pairs for elastic scattering reactions within simulated event samples.


\section{Signal significance}
The model includes assessments of the significance of observing a signal in the presence of background. It does not presently include background due to non-neutrino sources, such as radioactivity (i.e. $^{238}$U, $^{232}$Th) in the target liquid, cosmogenic neutron-rich isotopes (i.e. $^9$Li, $^{16}$N), fast neutrons, or random coincidences. The assessment statistic is
\begin{equation} \label{stats}
N_\sigma = \frac{S \sqrt {t}} {\sqrt {S + B}},
\end{equation}
with $S$ the signal rate, $B$ the background rate, and $t$ the detector live time. The denominator represents the Poisson error on the total number observed events.
With $S$ and $B$ in NIU, then $t$ is given in years of exposure of a detector with $10^{32}$ free proton or electron targets. To reach the same significance $N_\sigma$ with a $1$-kT water target, divide the time by the appropriate conversion factor (/NIU/kT(H$_2$O)/y) given in Table~\ref{tab:const}.

\section{Summary}
This paper presents the input data, formulae, and plots resulting from the calculations, which, in addition to the time-dependent reaction rates and energy spectra, model the directions of the antineutrinos from IAEA-registered nuclear power reactors and of the neutrinos from $^8$B decay in the Sun. The model includes estimates of the steady state reaction rates and energy spectra of the antineutrinos from the crust and mantle of the Earth. Results are available for any location near the surface of the Earth and comprise both quasi-elastic scattering on free protons and elastic scattering on atomic electrons. This paper compares model results for two underground locations, the Boulby Mine in the United Kingdom and the Morton Salt Mine in the United States. Operational nuclear power reactors are within about $20$ kilometers of these mines, making them candidate sites for hosting antineutrino detectors capable of identifying, monitoring, and locating remote nuclear activity. The model, which is implemented in a web application at https://geoneutrinos.org/reactors/,\footnote{Not all of the equations and plots in this paper are currently available in the web application. Work towards completing the implementation is ongoing.} provides references for the input data and the formulae, as well as an interactive calculator of the significance of the rate of any of the neutrino sources relative to other sources taken as background. Documentation of the implementation of the web application is planned for a separate, companion report.

\section*{Acknowledgments}
This work was supported in part by Lawrence Livermore National Security, LLC.

\newpage
\bibliographystyle{apsrev4-1}
\nocite{apsrev41control}
\bibliography{Nugeo_bib,revtex-custom}

\begin{thebibliography}{48}%
\makeatletter
\providecommand \@ifxundefined [1]{%
 \@ifx{#1\undefined}
}%
\providecommand \@ifnum [1]{%
 \ifnum #1\expandafter \@firstoftwo
 \else \expandafter \@secondoftwo
 \fi
}%
\providecommand \@ifx [1]{%
 \ifx #1\expandafter \@firstoftwo
 \else \expandafter \@secondoftwo
 \fi
}%
\providecommand \natexlab [1]{#1}%
\providecommand \enquote  [1]{``#1''}%
\providecommand \bibnamefont  [1]{#1}%
\providecommand \bibfnamefont [1]{#1}%
\providecommand \citenamefont [1]{#1}%
\providecommand \href@noop [0]{\@secondoftwo}%
\providecommand \href [0]{\begingroup \@sanitize@url \@href}%
\providecommand \@href[1]{\@@startlink{#1}\@@href}%
\providecommand \@@href[1]{\endgroup#1\@@endlink}%
\providecommand \@sanitize@url [0]{\catcode `\\12\catcode `\$12\catcode
  `\&12\catcode `\#12\catcode `\^12\catcode `\_12\catcode `\%12\relax}%
\providecommand \@@startlink[1]{}%
\providecommand \@@endlink[0]{}%
\providecommand \url  [0]{\begingroup\@sanitize@url \@url }%
\providecommand \@url [1]{\endgroup\@href {#1}{\urlprefix }}%
\providecommand \urlprefix  [0]{URL }%
\providecommand \Eprint [0]{\href }%
\providecommand \doibase [0]{http://dx.doi.org/}%
\providecommand \selectlanguage [0]{\@gobble}%
\providecommand \bibinfo  [0]{\@secondoftwo}%
\providecommand \bibfield  [0]{\@secondoftwo}%
\providecommand \translation [1]{[#1]}%
\providecommand \BibitemOpen [0]{}%
\providecommand \bibitemStop [0]{}%
\providecommand \bibitemNoStop [0]{.\EOS\space}%
\providecommand \EOS [0]{\spacefactor3000\relax}%
\providecommand \BibitemShut  [1]{\csname bibitem#1\endcsname}%
\let\auto@bib@innerbib\@empty
\bibitem [{\citenamefont {Usman}\ \emph {et~al.}(2015)\citenamefont {Usman}
  \emph {et~al.}}]{agm15}%
  \BibitemOpen
  \bibfield  {author} {\bibinfo {author} {\bibfnamefont {S.~M.}\ \bibnamefont
  {Usman}} \emph {et~al.},\ }\bibfield  {title} {\enquote {\bibinfo {title}
  {{AGM2015: Antineutrino Global Map 2015}},}\ }\href@noop {} {\bibfield
  {journal} {\bibinfo  {journal} {Sci. Rep.}\ }\textbf {\bibinfo {volume}
  {5}},\ \bibinfo {pages} {13945} (\bibinfo {year} {2015})}\BibitemShut
  {NoStop}%
\bibitem [{\citenamefont {Boireau}\ \emph {et~al.}(2015)\citenamefont {Boireau}
  \emph {et~al.}}]{nucifer15}%
  \BibitemOpen
  \bibfield  {author} {\bibinfo {author} {\bibfnamefont {G.}~\bibnamefont
  {Boireau}} \emph {et~al.},\ }\href@noop {} {\enquote {\bibinfo {title}
  {{Online Monitoring of the Osiris Reactor with the Nucifer Neutrino
  Detector}},}\ }\bibinfo {howpublished} {arXiv:1509.05610} (\bibinfo {year}
  {2015})\BibitemShut {NoStop}%
\bibitem [{\citenamefont {Bowden}\ \emph {et~al.}(2007)\citenamefont {Bowden}
  \emph {et~al.}}]{songs07}%
  \BibitemOpen
  \bibfield  {author} {\bibinfo {author} {\bibfnamefont {N.~S.}\ \bibnamefont
  {Bowden}} \emph {et~al.},\ }\bibfield  {title} {\enquote {\bibinfo {title}
  {{Experimental Results from an Antineutrino Detector for Cooperative
  Monitoring of Nuclear Reactors}},}\ }\href@noop {} {\bibfield  {journal}
  {\bibinfo  {journal} {Nucl. Inst. Meth.}\ }\textbf {\bibinfo {volume}
  {A572}},\ \bibinfo {pages} {985} (\bibinfo {year} {2007})}\BibitemShut
  {NoStop}%
\bibitem [{\citenamefont {Jocher}\ \emph {et~al.}(2013)\citenamefont {Jocher}
  \emph {et~al.}}]{nudar13}%
  \BibitemOpen
  \bibfield  {author} {\bibinfo {author} {\bibfnamefont {G.~R.}\ \bibnamefont
  {Jocher}} \emph {et~al.},\ }\bibfield  {title} {\enquote {\bibinfo {title}
  {{Theoretical antineutrino detection, direction, and ranging at long
  distances}},}\ }\href@noop {} {\bibfield  {journal} {\bibinfo  {journal}
  {Phys. Rep.}\ }\textbf {\bibinfo {volume} {527}},\ \bibinfo {pages} {131}
  (\bibinfo {year} {2013})}\BibitemShut {NoStop}%
\bibitem [{\citenamefont {Lasserre}\ \emph {et~al.}(2010)\citenamefont
  {Lasserre} \emph {et~al.}}]{snif10}%
  \BibitemOpen
  \bibfield  {author} {\bibinfo {author} {\bibfnamefont {T.}~\bibnamefont
  {Lasserre}} \emph {et~al.},\ }\href@noop {} {\enquote {\bibinfo {title}
  {{SNIF: A Futuristic Neutrino Probe for Undeclared Nuclear Fission
  Reactors}},}\ }\bibinfo {howpublished} {arXiv:1011.3850} (\bibinfo {year}
  {2010})\BibitemShut {NoStop}%
\bibitem [{\citenamefont {Bernstein}\ \emph {et~al.}(2010)\citenamefont
  {Bernstein} \emph {et~al.}}]{adam10}%
  \BibitemOpen
  \bibfield  {author} {\bibinfo {author} {\bibfnamefont {A.}~\bibnamefont
  {Bernstein}} \emph {et~al.},\ }\bibfield  {title} {\enquote {\bibinfo {title}
  {{Nuclear Security Applications of Antineutrino Detectors: Current
  Capabilities and Future Prospects}},}\ }\href@noop {} {\bibfield  {journal}
  {\bibinfo  {journal} {Sci. Glob. Sec.}\ }\textbf {\bibinfo {volume} {18}},\
  \bibinfo {pages} {127} (\bibinfo {year} {2010})}\BibitemShut {NoStop}%
\bibitem [{\citenamefont {Reines}\ and\ \citenamefont
  {Cowan}(1953)}]{reines53}%
  \BibitemOpen
  \bibfield  {author} {\bibinfo {author} {\bibfnamefont {F.}~\bibnamefont
  {Reines}}\ and\ \bibinfo {author} {\bibfnamefont {C.~L.}\ \bibnamefont
  {Cowan}},\ }\bibfield  {title} {\enquote {\bibinfo {title} {{Detection of the
  free neutrino}},}\ }\href@noop {} {\bibfield  {journal} {\bibinfo  {journal}
  {Phys. Rev.}\ }\textbf {\bibinfo {volume} {92}},\ \bibinfo {pages} {830 }
  (\bibinfo {year} {1953})}\BibitemShut {NoStop}%
\bibitem [{\citenamefont {Reines}\ \emph {et~al.}(1976)\citenamefont {Reines},
  \citenamefont {Gurr},\ and\ \citenamefont {Sobel}}]{reines76}%
  \BibitemOpen
  \bibfield  {author} {\bibinfo {author} {\bibfnamefont {F.}~\bibnamefont
  {Reines}}, \bibinfo {author} {\bibfnamefont {H.~S.}\ \bibnamefont {Gurr}}, \
  and\ \bibinfo {author} {\bibfnamefont {H.~W.}\ \bibnamefont {Sobel}},\
  }\bibfield  {title} {\enquote {\bibinfo {title} {{Detection of
  $\overline{\nu}_e$ - $e$ scattering}},}\ }\href@noop {} {\bibfield  {journal}
  {\bibinfo  {journal} {Phys. Rev. Lett.}\ }\textbf {\bibinfo {volume} {37}},\
  \bibinfo {pages} {315 } (\bibinfo {year} {1976})}\BibitemShut {NoStop}%
\bibitem [{\citenamefont {Learned}\ \emph {et~al.}(2008)\citenamefont {Learned}
  \emph {et~al.}}]{jgl08}%
  \BibitemOpen
  \bibfield  {author} {\bibinfo {author} {\bibfnamefont {J.~G.}\ \bibnamefont
  {Learned}} \emph {et~al.},\ }\bibfield  {title} {\enquote {\bibinfo {title}
  {{Determination of Neutrino Mass Hierarchy and $\theta_{13}$ with a Remote
  Detector of Reactor Antineutrinos}},}\ }\href@noop {} {\bibfield  {journal}
  {\bibinfo  {journal} {Phys. Rev.}\ }\textbf {\bibinfo {volume} {D78}},\
  \bibinfo {pages} {071302} (\bibinfo {year} {2008})}\BibitemShut {NoStop}%
\bibitem [{\citenamefont {Araki}\ \emph {et~al.}(2005)\citenamefont {Araki}
  \emph {et~al.}}]{kl05}%
  \BibitemOpen
  \bibfield  {author} {\bibinfo {author} {\bibfnamefont {T.}~\bibnamefont
  {Araki}} \emph {et~al.},\ }\bibfield  {title} {\enquote {\bibinfo {title}
  {{Experimental investigation of geologically produced antineutrinos with
  KamLAND}},}\ }\href@noop {} {\bibfield  {journal} {\bibinfo  {journal}
  {Nature}\ }\textbf {\bibinfo {volume} {436}},\ \bibinfo {pages} {499}
  (\bibinfo {year} {2005})}\BibitemShut {NoStop}%
\bibitem [{\citenamefont {Gando}\ \emph {et~al.}(2013)\citenamefont {Gando}
  \emph {et~al.}}]{gando13}%
  \BibitemOpen
  \bibfield  {author} {\bibinfo {author} {\bibfnamefont {A.}~\bibnamefont
  {Gando}} \emph {et~al.},\ }\bibfield  {title} {\enquote {\bibinfo {title}
  {{Reactor on-off antineutrino measurement with KamLAND}},}\ }\href@noop {}
  {\bibfield  {journal} {\bibinfo  {journal} {Phys. Rev. D}\ }\textbf {\bibinfo
  {volume} {88}},\ \bibinfo {pages} {033001} (\bibinfo {year}
  {2013})}\BibitemShut {NoStop}%
\bibitem [{\citenamefont {Agostini}\ \emph {et~al.}(2015)\citenamefont
  {Agostini} \emph {et~al.}}]{agostini15}%
  \BibitemOpen
  \bibfield  {author} {\bibinfo {author} {\bibfnamefont {M.}~\bibnamefont
  {Agostini}} \emph {et~al.},\ }\bibfield  {title} {\enquote {\bibinfo {title}
  {{Spectroscopy of geo-neutrinos from 2056 days of Borexino data}},}\
  }\href@noop {} {\bibfield  {journal} {\bibinfo  {journal} {Phys. Rev.}\
  }\textbf {\bibinfo {volume} {D92}},\ \bibinfo {pages} {031101} (\bibinfo
  {year} {2015})}\BibitemShut {NoStop}%
\bibitem [{\citenamefont {Dye}\ \emph {et~al.}(2015)\citenamefont {Dye} \emph
  {et~al.}}]{dye_etal15}%
  \BibitemOpen
  \bibfield  {author} {\bibinfo {author} {\bibfnamefont {S.~T.}\ \bibnamefont
  {Dye}} \emph {et~al.},\ }\bibfield  {title} {\enquote {\bibinfo {title}
  {{Geo-neutrinos and Earth models}},}\ }\href@noop {} {\bibfield  {journal}
  {\bibinfo  {journal} {Phys. Proc.}\ }\textbf {\bibinfo {volume} {61}},\
  \bibinfo {pages} {310 } (\bibinfo {year} {2015})}\BibitemShut {NoStop}%
\bibitem [{\citenamefont {Baldoncini}\ \emph {et~al.}(2015)\citenamefont
  {Baldoncini} \emph {et~al.}}]{baldoncini}%
  \BibitemOpen
  \bibfield  {author} {\bibinfo {author} {\bibfnamefont {M.}~\bibnamefont
  {Baldoncini}} \emph {et~al.},\ }\bibfield  {title} {\enquote {\bibinfo
  {title} {{Reference worldwide model for antineutrinos from reactors}},}\
  }\href@noop {} {\bibfield  {journal} {\bibinfo  {journal} {Phys. Rev.}\
  }\textbf {\bibinfo {volume} {D91}},\ \bibinfo {pages} {065002} (\bibinfo
  {year} {2015})}\BibitemShut {NoStop}%
\bibitem [{\citenamefont {Fischbach}\ \emph {et~al.}(1977)\citenamefont
  {Fischbach} \emph {et~al.}}]{fisch77}%
  \BibitemOpen
  \bibfield  {author} {\bibinfo {author} {\bibfnamefont {E.}~\bibnamefont
  {Fischbach}} \emph {et~al.},\ }\bibfield  {title} {\enquote {\bibinfo {title}
  {{Elastic neutrino-proton scattering}},}\ }\href@noop {} {\bibfield
  {journal} {\bibinfo  {journal} {{Phys. Rev. D}}\ }\textbf {\bibinfo {volume}
  {15}},\ \bibinfo {pages} {97} (\bibinfo {year} {1977})}\BibitemShut {NoStop}%
\bibitem [{\citenamefont {Vogel}\ and\ \citenamefont {Beacom}(1999)}]{vogel99}%
  \BibitemOpen
  \bibfield  {author} {\bibinfo {author} {\bibfnamefont {P.}~\bibnamefont
  {Vogel}}\ and\ \bibinfo {author} {\bibfnamefont {J.~F.}\ \bibnamefont
  {Beacom}},\ }\bibfield  {title} {\enquote {\bibinfo {title} {{Angular
  distribution of inverse neutron decay $\overline{\nu}_e + p \rightarrow e^+ +
  n$}},}\ }\href@noop {} {\bibfield  {journal} {\bibinfo  {journal} {Phys. Rev.
  D}\ }\textbf {\bibinfo {volume} {60}},\ \bibinfo {pages} {053003} (\bibinfo
  {year} {1999})}\BibitemShut {NoStop}%
\bibitem [{\citenamefont {Boehm}\ \emph {et~al.}(2000)\citenamefont {Boehm}
  \emph {et~al.}}]{paloverde00}%
  \BibitemOpen
  \bibfield  {author} {\bibinfo {author} {\bibfnamefont {F.}~\bibnamefont
  {Boehm}} \emph {et~al.},\ }\bibfield  {title} {\enquote {\bibinfo {title}
  {{Results from the Palo Verde Neutrino Oscillation Experiment}},}\
  }\href@noop {} {\bibfield  {journal} {\bibinfo  {journal} {Phys. Rev. D}\
  }\textbf {\bibinfo {volume} {62}},\ \bibinfo {pages} {072002} (\bibinfo
  {year} {2000})}\BibitemShut {NoStop}%
\bibitem [{\citenamefont {Apollonio}\ \emph {et~al.}(2000)\citenamefont
  {Apollonio} \emph {et~al.}}]{chooz00}%
  \BibitemOpen
  \bibfield  {author} {\bibinfo {author} {\bibfnamefont {M.}~\bibnamefont
  {Apollonio}} \emph {et~al.},\ }\bibfield  {title} {\enquote {\bibinfo {title}
  {{Determination of neutrino incoming direction in the CHOOZ experiment and
  Supernova explosion location by scintillator detectors}},}\ }\href@noop {}
  {\bibfield  {journal} {\bibinfo  {journal} {Phys. Rev. D}\ }\textbf {\bibinfo
  {volume} {61}},\ \bibinfo {pages} {012001} (\bibinfo {year}
  {2000})}\BibitemShut {NoStop}%
\bibitem [{\citenamefont {Leyton}\ \emph {et~al.}(2017)\citenamefont {Leyton},
  \citenamefont {Dye},\ and\ \citenamefont {Monroe}}]{gnudir}%
  \BibitemOpen
  \bibfield  {author} {\bibinfo {author} {\bibfnamefont {M.}~\bibnamefont
  {Leyton}}, \bibinfo {author} {\bibfnamefont {S.}~\bibnamefont {Dye}}, \ and\
  \bibinfo {author} {\bibfnamefont {J.}~\bibnamefont {Monroe}},\ }\bibfield
  {title} {\enquote {\bibinfo {title} {{Exploring the hidden interior of the
  Earth with directional neutrino measurements}},}\ }\href@noop {} {\bibfield
  {journal} {\bibinfo  {journal} {Nat. Commun.}\ }\textbf {\bibinfo {volume}
  {8}},\ \bibinfo {pages} {15989} (\bibinfo {year} {2017})}\BibitemShut
  {NoStop}%
\bibitem [{\citenamefont {Strumia}\ and\ \citenamefont
  {Vissani}(2003)}]{strumia03}%
  \BibitemOpen
  \bibfield  {author} {\bibinfo {author} {\bibfnamefont {A.}~\bibnamefont
  {Strumia}}\ and\ \bibinfo {author} {\bibfnamefont {F.}~\bibnamefont
  {Vissani}},\ }\bibfield  {title} {\enquote {\bibinfo {title} {{Precise
  quasielastic neutrino/nucleon cross-section}},}\ }\href@noop {} {\bibfield
  {journal} {\bibinfo  {journal} {Phys. Lett. B}\ }\textbf {\bibinfo {volume}
  {564}},\ \bibinfo {pages} {42} (\bibinfo {year} {2003})}\BibitemShut
  {NoStop}%
\bibitem [{pdg(2020{\natexlab{a}})}]{pdg2020}%
  \BibitemOpen
  \href@noop {} {}\bibinfo {howpublished}
  {https://pdg.lbl.gov/2020/reviews/rpp2020-rev-phys-constants.pdf} (\bibinfo
  {year} {2020}{\natexlab{a}})\BibitemShut {NoStop}%
\bibitem [{\citenamefont {Fukugita}\ and\ \citenamefont
  {Yanagida}(2003)}]{fukuyana}%
  \BibitemOpen
  \bibfield  {author} {\bibinfo {author} {\bibfnamefont {M.}~\bibnamefont
  {Fukugita}}\ and\ \bibinfo {author} {\bibfnamefont {T.}~\bibnamefont
  {Yanagida}},\ }\href@noop {} {\emph {\bibinfo {title} {Physics of
  Neutrinos}}}\ (\bibinfo  {publisher} {Springer-Verlag},\ \bibinfo {address}
  {Berlin Heidelberg},\ \bibinfo {year} {2003})\BibitemShut {NoStop}%
\bibitem [{\citenamefont {Erler}\ and\ \citenamefont
  {Ramsey-Musolf}(2005)}]{erler05}%
  \BibitemOpen
  \bibfield  {author} {\bibinfo {author} {\bibfnamefont {J.}~\bibnamefont
  {Erler}}\ and\ \bibinfo {author} {\bibfnamefont {M.}~\bibnamefont
  {Ramsey-Musolf}},\ }\bibfield  {title} {\enquote {\bibinfo {title} {{Weak
  mixing angle at low energies}},}\ }\href@noop {} {\bibfield  {journal}
  {\bibinfo  {journal} {{Phys. Rev. D}}\ }\textbf {\bibinfo {volume} {72}},\
  \bibinfo {pages} {073003} (\bibinfo {year} {2005})}\BibitemShut {NoStop}%
\bibitem [{\citenamefont {Patrignani}\ \emph {et~al.}(2016)\citenamefont
  {Patrignani} \emph {et~al.}}]{pdg2016}%
  \BibitemOpen
  \bibfield  {author} {\bibinfo {author} {\bibfnamefont {C.}~\bibnamefont
  {Patrignani}} \emph {et~al.},\ }\bibfield  {title} {\enquote {\bibinfo
  {title} {{2017 Review of Particle Physics}},}\ }\href@noop {} {\bibfield
  {journal} {\bibinfo  {journal} {Chin. Phys. C}\ }\textbf {\bibinfo {volume}
  {40}},\ \bibinfo {pages} {100001} (\bibinfo {year} {2016})}\BibitemShut
  {NoStop}%
\bibitem [{pdg(2020{\natexlab{b}})}]{pdgnuosc2020}%
  \BibitemOpen
  \href@noop {} {}\bibinfo {howpublished}
  {https://pdg.lbl.gov/2020/reviews/rpp2020-rev-neutrino-mixing.pdf} (\bibinfo
  {year} {2020}{\natexlab{b}})\BibitemShut {NoStop}%
\bibitem [{\citenamefont {Dye}(2009)}]{dye09}%
  \BibitemOpen
  \bibfield  {author} {\bibinfo {author} {\bibfnamefont {S.~T.}\ \bibnamefont
  {Dye}},\ }\bibfield  {title} {\enquote {\bibinfo {title} {{Neutrino Mixing
  Discriminates Geo-reactor Models}},}\ }\href@noop {} {\bibfield  {journal}
  {\bibinfo  {journal} {Phys. Lett. B}\ }\textbf {\bibinfo {volume} {679}},\
  \bibinfo {pages} {15} (\bibinfo {year} {2009})}\BibitemShut {NoStop}%
\bibitem [{\citenamefont {{Mueller, Th. A.}}\ \emph {et~al.}(2011)\citenamefont
  {{Mueller, Th. A.}} \emph {et~al.}}]{mueller}%
  \BibitemOpen
  \bibfield  {author} {\bibinfo {author} {\bibnamefont {{Mueller, Th. A.}}}
  \emph {et~al.},\ }\bibfield  {title} {\enquote {\bibinfo {title} {{Improved
  predictions of reactor antineutrino spectra}},}\ }\href@noop {} {\bibfield
  {journal} {\bibinfo  {journal} {Phys. Rev. C}\ }\textbf {\bibinfo {volume}
  {83}},\ \bibinfo {pages} {054615} (\bibinfo {year} {2011})}\BibitemShut
  {NoStop}%
\bibitem [{\citenamefont {Huber}(2011)}]{huber}%
  \BibitemOpen
  \bibfield  {author} {\bibinfo {author} {\bibfnamefont {P.}~\bibnamefont
  {Huber}},\ }\bibfield  {title} {\enquote {\bibinfo {title} {{Determination of
  antineutrino spectra from nuclear reactors}},}\ }\href@noop {} {\bibfield
  {journal} {\bibinfo  {journal} {Phys. Rev. C}\ }\textbf {\bibinfo {volume}
  {84}},\ \bibinfo {pages} {024617} (\bibinfo {year} {2011})}\BibitemShut
  {NoStop}%
\bibitem [{\citenamefont {Kopeikin}\ \emph {et~al.}(2004)\citenamefont
  {Kopeikin} \emph {et~al.}}]{kopeikin_etal}%
  \BibitemOpen
  \bibfield  {author} {\bibinfo {author} {\bibfnamefont {V.~I.}\ \bibnamefont
  {Kopeikin}} \emph {et~al.},\ }\bibfield  {title} {\enquote {\bibinfo {title}
  {{Reactor as a Source of Antineutrinos: Thermal Fission Energy}},}\
  }\href@noop {} {\bibfield  {journal} {\bibinfo  {journal} {Phys. Atom.
  Nucl.}\ }\textbf {\bibinfo {volume} {67}},\ \bibinfo {pages} {1892} (\bibinfo
  {year} {2004})}\BibitemShut {NoStop}%
\bibitem [{pri({\natexlab{a}})}]{prisgloss}%
  \BibitemOpen
  \href@noop {} {}\bibinfo {howpublished}
  {https://www.iaea.org/PRIS/Glossary.aspx} ({\natexlab{a}})\BibitemShut
  {NoStop}%
\bibitem [{mil()}]{mills}%
  \BibitemOpen
  \href@noop {} {}\bibinfo {howpublished} {R. Mills, private communication,
  Dec. 12, 2018}\BibitemShut {NoStop}%
\bibitem [{mch()}]{mchen}%
  \BibitemOpen
  \href@noop {} {}\bibinfo {howpublished} {M. Chen, private communication, Aug.
  14, 2017}\BibitemShut {NoStop}%
\bibitem [{pri({\natexlab{b}})}]{pris}%
  \BibitemOpen
  \href@noop {} {}\bibinfo {howpublished}
  {https://www.iaea.org/resources/databases/power-reactor-information-system-pris}
  ({\natexlab{b}})\BibitemShut {NoStop}%
\bibitem [{\citenamefont {Jaffke}\ and\ \citenamefont
  {Huber}(2017)}]{jaffke_huber17}%
  \BibitemOpen
  \bibfield  {author} {\bibinfo {author} {\bibfnamefont {P.}~\bibnamefont
  {Jaffke}}\ and\ \bibinfo {author} {\bibfnamefont {P.}~\bibnamefont {Huber}},\
  }\bibfield  {title} {\enquote {\bibinfo {title} {{Determining reactor fuel
  type from continuous antineutrino monitoring}},}\ }\href@noop {} {\bibfield
  {journal} {\bibinfo  {journal} {Phys. Rev. Applied}\ }\textbf {\bibinfo
  {volume} {8}},\ \bibinfo {pages} {034005} (\bibinfo {year}
  {2017})}\BibitemShut {NoStop}%
\bibitem [{\citenamefont {Mantovani}\ \emph {et~al.}(2004)\citenamefont
  {Mantovani} \emph {et~al.}}]{mantovani}%
  \BibitemOpen
  \bibfield  {author} {\bibinfo {author} {\bibfnamefont {F.}~\bibnamefont
  {Mantovani}} \emph {et~al.},\ }\bibfield  {title} {\enquote {\bibinfo {title}
  {{Antineutrinos from Earth: A reference model and its uncertainties}},}\
  }\href@noop {} {\bibfield  {journal} {\bibinfo  {journal} {Phys. Rev. D}\
  }\textbf {\bibinfo {volume} {69}},\ \bibinfo {pages} {013001} (\bibinfo
  {year} {2004})}\BibitemShut {NoStop}%
\bibitem [{inf()}]{infn}%
  \BibitemOpen
  \href@noop {} {}\bibinfo {howpublished}
  {https://www.fe.infn.it/antineutrino/}\BibitemShut {NoStop}%
\bibitem [{\citenamefont {Kopeikin}()}]{kopeikin12}%
  \BibitemOpen
  \bibfield  {author} {\bibinfo {author} {\bibfnamefont {t.~.}\ \bibnamefont
  {Kopeikin}, \bibfnamefont {V.~I.}},\ }\href@noop {} {\ }\BibitemShut
  {NoStop}%
\bibitem [{nis()}]{nist}%
  \BibitemOpen
  \href@noop {} {}\bibinfo {howpublished}
  {http://physics.nist.gov/cuu/Constants/index.html}\BibitemShut {NoStop}%
\bibitem [{\citenamefont {Preston}(1962)}]{pres62}%
  \BibitemOpen
  \bibfield  {author} {\bibinfo {author} {\bibfnamefont {M.~A.}\ \bibnamefont
  {Preston}},\ }\href@noop {} {\emph {\bibinfo {title} {Physics of the
  Nucleus}}}\ (\bibinfo  {publisher} {Addison-Wesley},\ \bibinfo {address}
  {Reading, Mass.},\ \bibinfo {year} {1962})\BibitemShut {NoStop}%
\bibitem [{\citenamefont {Enomoto}()}]{sanshiro_spec}%
  \BibitemOpen
  \bibfield  {author} {\bibinfo {author} {\bibfnamefont {S.}~\bibnamefont
  {Enomoto}},\ }\href@noop {} {\enquote {\bibinfo {title} {{Geoneutrino Spectra
  and Luminosity}},}\ }\bibinfo {howpublished}
  {https://www.awa.tohoku.ac.jp/~sanshiro/research/geoneutrino/spectrum/}\BibitemShut
  {NoStop}%
\bibitem [{\citenamefont {McDonough}\ and\ \citenamefont {Sun}(1995)}]{mcd95}%
  \BibitemOpen
  \bibfield  {author} {\bibinfo {author} {\bibfnamefont {W.~F.}\ \bibnamefont
  {McDonough}}\ and\ \bibinfo {author} {\bibfnamefont {S.-s.}\ \bibnamefont
  {Sun}},\ }\bibfield  {title} {\enquote {\bibinfo {title} {{The composition of
  the Earth}},}\ }\href@noop {} {\bibfield  {journal} {\bibinfo  {journal}
  {Chem. Geol.}\ }\textbf {\bibinfo {volume} {120}},\ \bibinfo {pages} {223}
  (\bibinfo {year} {1995})}\BibitemShut {NoStop}%
\bibitem [{\citenamefont {Huang}\ \emph {et~al.}(2013)\citenamefont {Huang}
  \emph {et~al.}}]{huang13}%
  \BibitemOpen
  \bibfield  {author} {\bibinfo {author} {\bibfnamefont {Y.}~\bibnamefont
  {Huang}} \emph {et~al.},\ }\bibfield  {title} {\enquote {\bibinfo {title} {{A
  reference Earth model for the heat producing elements and associated
  geoneutrino flux}},}\ }\href@noop {} {\bibfield  {journal} {\bibinfo
  {journal} {Geochem., Geophys., Geosyst.}\ }\textbf {\bibinfo {volume} {14}},\
  \bibinfo {pages} {2003} (\bibinfo {year} {2013})}\BibitemShut {NoStop}%
\bibitem [{\citenamefont {Dziewonski}\ and\ \citenamefont
  {Anderson}(1981)}]{dziew81}%
  \BibitemOpen
  \bibfield  {author} {\bibinfo {author} {\bibfnamefont {A.~M.}\ \bibnamefont
  {Dziewonski}}\ and\ \bibinfo {author} {\bibfnamefont {D.~L.}\ \bibnamefont
  {Anderson}},\ }\bibfield  {title} {\enquote {\bibinfo {title} {{Preliminary
  Reference Earth Model (PREM)}},}\ }\href@noop {} {\bibfield  {journal}
  {\bibinfo  {journal} {Phys. Earth Planet. Inter.}\ }\textbf {\bibinfo
  {volume} {25}},\ \bibinfo {pages} {297} (\bibinfo {year} {1981})}\BibitemShut
  {NoStop}%
\bibitem [{\citenamefont {Krauss}\ \emph {et~al.}(1984)\citenamefont {Krauss},
  \citenamefont {Glashow},\ and\ \citenamefont {Schramm}}]{kgs84}%
  \BibitemOpen
  \bibfield  {author} {\bibinfo {author} {\bibfnamefont {L.}~\bibnamefont
  {Krauss}}, \bibinfo {author} {\bibfnamefont {S.}~\bibnamefont {Glashow}}, \
  and\ \bibinfo {author} {\bibfnamefont {D.}~\bibnamefont {Schramm}},\
  }\bibfield  {title} {\enquote {\bibinfo {title} {{Antineutrino Astronomy and
  Geophysics}},}\ }\href@noop {} {\bibfield  {journal} {\bibinfo  {journal}
  {{Nature}}\ }\textbf {\bibinfo {volume} {310}},\ \bibinfo {pages} {191}
  (\bibinfo {year} {1984})}\BibitemShut {NoStop}%
\bibitem [{\citenamefont {Jocher}()}]{jocher}%
  \BibitemOpen
  \bibfield  {author} {\bibinfo {author} {\bibfnamefont {G.}~\bibnamefont
  {Jocher}},\ }\href@noop {} {\enquote {\bibinfo {title} {{Ultralytics
  Worldwide Reactor Database}},}\ }\bibinfo {howpublished}
  {http://www.ultralytics.com/agm2015}\BibitemShut {NoStop}%
\bibitem [{\citenamefont {Winter}\ \emph {et~al.}(2006)\citenamefont {Winter}
  \emph {et~al.}}]{winter06}%
  \BibitemOpen
  \bibfield  {author} {\bibinfo {author} {\bibfnamefont {W.~T.}\ \bibnamefont
  {Winter}} \emph {et~al.},\ }\bibfield  {title} {\enquote {\bibinfo {title}
  {{The $^8$B neutrino spectrum}},}\ }\href@noop {} {\bibfield  {journal}
  {\bibinfo  {journal} {Phys. Rev. C}\ }\textbf {\bibinfo {volume} {73}},\
  \bibinfo {pages} {025503} (\bibinfo {year} {2006})}\BibitemShut {NoStop}%
\bibitem [{\citenamefont {Abe}\ \emph {et~al.}(2016)\citenamefont {Abe} \emph
  {et~al.}}]{abe2016}%
  \BibitemOpen
  \bibfield  {author} {\bibinfo {author} {\bibfnamefont {K.}~\bibnamefont
  {Abe}} \emph {et~al.},\ }\bibfield  {title} {\enquote {\bibinfo {title}
  {{Solar neutrino measurements in Super-Kamiokande-IV}},}\ }\href@noop {}
  {\bibfield  {journal} {\bibinfo  {journal} {Phys. Rev. D}\ }\textbf {\bibinfo
  {volume} {94}},\ \bibinfo {pages} {052010} (\bibinfo {year}
  {2016})}\BibitemShut {NoStop}%
\bibitem [{noa()}]{noaa}%
  \BibitemOpen
  \href@noop {} {}\bibinfo {howpublished}
  {https://www.esrl.noaa.gov/gmd/grad/solcalc/calcdetails.html}\BibitemShut
  {NoStop}%
\end{thebibliography}%

\end{document}